\documentclass[11pt,a4paper]{article}

\usepackage{fancyhdr, graphicx}

\usepackage{graphicx}
\usepackage{subfigure}
\usepackage{amsmath}
\usepackage{psfrag}
\usepackage[Gray,squaren]{SIunits}
\usepackage{cite}
\usepackage{amssymb}
\usepackage{xspace}
\usepackage{hyperref}
\usepackage{url}
\usepackage{mathrsfs}

\newcommand{\mc}{\cal{ }}

\newcommand{\bef}{\begin{figure}[:ht]}
\newcommand{\enf}{\end{figure}}
\newcommand{\bec}{\begin{center}}
\newcommand{\enc}{\end{center}}
\newcommand{\beq}{\begin{equation}}
\newcommand{\eeq}{\end{equation}}

\newcommand{\etal}{{\it et al.}}

\def\MeV{\ifmmode {\mathrm{\ Me\kern -0.1em V}}\else
                   \textrm{Me\kern -0.1em V}\fi}
\def\GeV{\ifmmode {\mathrm{\ Ge\kern -0.1em V}}\else
                   \textrm{Ge\kern -0.1em V}\fi}
\def\TeV{\ifmmode {\mathrm{\ Te\kern -0.1em V}}\else
                   \textrm{Te\kern -0.1em V}\fi}

\newcommand\units{\,\mathrm}

\newcommand{\ele}{1}
\newcommand{\pos}{2}
\newcommand{\varEv}{\ensuremath{i}\xspace}
\newcommand{\parset}[1]{\ensuremath{[p]^{#1}}}
\newcommand{\chisquare}{\ensuremath{\chi^{2}}\xspace}

\newcommand{\conv}{\ensuremath{\otimes}}

\newcommand{\Eele}{\ensuremath{E_{\ele}}\xspace}
\newcommand{\Epos}{\ensuremath{E_{\pos}}\xspace}

\newcommand{\epem}{\ensuremath{\mathrm{e}^{+}\mathrm{e}^{-}}\xspace}

\newcommand{\lumispec}[1]{\ensuremath{\mathcal{L}\bigl(#1\bigr)}\xspace}
\newcommand{\lumispecNoArg}{\ensuremath{\mathcal{L}}\xspace}
\newcommand{\rootsnom}{\ensuremath{\sqrt{\smash[b]{s_{\mathrm{nom}}}}}\xspace}
\newcommand{\rootsaco}{\ensuremath{\sqrt{\smash[b]{s^{\raisebox{-0.8pt}{$\tiny\smash\prime$}}_{\smash[t]{\mathrm{acol}}}}}}\xspace}

\newcommand{\Det}[1]{\ensuremath{\mathrm{D}\bigl(#1\bigr)}\xspace}
\newcommand{\ISR}[1]{\ensuremath{\mathrm{ISR}\bigl(#1\bigr)}\xspace}
\newcommand{\FSR}[1]{\ensuremath{\mathrm{FSR}\bigl(#1\bigr)}\xspace}

\newcommand{\guineapig}{\textsc{GuineaPig}\xspace}
\newcommand{\bhwide}{\textsc{BHWide}\xspace}
\newcommand{\geant}{\textsc{Geant4}\xspace}

\newcommand{\fbinv}{\ensuremath{\mathrm{fb}^{-1}}\xspace}

\def\MeV{\ifmmode {\mathrm{\ Me\kern -0.1em V}}\else
                   \textrm{Me\kern -0.1em V}\fi}
\def\GeV{\ifmmode {\mathrm{\ Ge\kern -0.1em V}}\else
                   \textrm{Ge\kern -0.1em V}\fi}
\def\TeV{\ifmmode {\mathrm{\ Te\kern -0.1em V}}\else
                   \textrm{Te\kern -0.1em V}\fi}

\usepackage{epsfig}
\date{\today}

\begin{document}

\nopagebreak

\newcommand{\prepnum}[1]{\gdef\@prepnum{#1}}
\prepnum{\bf LC-DET-2013-029}

\title{
  \includegraphics[width=0.15\textwidth]{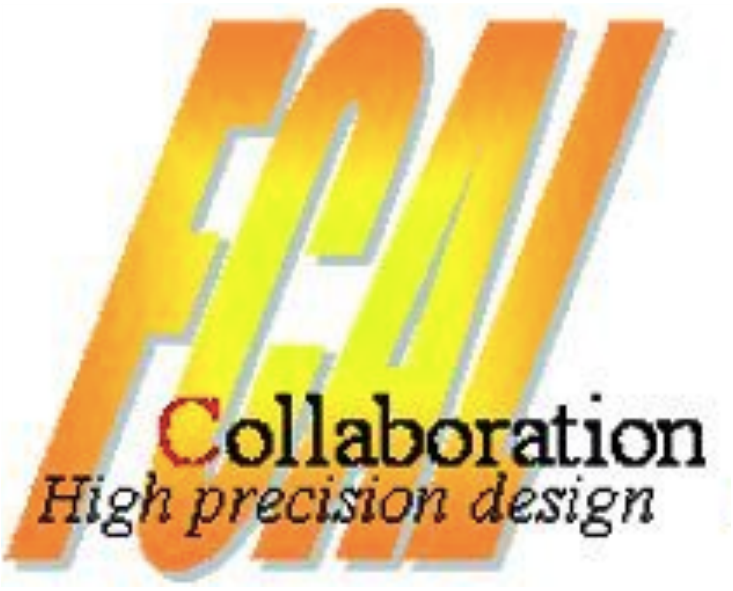}
  \hfill\@prepnum\\[2ex]
  \Huge  \bf {ECFA Detector R\&D Panel \\ Review Report }
}

\author{\Large \bf {The FCAL Collaboration}}

\date{\bf June 2013}

\maketitle
\thispagestyle{empty}

\abstract{Two special calorimeters are foreseen for the instrumentation of
the very forward region of an ILC or CLIC detector;  
a luminometer (LumiCal) designed to measure the rate of
low angle Bhabha scattering events with a precision better than 10$^{-3}$ at the ILC and 10$^{-2}$ at CLIC, 
and a low polar-angle calorimeter (BeamCal). The latter will be hit by a large amount 
of beamstrahlung remnants. The intensity and the spatial shape of these
depositions will provide a fast luminosity estimate,
as well as determination of beam parameters. The sensors
of this calorimeter must be radiation-hard.
Both devices will improve the e.m. hermeticity of the detector in the search for new particles.
Finely segmented and very compact electromagnetic calorimeters will match these
requirements. Due to the high occupancy, fast front-end electronics will be needed.
Monte Carlo studies were performed to investigate the impact of beam-beam interactions and 
physics background processes
on the luminosity measurement, and of  beamstrahlung on the performance of BeamCal, as well as to optimise the design of both calorimeters. 
Dedicated sensors, front-end and ADC ASICs have been designed for the ILC
and prototypes are available. Prototypes of sensor planes fully assembled with readout electronics have been 
studied in electron beams. }

\newpage

\begin{center}
{        \Large \bf {The FCAL Collaboration}      }
\end{center}

H. Abramowicz$^a$, A. Abusleme$^b$, K. Afanaciev$^c$, J. Aguilar$^d$,  E. Alvarez$^b$, P. Bambade$^e$, L. Bortko$^{f,1}$, I. Bozovic-Jelisavcic$^g$,
E. Castro$^f$, G. Chelkov$^h$, C. Coca$^i$, W. Daniluk$^j$, A. Dragone$^k$, L. Dumitru$^i$, K. Elsener$^l$, I. Emeliantchik$^c$, E. Firu$^r$, J. Fischer$^{f,5}$, T. Fiutowski$^d$, V. Ghenescu$^r$, M. Gostkin$^h$, 
G. Grzelak$^{j,3}$, G. Haller$^k$, H. Henschel$^f$, A. Ignatenko$^{c,1,4}$, M. Idzik$^d$, K. Ito$^m$,  S. Kananov$^a$, E. Kielar$^j$, S. Kollowa$^{f,1}$, J. Kotula$^j$, Z. Krumstein$^h$, B. Krupa$^j$, S. Kulis$^{d,l}$, W. Lange$^f$, A. Levy$^a$, I. Levy$^a$, L. Linssen$^l$, 
W. Lohmann$^{f,1}$,  S. Lukic$^g$, J. Moron$^d$, A. Moszczynski$^j$, U. Nauenberg$^n$, A. Neagu$^r$, O. Novgorodova$^{f,1}$, F.-X. Nuiry$^l$,
M. Ohlerich$^{f,1}$, M. Orlandea$^i$, G. Oleinik$^n$, K. Oliwa$^j$,
A. Olshevski$^h$, M. Pandurovic$^g$, B. Pawlik$^j$, T. Preda$^r$, D. Przyborowski$^d$, Y. Sato$^m$, I. Sadeh$^a$, A. Sailer$^l$, B. Schumm$^o$, S. Schuwalow$^{f,6}$, R. Schwartz$^a$, 
I. Smiljanic$^g$, K. Swientek$^d$, Y. Takubo$^m$, E. Teodorescu$^i$, W. Wierba$^j$, H. Yamamoto$^m$, L. Zawiejski$^j$, T.-S. Zgura$^r$  and J. Zhang$^p$
\\
\\
\noindent
a Tel Aviv University, Tel Aviv, Israel \\
b Pontifica Universidad Catolica de Chile, Chile \\
c NCPHEP, Minsk, Belarus \\
d AGH University of Science \& Technology, Cracow, Poland \\
e Laboratoire de l Accelerateur Lineaire, Orsay, France \\
f DESY, Zeuthen, Germany \\
g Vinca Institute of Nuclear Sciences, University of Belgrade, Serbia \\
h JINR, Dubna, Russia \\
i IFIN-HH, Bucharest, Romania \\
j INP PAN, Cracow, Poland \\
k SLAC, Menlo Park, U.S.A. \\
l CERN, Geneva, Switzerland \\
m Tohoku University, Sendai, Japan \\
n University of Colorado, Boulder, U.S.A. \\
o UC California, Santa Cruz, U.S.A. \\
p ANL, Argonne, U.S.A. \\
r ISS, Bucharest, Romania \\
\\
1 Also at Brandenburg University of Technology, Cottbus, Germany \\
2 Now at BTO Consulting AG, Berlin, Germany \\
3 Also at University of Warsaw, Poland \\
4  Now at DESY, Hamburg, Germany \\
5 Also Humboldt University, Berlin, Germany \\
6 Also Hamburg University, Hamburg, Germany

\newpage

\tableofcontents

\section{Introduction and challenges}

A high energy e$^+$e$^-$ linear collider 
is considered to be the future research facility complementary to the LHC collider.
Whereas LHC has "a priori" a higher potential for discoveries, an $e^+e^-$ collider will allow
precision measurements to explore in detail the mechanism of 
electroweak symmetry breaking and the properties of the physics beyond the Standard Model, should it be found at the LHC.
Two concepts of an $e^+e^-$ linear collider are presently considered, the ILC~\cite{ILC_pub}
and CLIC~\cite{clic_info}.
The centre-of-mass energy of the ILC, using superconducting cavities, is expected to be
500~GeV \footnote{Currently discussions are ongoing to operate the ILC as a Higgs factory at lower energy} , with a possibility of an upgrade to 1~TeV. CLIC is based on 
normal-conducting cavities where the RF power for the acceleration of the colliding beams is extracted from a high-current drive beam that runs parallel with the main linac. CLIC will allow to collide electrons and positrons up to  energies of 3~TeV.  
An R\&D program is ongoing to develop detector technologies
for precision measurements in this new energy domain. 

Two detectors, 
the ILD~\cite{ILC_pub,ILD_pub} and the
SiD~\cite{ILC_pub,SiD_pub}, were reviewed and validated for the ILC. Detailed Baseline Design Reports have recently been published.
Similar detector designs are proposed for CLIC. 
In both detector designs,
two specialised calorimeters are foreseen in the very forward region; LumiCal for the precise measurement
of the luminosity, and BeamCal for a fast estimate of the 
luminosity and for the control of beam parameters~\cite{ieee1}. 
Both will also improve the hermeticity of the detector, important e.g. for new-particle searches with 
missing energy signature~\cite{drugakov}. 
To support beam-tuning, an additional pair-monitor will be positioned just 
in front of BeamCal. In the following, mostly the design of the ILC option at 500 GeV will be presented.

The luminosity is a key parameter of the colliders. The precision of its measurement translates directly into
the uncertainty for all cross section measurements. 
With LumiCal, 
the luminosity will be measured using
Bhabha 
scattering, $e^+e^- \rightarrow e^+e^- (\gamma)$, as a gauge process.
Bhabha scattering at low polar angles is dominated by QED processes and can be calculated precisely~\cite{new_Bhabha}. 
To match the physics benchmarks,
an 
accuracy of better than
10$^{-3}$ is needed at a centre-of-mass energy of 500 GeV and better than 10$^{-2}$ at 3 TeV~\cite{ILD_pub}. 
To reach these accuracies,
a precision device is needed,
with particularly challenging  
requirements on the mechanics, sensors, read-out 
and position control.

The BeamCal is positioned
at the edge of the detector adjacent to the beam pipe. 
At the energies of these linear colliders,  there is a new phenomenon to tackle -- the
beamstrahlung.
When electron and positron bunches
collide, 
the particles are accelerated in the magnetic field of the bunches
towards the bunch centres. This is the so-called pinch effect that enhances the 
luminosity. However, 
electrons and positrons will radiate photons. A fraction of these 
photons convert in the Coulomb field of
the bunch particles creating low energy $e^+e^-$ pairs. 
A considerable fraction of these pairs 
will deposit their energy after each bunch crossing
in the BeamCal. These deposits, useful for a 
bunch-by-bunch luminosity estimate and determination of 
beam parameters~\cite{grah1}, 
will lead, however, to a 
radiation dose of about one MGy per year in the sensors
at low polar angles.
Hence radiation-hard sensors are needed for the instrumentation of the
BeamCal. 

Optionally, the BeamCal may be supplemented by a pair monitor, consisting of a layer 
of pixel sensors positioned 
just in front of it to measure the density of 
beamstrahlung pairs and
give additional information for the beam parameter determination.

All detectors in the very forward region have to tackle relatively high 
occupancy, requiring special front-end electronics.

A small Moli\`{e}re radius is of importance for both calorimeters. 
It ensures high energy
electron veto capability for BeamCal even
at small polar angles. 
In LumiCal the precise reconstruction of electron, positron and photon 
showers in Bhabha events
is facilitated.
Both calorimeters also shield the inner tracking detectors from back-scattered particles
induced by beamstrahlung pairs hitting the downstream beam-pipe and magnets.
\begin{figure}[h]
\centerline{\includegraphics[width=0.8\columnwidth]{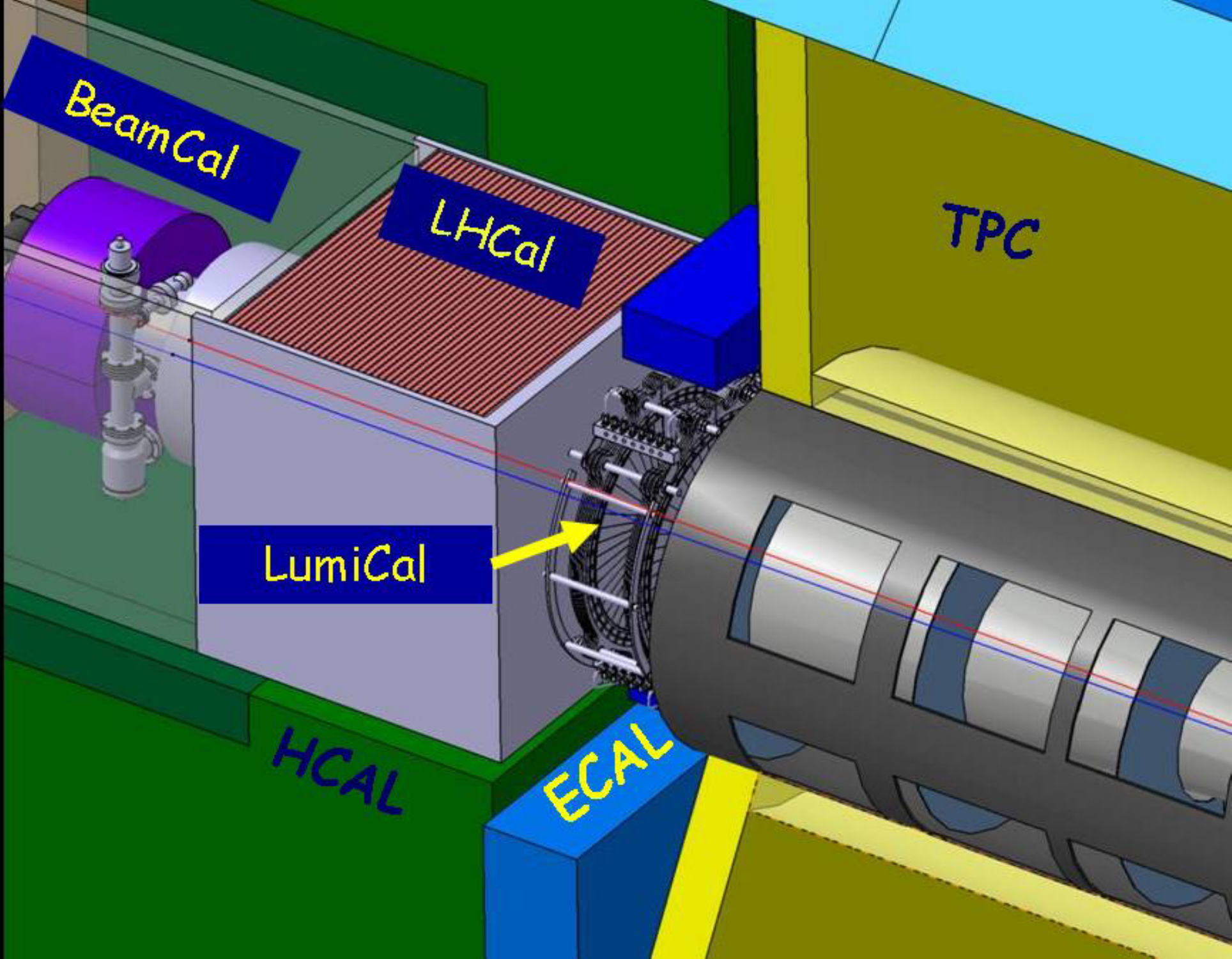}}
\caption{The very forward region of the ILD detector. LumiCal, BeamCal and LHCal are carried by 
the support tube for the final focusing quadrupole and the beam-pipe. 
LHCal  extends the coverage of the hadron calorimeter 
to the polar angle range of LumiCal.
TPC denotes the central track chamber, ECAL the electromagnetic and 
HCAL the hadron calorimeter.\label{fig:Forward_structure}}
\end{figure}

A sketch of the very forward region of the ILD detector~\cite{ILD_pub}
is 
shown in 
Figure~\ref{fig:Forward_structure}. 
LumiCal and BeamCal are designed as cylindrical 
sensor-tungsten sandwich
electromagnetic calorimeters.
Both consist of 30 absorber disks of 3.5 mm thickness, each corresponding to
one radiation length, interspersed with sensor 
layers.
Each sensor layer is segmented radially and azimuthally 
into pads.
 Front-end ASICs are positioned at the outer radius
of the calorimeters.
 LumiCal is positioned inside and aligned with the 
end-cap electromagnetic calorimeter.
BeamCal is placed just in front of the final focus quadrupole.
BeamCal covers polar angles between 5 and 40 mrad and LumiCal between
31 and 77 mrad. 

Colliding beams enter the interaction point, IP, with a crossing angle of 
14 mrad at the ILC and 20 mrad at CLIC Both calorimeters are centred around the outgoing beam.

\section{LumiCal simulation studies} 

The differential cross section of 
Bhabha scattering,
$\frac{d\sigma_{\mathrm{B}}}{d\theta}$,
can be calculated precisely 
from theory~\cite{new_Bhabha}.
In leading order it reads,
\begin{equation}{
\frac{d\sigma_{\mathrm{B}}}{d\theta} =
\frac{2\pi \alpha^{2}_{\rm{em}}}{s} \frac{\sin \theta}{\sin ^{4}(\theta / 2)} \approx 
\frac{32\pi \alpha^{2}_{\rm{em}}}{s} \frac{1}{\theta^{3}} \; ,
}\label{bhabhaXs2EQ} \end{equation}
where $\theta$ is the polar angle of the scattered electron with respect to the beam.
The approximation holds at small $\theta$.

\begin{figure}[h]
\centering
\subfigure[]{
\label{fig:bhabhaXSFIG1}
\includegraphics[width=.45\textwidth]{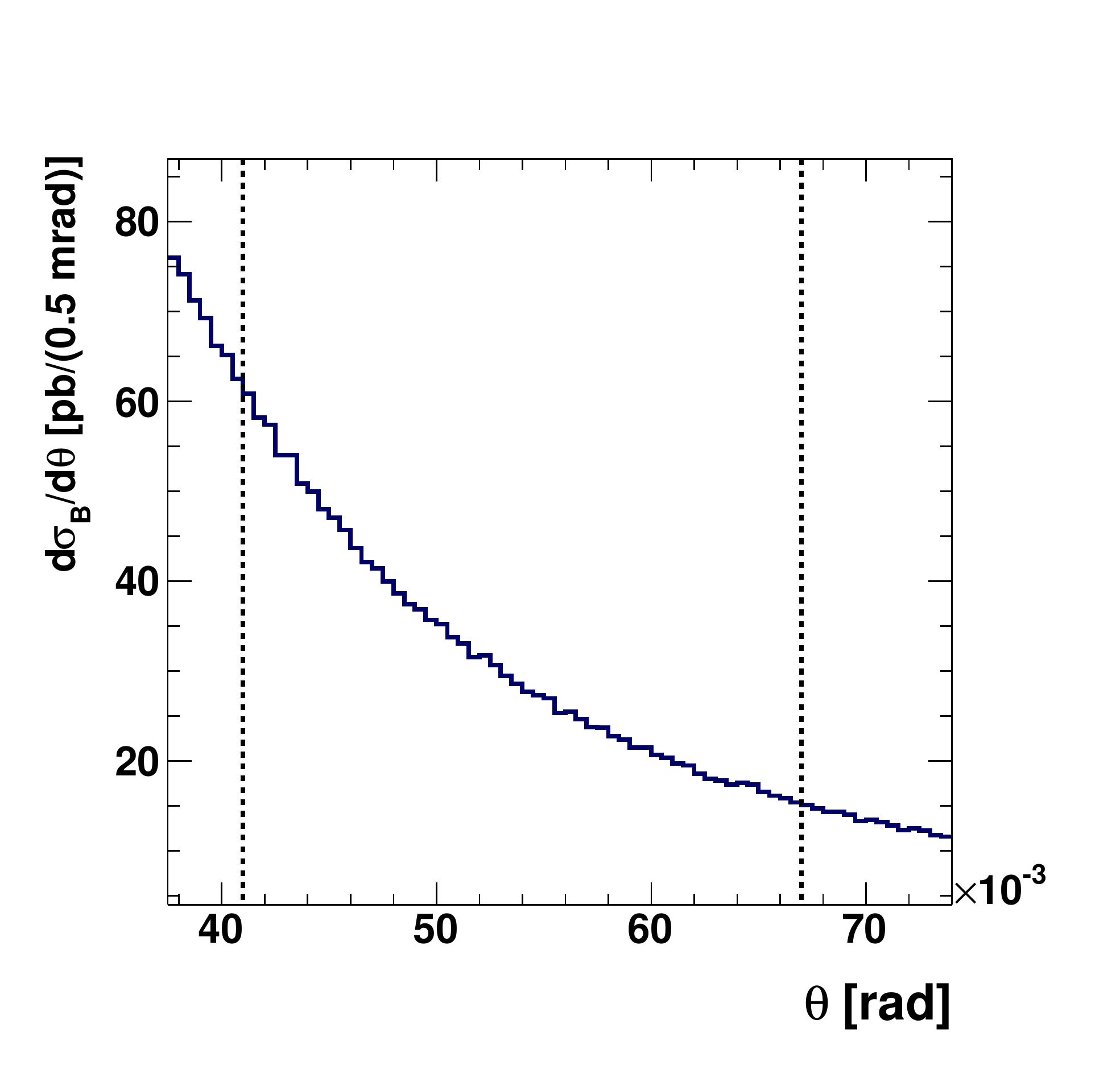}}
\subfigure[]{
\label{fig:bhabhaXSFIG2}
\includegraphics[width=.45\textwidth]{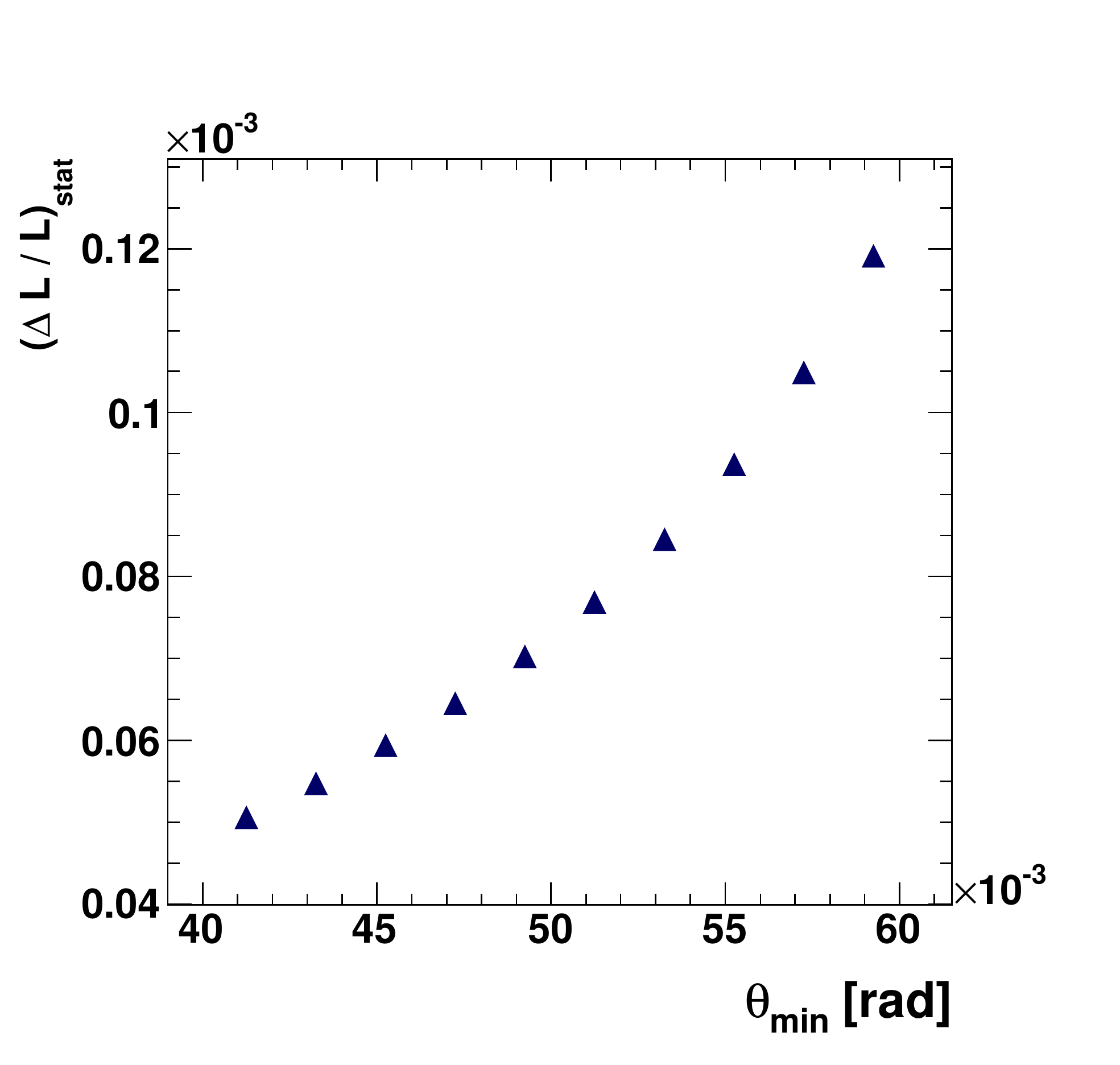}}
\caption{\Subref{fig:bhabhaXSFIG1} Dependence of $d\sigma_{\mathrm{B}}/d\theta$, the differential Bhabha 
cross-section, on the polar angle, $\theta$, at $\sqrt{s}=500$~GeV. 
The dashed lines mark the fiducial volume of LumiCal, $41<\theta^{f}<67$~mrad. 
\Subref{fig:bhabhaXSFIG2} Dependence of the statistical uncertainty in counting the number of 
 Bhabha events, $(\Delta \sf{L} / \sf{L})_{stat}$, on the minimal polar angle, $\theta_{\rm{min}}$, 
which defines the lower edge of the fiducial volume.
The upper limit is kept at 67 mrad. An integrated luminosity of 500 fb$^{-1}$ is assumed.}
\label{fig:bhabhaXSFIG}
\end{figure}
For a given rate of Bhabha events, $N_{\mathrm{B}}$, determined in a certain 
$\theta$-range,
the luminosity, $\sf{L}$, is obtained as
\begin{equation}{
{\sf{L}} = \frac{{N}_{\mathrm{B}}}{\sigma_{\mathrm{B}}},
}\label{lumiDefEQ} \end{equation}
where $\sigma_{\mathrm{B}}$ is the integral of the differential cross section, eqn.~(\ref{bhabhaXs2EQ}),
over the considered $\theta$-range. 
Because of the steep $\theta$ dependence of the cross section, 
as illustrated in   
Figure~\ref{fig:bhabhaXSFIG1}, 
the most critical quantity to control when counting 
Bhabha events 
is the inner acceptance radius 
of the calorimeter, defined as the lower cut in the polar 
angle, $\theta_{\rm{min}}^{\rm{f}}$. 
Hence a very precise  $\theta$ measurement is needed.
Furthermore, the $\theta$-range must be chosen such that
the
number of Bhabha events measured
provides the required relative
statistical uncertainty of 10$^{-3}$. 
By choosing the lower bound of the polar angle between 
40 and 60 mrad the latter requirement can be easily reached
as illustrated in Figure~\ref{fig:bhabhaXSFIG2}. 
Here a Bhabha event sample generated with the BHWIDE generator~\cite{BHWIDE}
was used. 
The generated sample corresponds to an integrated luminosity of 500 fb$^{-1}$.

Electromagnetic showers are simulated in LumiCal using the GEANT4~\cite{g4} based package 
Mokka~\cite{mokka}. Sensors consist of 300 $\mu$m thick silicon sectors covering an azimuthal 
angle of 30$^\circ$.
The depositions in each sensor pad are recorded, and a reconstruction of the 
shower is performed.
The position of an electromagnetic shower in LumiCal is reconstructed by performing 
a weighted average over the 
energy deposits in
 individual pads. 
The weight, ${\mc{W}}_{{i}}$, of a given detector pad i is 
determined by logarithmic weighting~\cite{bib17},  for which ${\mc{W}}_{{i}} = \mathrm{max} \{~ 0 ~,~ \mc{C} + \ln ({\mc{E}}_{{i}}
/ {{E}}_{\rm{tot}} ~) \}$. 
Here ${\mc{E}}_{{i}}$ refers to the individual pad energy, 
${\mc{E}}_{\rm{tot}}$ is the total energy in all pads, and $\mc{C}$ is a constant. 
In this way,
only pads which contain a sufficient fraction of the shower energy 
contribute to the reconstruction. 
The polar angle resolution, $\sigma_{\theta}$, and a polar angle measurement bias, 
$\Delta \theta$, are defined as the Gaussian width and 
the central value of the difference 
between the reconstructed and the generated polar angles. 
There is an optimal value 
for $\mc{C}$, for which $\sigma_{\theta}$ is minimal~\cite{bib20,bib23}. 

Non-zero values of $\Delta \theta$ are due to the non-linear
signal sharing on finite size pads  with gaps between them. 
The bias and the resolution in the polar angle measurement depend on the polar angle pad size. 
The bias causes a shift in the 
luminosity measurement, since events may migrate into or out 
of the fiducial volume. This shift reads as
\begin{equation}
\left( \frac{\Delta\sf{L}}{\sf{L}} \right)_{\rm{rec}} \approx 2 \frac{\Delta \theta}{\theta_{\rm{min}}^{\rm{f}}} \;.
\label{luminosityRelativeErrRec2EQ} 
\end{equation}
\begin{figure}[htpb]
\begin{center}
\subfigure[]{
\label{lumiBiasThetaRecFIG2}
\includegraphics[clip,width=.45\textwidth]{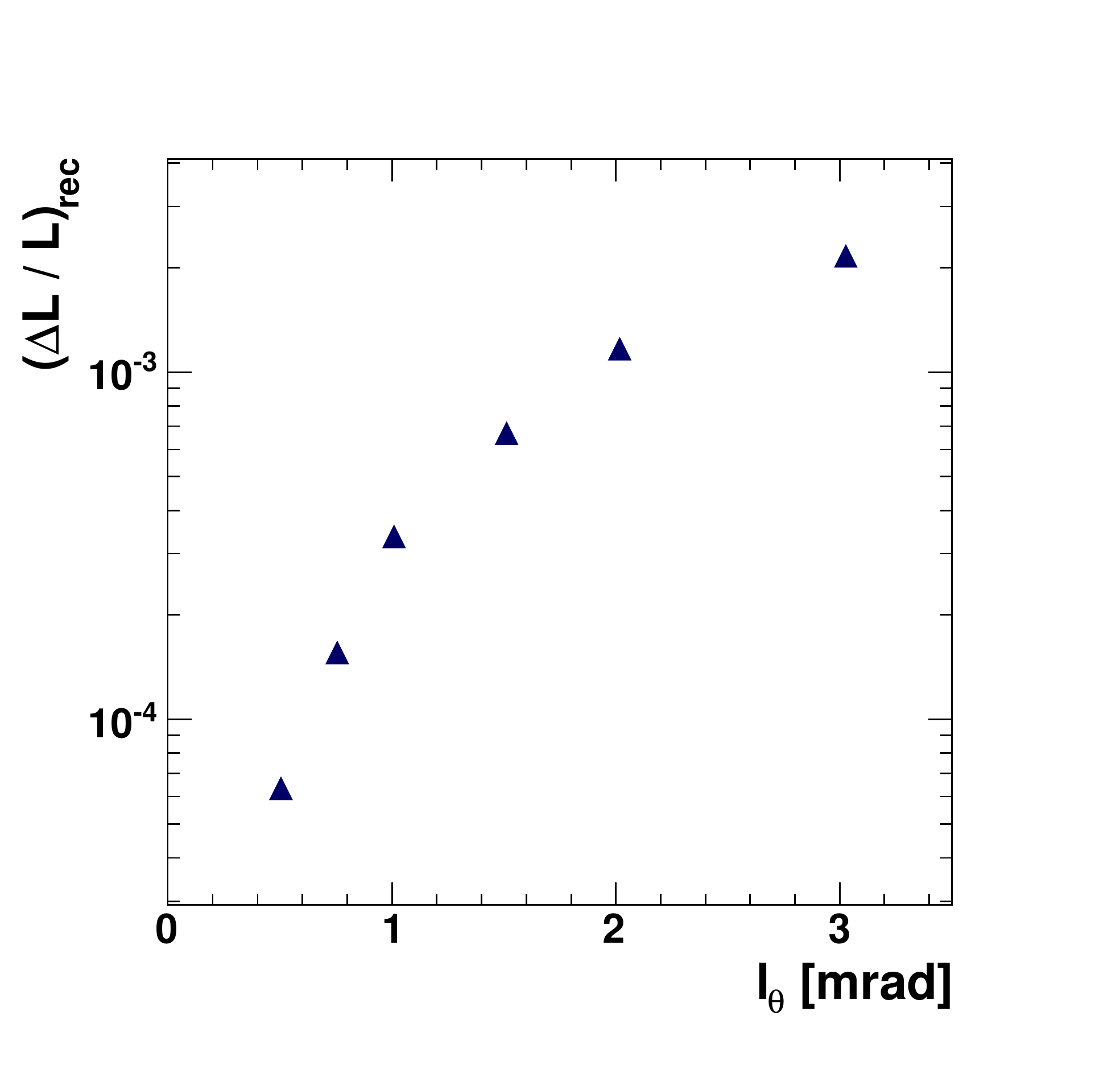}} 
\subfigure[]{
\label{engyResThetaMinMaxFIG1}
\includegraphics[clip,width=.5\textwidth,height=.43\textwidth]{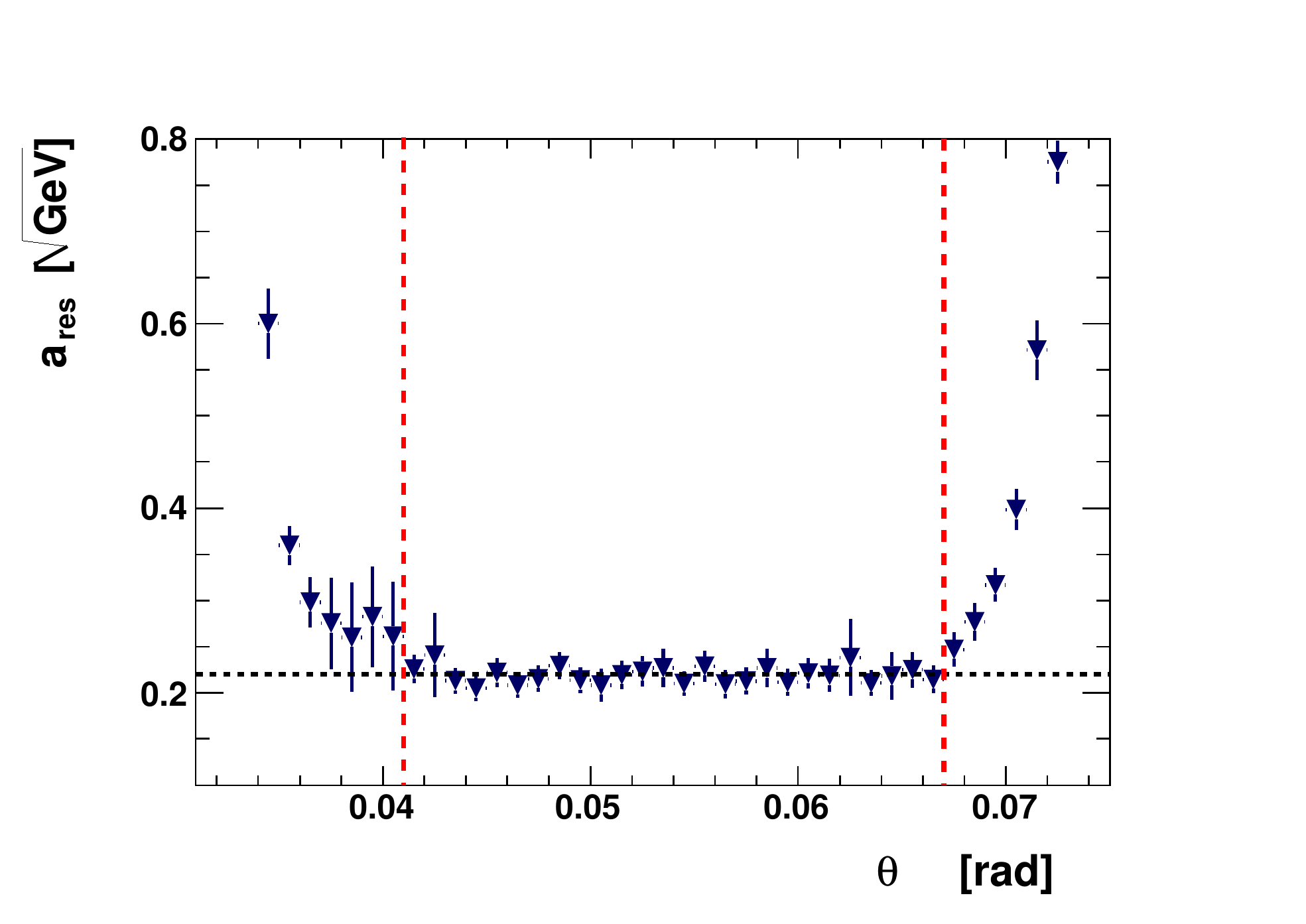}}
\caption{\Subref{lumiBiasThetaRecFIG2} 
Dependence of $\left( \Delta\sf{L} / \sf{L} \right)_{\rm{rec}}$, 
as defined in eqn.~(2.3), on 
the polar angle pad size, l$_{\theta}$. 
\Subref{engyResThetaMinMaxFIG1} 
The energy resolution, a$_{\rm{res}}$, for 250~GeV 
electrons as a function of the polar angle, $\theta$,
covering the polar angle range of the LumiCal.}
\end{center}
\end{figure}
Figure~\ref{lumiBiasThetaRecFIG2} 
shows the relative shift in the luminosity as a function 
of the polar 
angular pad size, l$_{\theta}$, using the optimal value of $\mc{C}$. 
For l$_{\theta} < 2$~mrad 
the shift in the luminosity measurement is smaller than $10^{-3}$. 
As the baseline for the design we have chosen
l$_{\theta} = 0.8$~mrad, 
which corresponds to 64~radial divisions of the sensor. 
For this segmentation
the polar angle
resolution and bias amount 
to $\sigma_{\theta} = (2.2\, \pm\, 0.01) \times 10^{-2}$
and $\Delta \theta = (3.2\, \pm\, 0.1) \times 10^{-3}$~mrad, respectively.
The relative shift in 
the luminosity is $\left( \Delta\sf{L} / \sf{L} \right)_{\rm{rec}} = 1.6 \times 10^{-4}$.

\begin{figure}[h]
\centerline{\includegraphics[width=0.8\columnwidth]{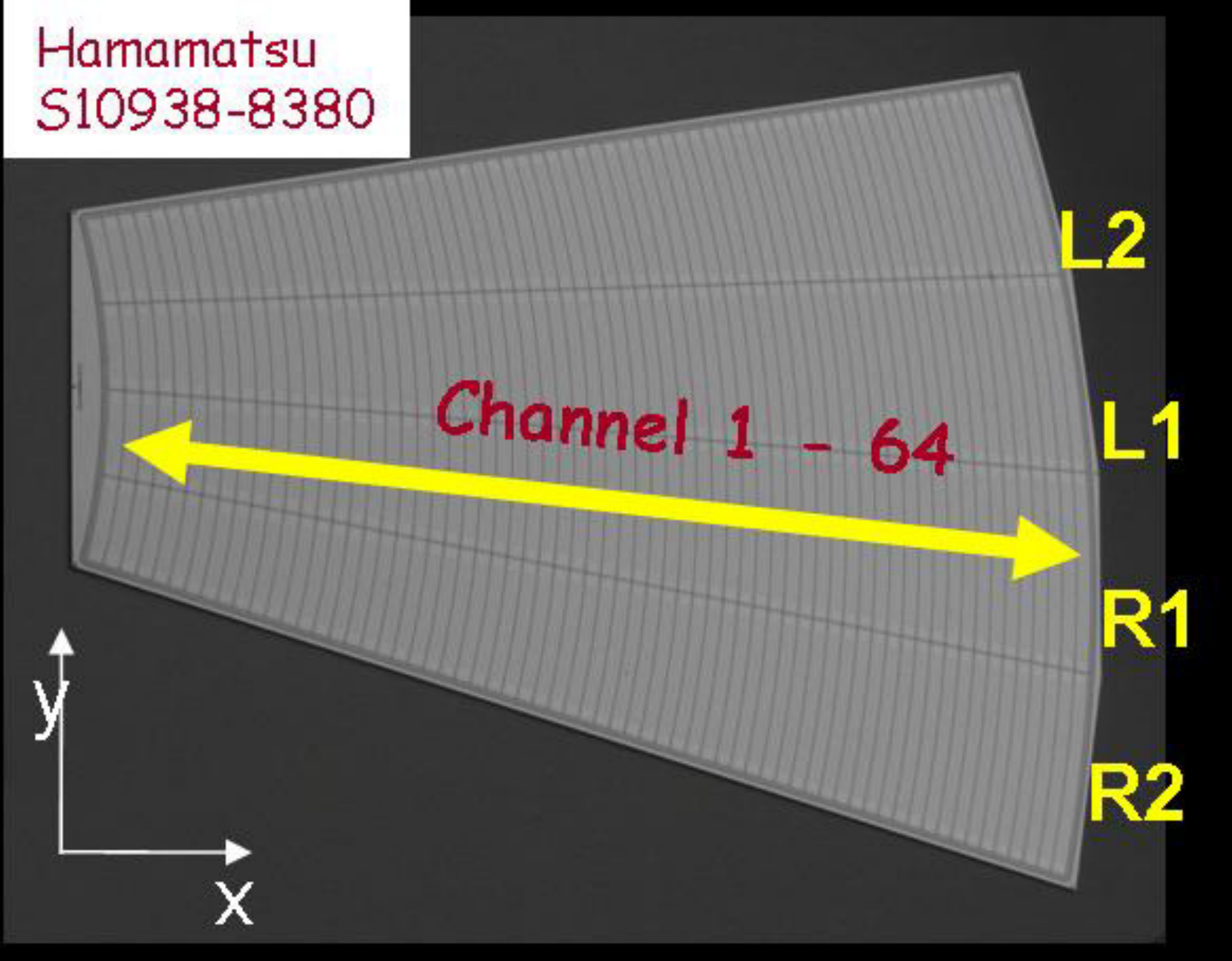}}
\caption{The layout of a sensor plane for LumiCal.\label{fig:sensor layout}}
\end{figure}

The azimuthal segmentation is less crucial for the measurement of luminosity and was set to 48 sectors. The layout of a sensor plane is shown in Figure~\ref{fig:sensor layout}. Mechanically, only four azimuthal sectors fit on a single silicon-tile and the tiles are separated by an uninstrumented gap of $2.5 \units{mm}$. 

The polar angle bias needs careful understanding
in
test-beam measurements 
with
sensors finally chosen for the calorimeter. 
Once its value is known,  a correction can be applied to the luminosity 
measurement.
The 
uncertainty of the luminosity measurement is then given by the uncertainty of the 
measured bias which may be smaller than the shift itself.
The value of $1.6 \times 10^{-4}$ can therefore be considered as an upper 
bound on the relative luminosity bias.

With
30 radiation lengths of tungsten as
absorber, high energy 
electrons and photons deposit almost all of their energy 
in the detector. 
Fiducial
cuts on the minimal 
and maximal reconstructed polar angles of the particles
used for the luminosity measurement
reject
events with shower leakage through the edges of  
LumiCal. 

The relative energy 
resolution, $\sigma_{\rm{E}} /{\rm{E}}$, is parametrised as
\begin{equation}{
\frac{\sigma_{{E}}}{{E}} = \frac{{{a}}_{\rm{res}}}{\sqrt{{{E}}_{\rm{beam}}~\mathrm{(GeV)}}},
}\label{engyResEQ} \end{equation}
\noindent where ${\rm{E}}$ and $\sigma_{\rm{E}}$ are, respectively, 
the central value and the standard deviation of 
the distribution of the energy deposited in the sensors for a beam of electrons with 
energy ${E}_{\rm{beam}}$. The parameter ${{a}}_{\rm{res}}$ is 
usually quoted as the energy resolution, a convention which will be followed here.

Figure~\ref{engyResThetaMinMaxFIG1} shows the energy resolution 
as a function of the polar angle $\theta$ for electron showers with energy 
250~GeV. 
The energy resolution parameter approaches minimal constant values between  
$\theta_{\rm{min}}$ = 41~mrad and $\theta_{\rm{max}}$ = 67~mrad, where the shower is fully contained
inside the calorimeter. 
The fiducial volume of LumiCal is thus defined to be the polar angular range 

\begin{equation}{
41 < \theta^{\rm{f}} < 67 ~ \mathrm{mrad} .
}\label{fiducialVlomueEQ} \end{equation}
For
electron showers located inside the fiducial volume of LumiCal,
the energy resolution is estimated to be  
${{a}}_{\rm{res}} = (0.21 \pm 0.02)~ \sqrt{\mathrm{GeV}}$. No dependence on the electron energy is found in the
energy range from 50 to 300 GeV.
In order to determine the energy of showering particles, 
the 
integrated deposited energy in the detector has to be multiplied by a calibration factor. 
The
calibration factor is found to be constant in the same energy range. 

\begin{figure}[h]
\begin{center}
\subfigure[]{\label{fig:bogdan-b(s)1}
\includegraphics[width=.52\textwidth]{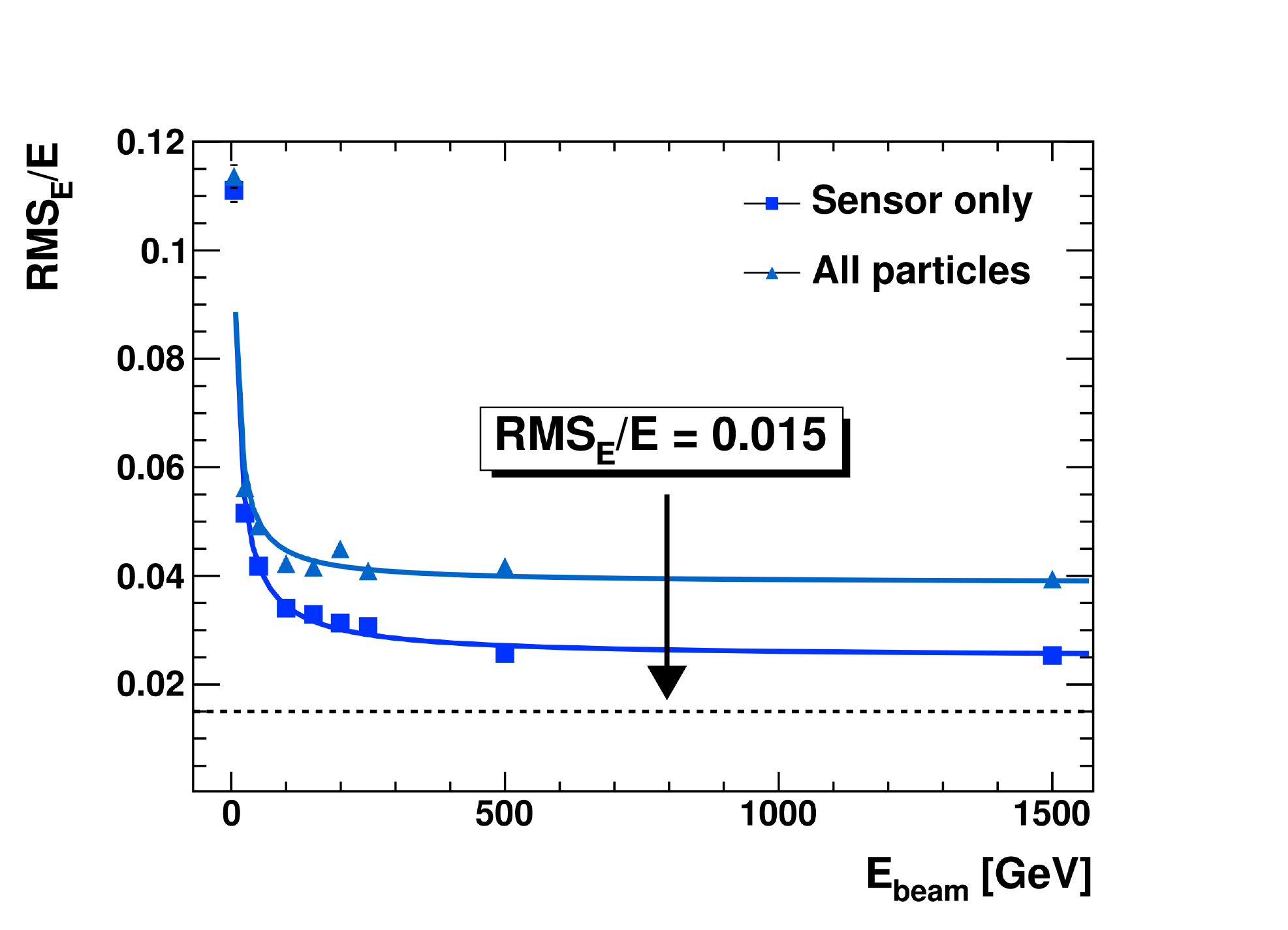}}
\subfigure[]{\label{fig:bogdan-b(s)2}
\includegraphics[width=.45\textwidth]{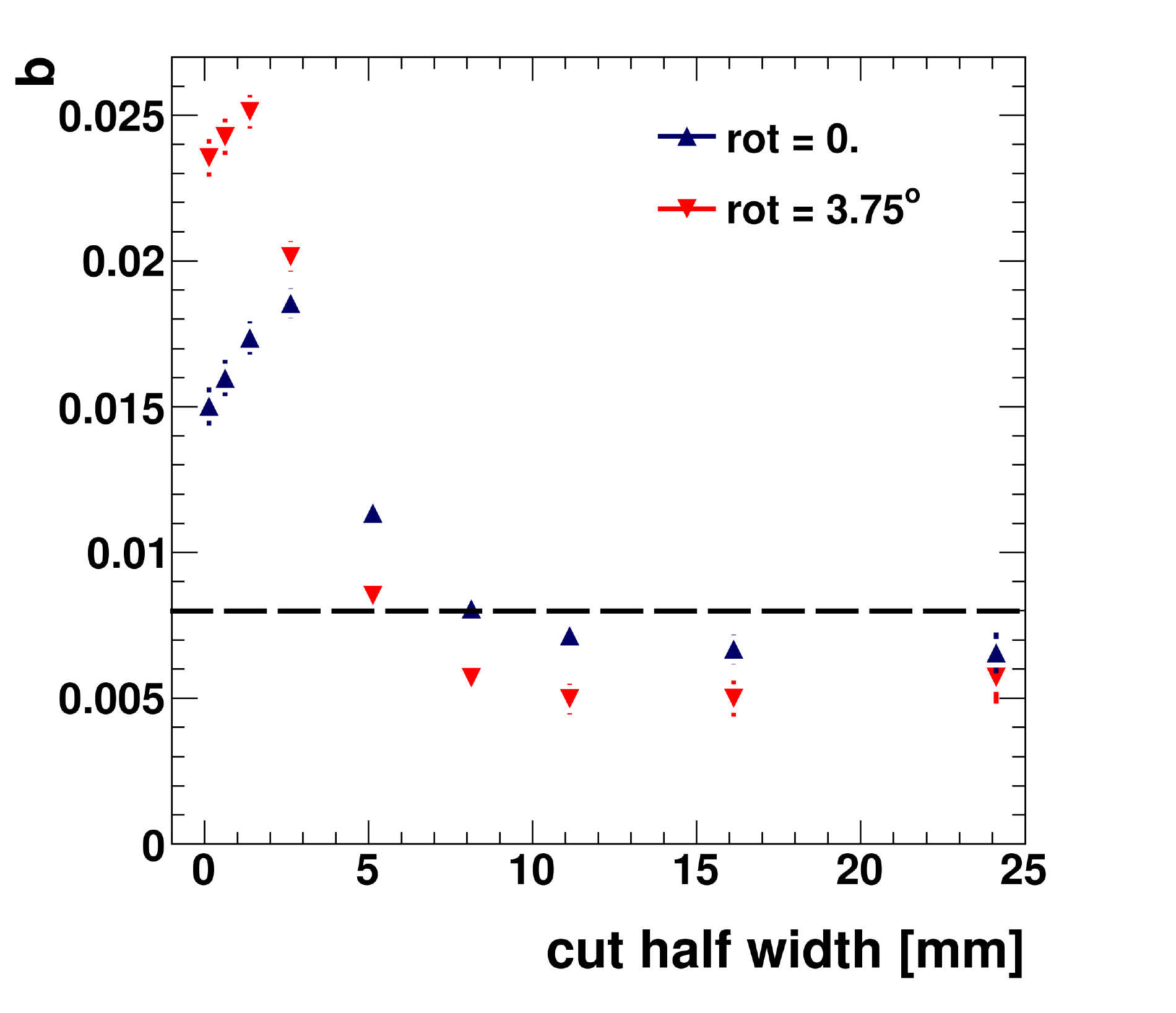}}
\caption{
\Subref{fig:bogdan-b(s)1} The expected LumiCal relative resolution $\frac{\mathrm{RMS}_E}{E}$ as a function of the electron energy $E$, for showers uniformly distributed over the fiducial volume of LumiCal without and with a requirement that the shower be started within the sensors area.
\Subref{fig:bogdan-b(s)2} The expected value of the resolution constant term, $b$, as a function of the minimal required distance of the reconstructed shower position from the middle of the tole gap.
}
\end{center}
\end{figure}

The above study did not take into account the presence of azimuthal gaps in the structure of the silicon sensors. The effect of these gaps on the energy resolution is to generate a constant term ($b$) due to energy leakage of the electromagnetic shower,
\begin{equation}
\frac{\sigma_E}{E} = \sqrt{\left( \frac{a_{\rm{res}}^2}{E_{\rm{beam}}(\rm{GeV})+ b^2}\right)}. 
\label{engyResEQconst} 
\end{equation}
For a realistic simulation of the gap structure, the constant term comes out to be $b=0.04$. This is shown in figure~\ref{fig:bogdan-b(s)1} in which $\frac{\sigma_E}{E}$ is plotted as a function of $E$ for a beam of particles uniformly spread over the face of the calorimeter (fiducial volume) and after selecting only the showers that start within the region of a sensor. In the latter case the constant term is reduced by almost a factor two. The constant term can be further reduced by rejecting showers reconstructed too close to the tile gap and even further by rotating two subsequent sensor planes as shown in figure~\ref{fig:bogdan-b(s)2}, in which $b$ is plotted as a function of the minimal required distance of the shower location from the middle of the tile gap. 
With the requirement of $1 \units{cm}$ distance to the gap edge and a rotation by 3.75\degree,  the constant term may be lowered to $b=0.005$ at the price of losing $30\%$ of the statistics~\cite{pawlik}.

\begin{figure}[h]
\begin{center}
\subfigure[]{\label{fig:electronicSignalFIG1}
\includegraphics[width=.45\textwidth]{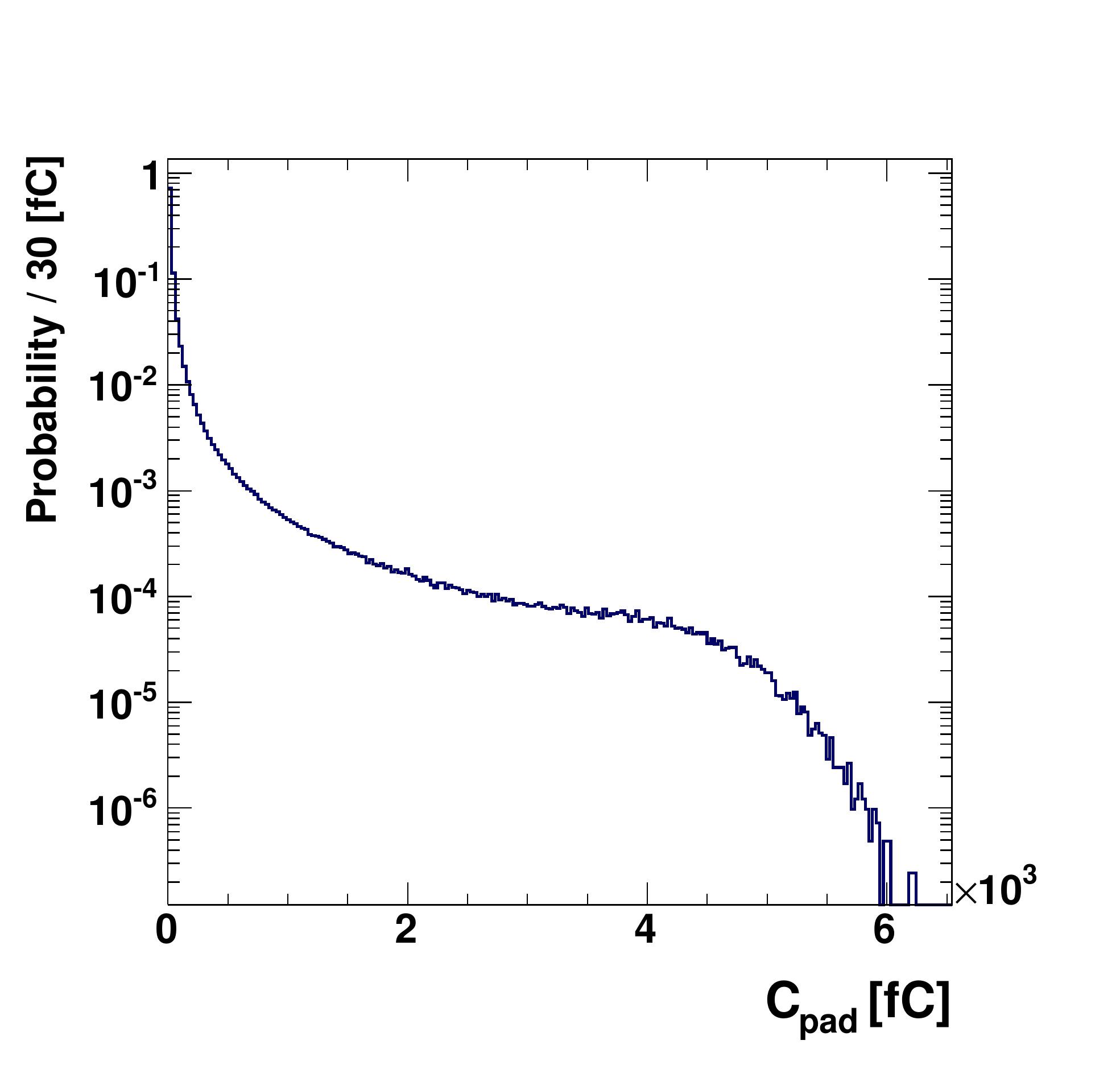}}
\subfigure[]{\label{fig:electronicSignalFIG2}
\includegraphics[width=.45\textwidth]{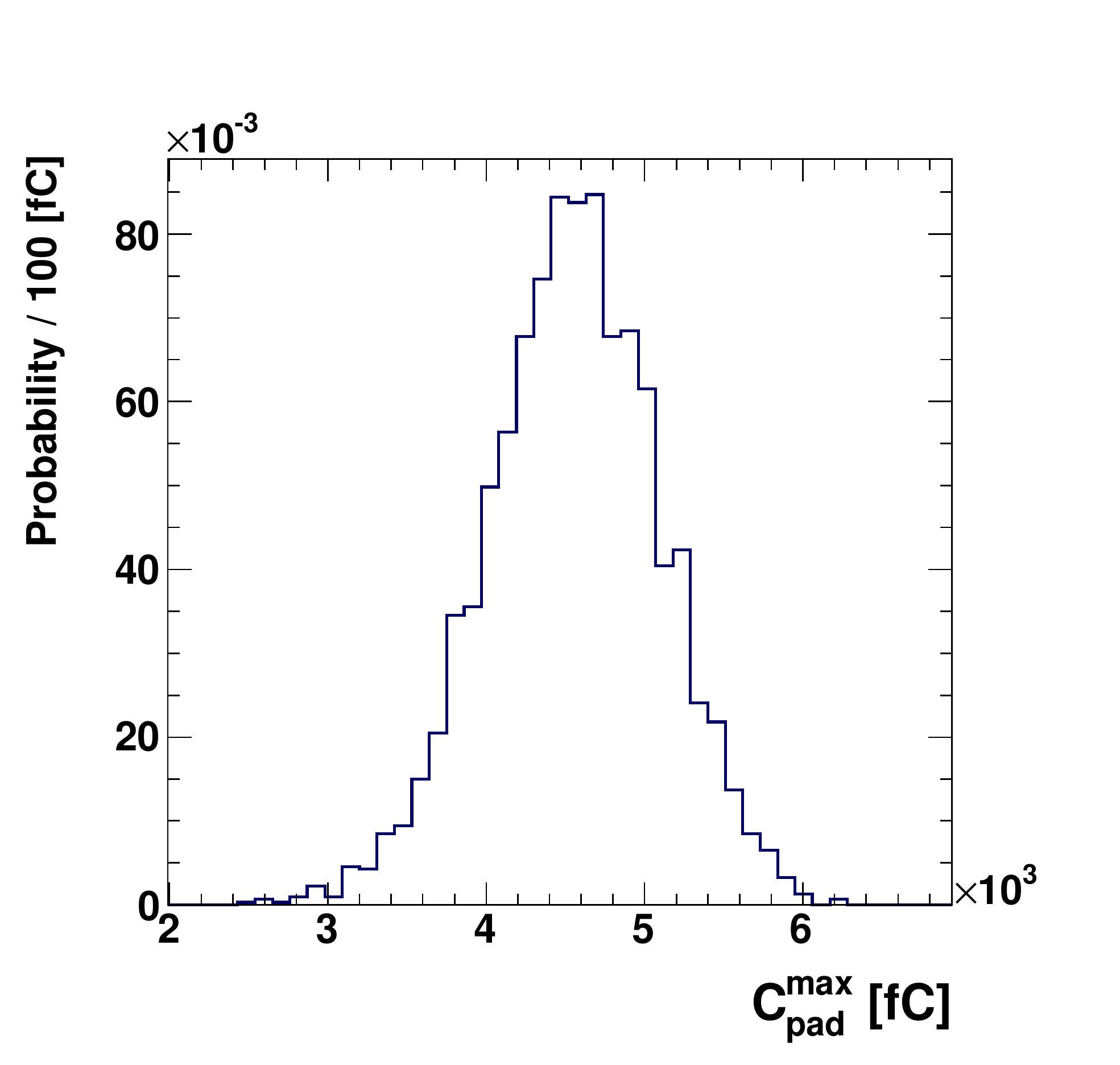}}
\caption{\Subref{fig:electronicSignalFIG1} Normalised
distribution of the charge deposited in a detector pad, ${{C}}_{\rm{pad}}$, by 250~GeV electron
showers. \Subref{fig:electronicSignalFIG2} Normalised distribution of the maximal charge
collected in a single pad per shower, ${{C}}_{\rm{pad}}^{\rm{max}}$, for 250~GeV electron showers.}
\end{center}
\end{figure}

The expected range of energy depositions
in the pads has been studied
for the passage of minimum ionising
particles, hereafter denoted as MIPs, and for showers of 250~GeV electrons~\cite{bib25}. 
The energy deposition in silicon is converted to released ionisation charge.
The distribution of the charge in a single pad ${C}_{\rm{pad}}$, is 
shown in Figure~\ref{fig:electronicSignalFIG1}.
It ranges between
$4 <{{C}}_{\rm{pad}} < 6 \times 10^{3}$~fC.
The 
distribution of the maximal charge collected in a single pad
is shown in  Figure~\ref{fig:electronicSignalFIG2}.
About 95~\% of electron shower signals are less than $5.4\times 10^{3}$~fC.

The impact of the digitisation of the detector signal on the LumiCal performance is investigated 
in  Ref.~\cite{bib25}.
It is shown that an ADC with 8~bit resolution is sufficient to keep the energy resolution 
quoted above. No bias in the energy measurement is found.

\section{BeamCal Design Studies}

BeamCal will be hit after each bunch-crossing 
by a large amount of beamstrahlung pairs.
Their number, energy and spatial distribution 
depend
on the beam parameters and the magnetic field 
inside the detector. 
For
different beam-parameter sets at centre-of-mass energies of 0.5, 1 and 3 TeV 
\cite{ILC_pub,clic_info},
beamstrahlung pairs are generated with the GUINEA-PIG program~\cite{guinea}. 
In both the ILD and the CLIC detectors a solenoidal field of 3.5 Tesla is assumed. In the ILC detector
in addition an
anti-DID field~\cite{anti-DID} is superimposed. 
Beamstrahlung pairs are simulated 
in the detector, using a program based on 
GEANT4. 

\begin{figure}[h]
\subfigure[]{\label{fig:beam_deposits}
\includegraphics[width=0.45\columnwidth,height=0.45\columnwidth]{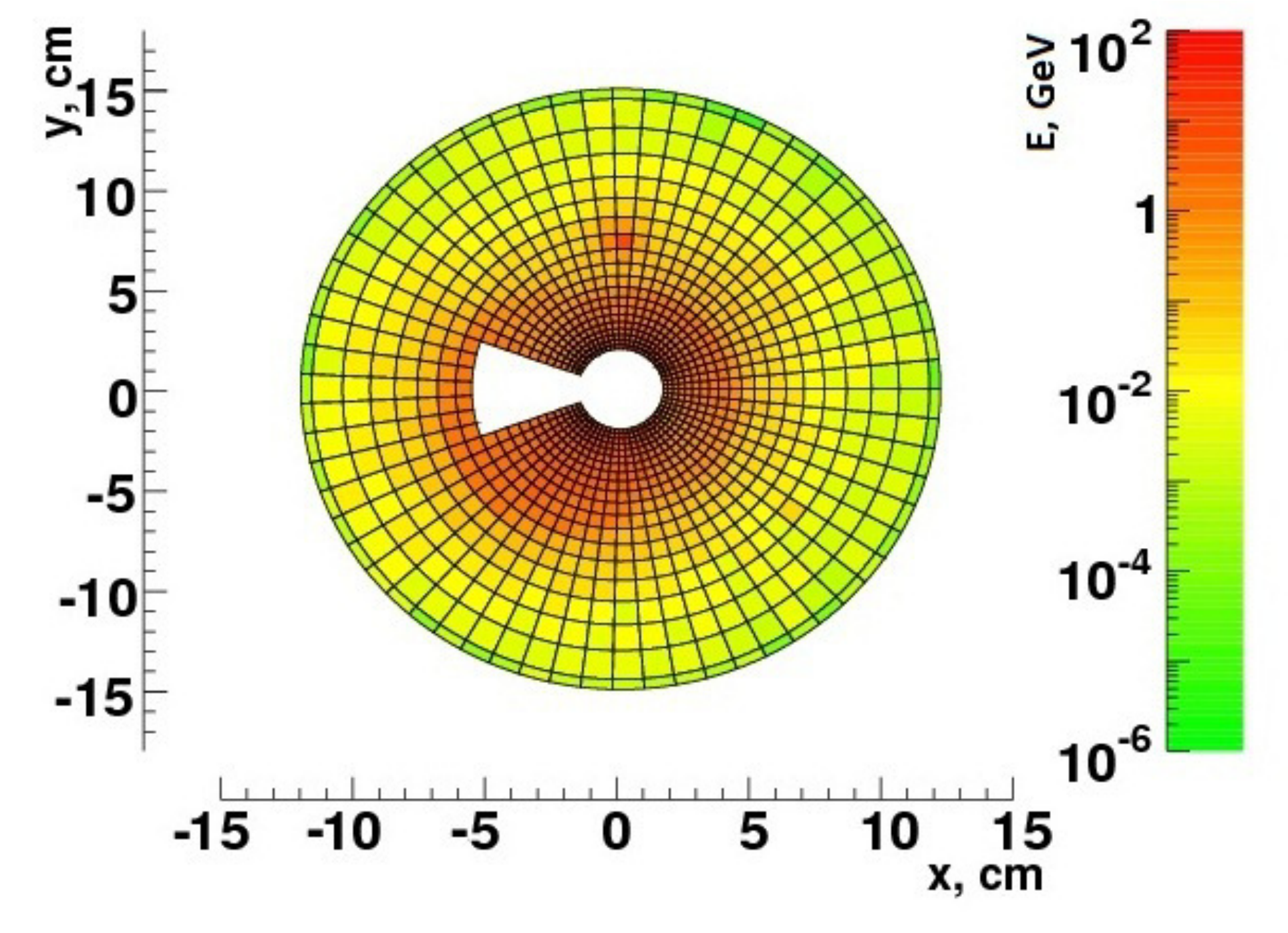}
}
\subfigure[]{
\label{fig:clic_deposits}
\includegraphics[width=0.45\columnwidth,height=0.45\columnwidth]{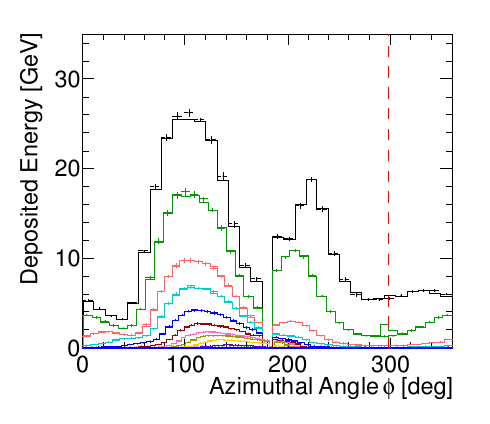}}
 \caption{\Subref{fig:beam_deposits}
The distribution of the energy 
deposited by beamstrahlung pairs from the ILC after one bunch
crossing in the sensors of BeamCal. The centre-of-mass energy is 1 TeV.
Superimposed is the deposition of a single high energy electron (red spot in the upper part).  
The depositions are integrated over the pads and summed up for each pad over all sensors layers. 
The white area in the centre 
allows space for the beam-pipes.
\Subref{fig:clic_deposits}   
The energy deposited per ring of BeamCal as a function of the azimuthal angle at 3 TeV.} 
\end{figure}

As an example, 
the energy, per beam crossing,
deposited in the sensors
of BeamCal at a centre-of-mass energy of 1 TeV, 
is about 150 GeV and distributed over the calorimeter, as shown in Figure~\ref{fig:beam_deposits}. 
The shape of 
these depositions
allows a bunch-by-bunch 
luminosity estimate and the determination of beam parameters
with accuracies better than 10\%~\cite{grah1} .

In CLIC, due to the short time between 
two bunches, the depositions of 40 bunch-crossings are added before readout. The energy deposited is shown in 
Figure~\ref{fig:clic_deposits} as a function of the azimuthal angle.

ƒ
For search experiments
it is important 
to detect  
single high energy 
electrons on top of the
wider-spread beamstrahlung pairs. This feature allows to suppress the
background from two-photon processes in a search e.g. 
for super-symmetric tau-leptons~\cite{drugakov} 
in a large fraction of the parameter space. 
Superimposed on the pair depositions 
in Figure~\ref{fig:beam_deposits}
is the deposition of an electron of 200 GeV, seen as the red spot 
in
the upper part of the calorimeter.

\begin{figure}[h]
\begin{center}
\includegraphics[width=0.45\columnwidth,height=0.45\columnwidth]{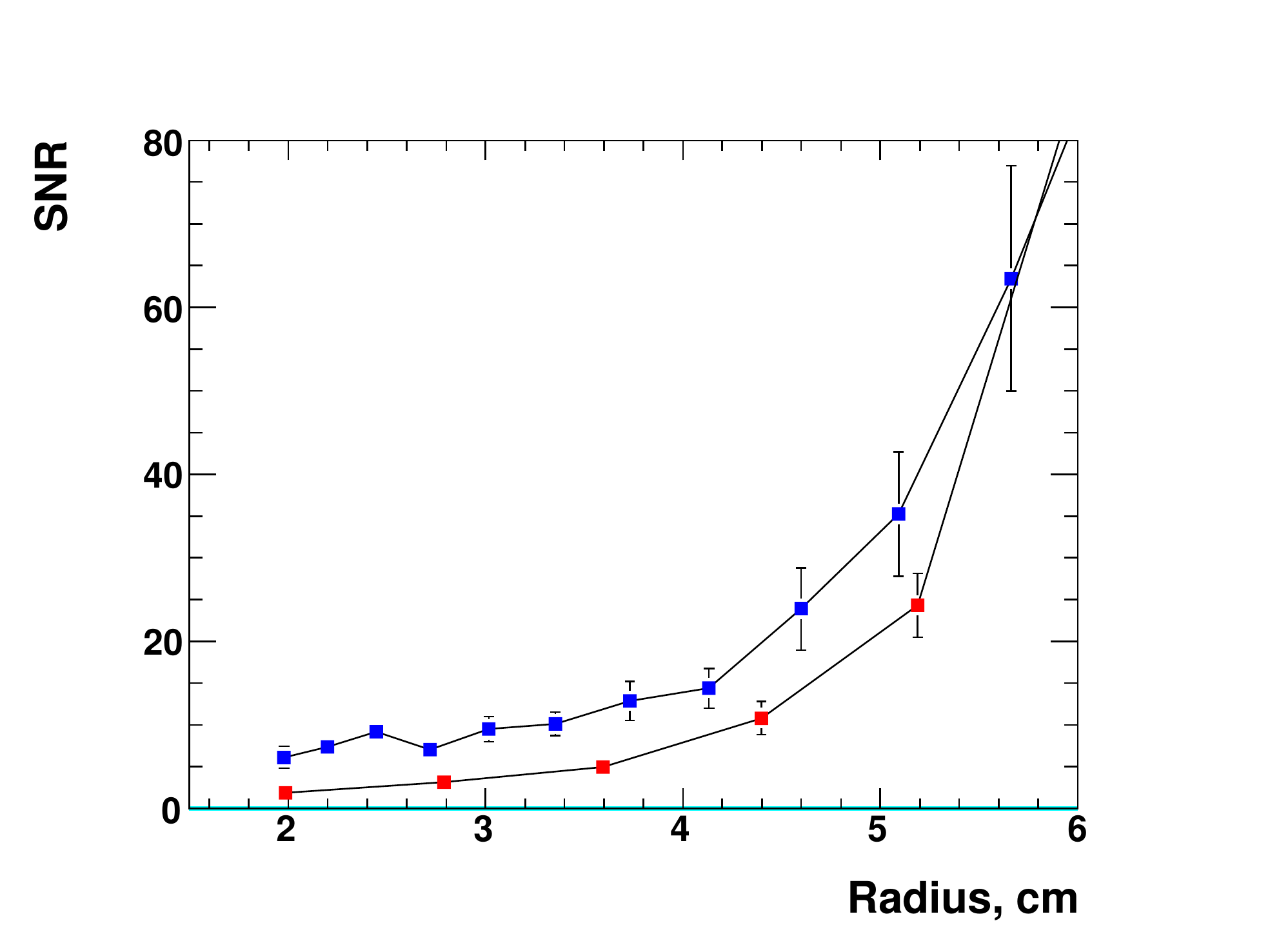}
\end{center}
\caption[]{\label{fig:SN_equal}
The signal-to-noise ratio in BeamCal for equal size pads (red dots) and pads 
proportional to the radius (blue dots) as a function of the radius.
}
\end{figure}
To obtain a high detection efficiency, two different 
sensor segmentations have been studied at 1 TeV. 
In one segmentation the pads are of almost equal size 
and in the other, the pad size is growing with the radius, as shown in Figure~\ref{fig:beam_deposits}.
The number of pads is kept constant.
The average deposition from pairs on each pad is obtained by simulating
10 bunch crossings. A signal-to-noise ratio is defined as the ratio between the average deposition 
from a single high energy electron and the standard deviation of the average depositions from 
beamstrahlung pairs only.
This signal-to-noise ratio is shown in Figure~\ref{fig:SN_equal} for 
single electrons of 100 GeV at a centre-of mass energy of 1 TeV.
A clear improvement of the signal-to-noise ratio is found for the case of growing pad sizes with the radius.

\begin{figure}[h]
\subfigure[]{\label{fig:efficiency}
\includegraphics[width=0.55\columnwidth,height=0.45\columnwidth]{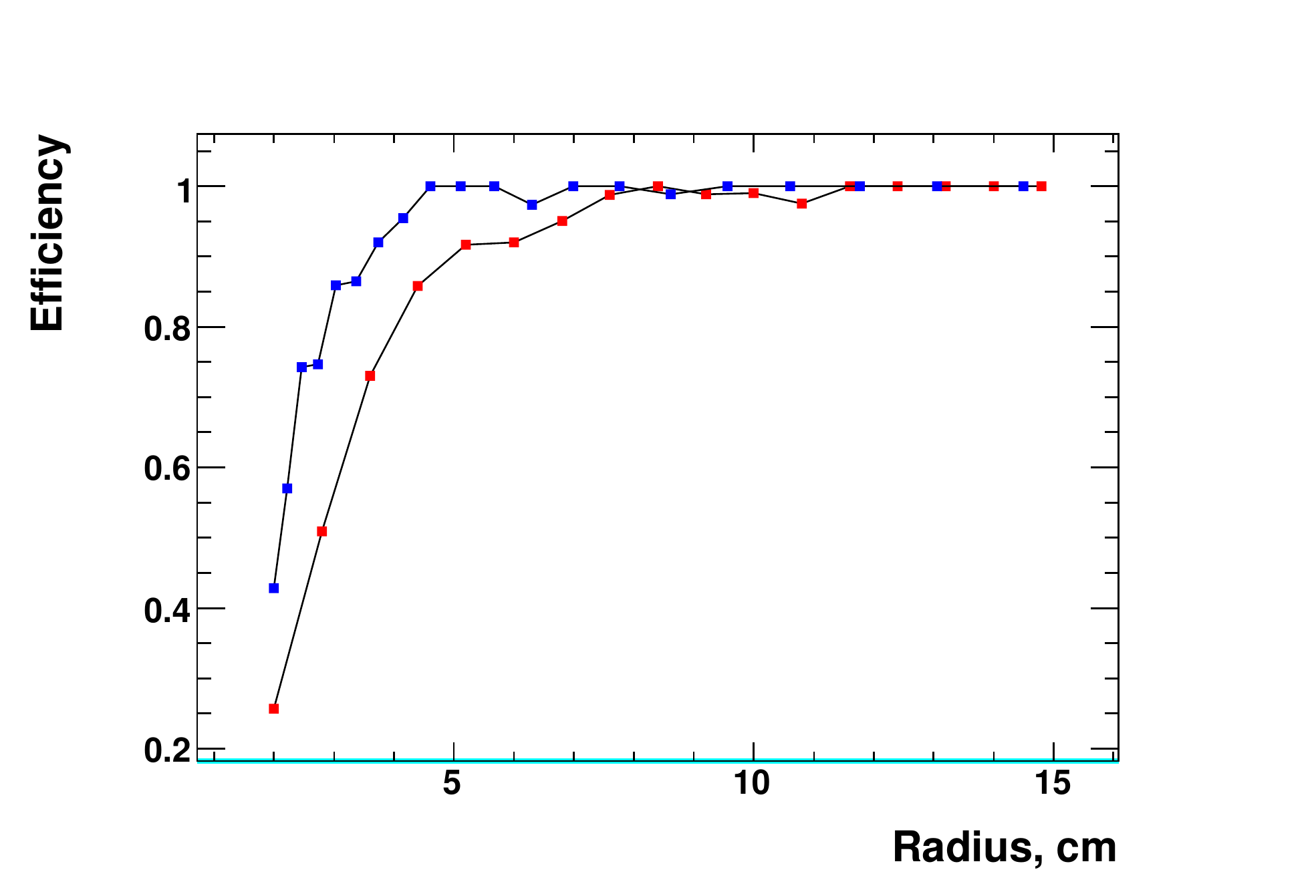}
}
\subfigure[]{\label{fig:clic-efficiency}
\includegraphics[width=0.45\columnwidth,height=0.45\columnwidth]{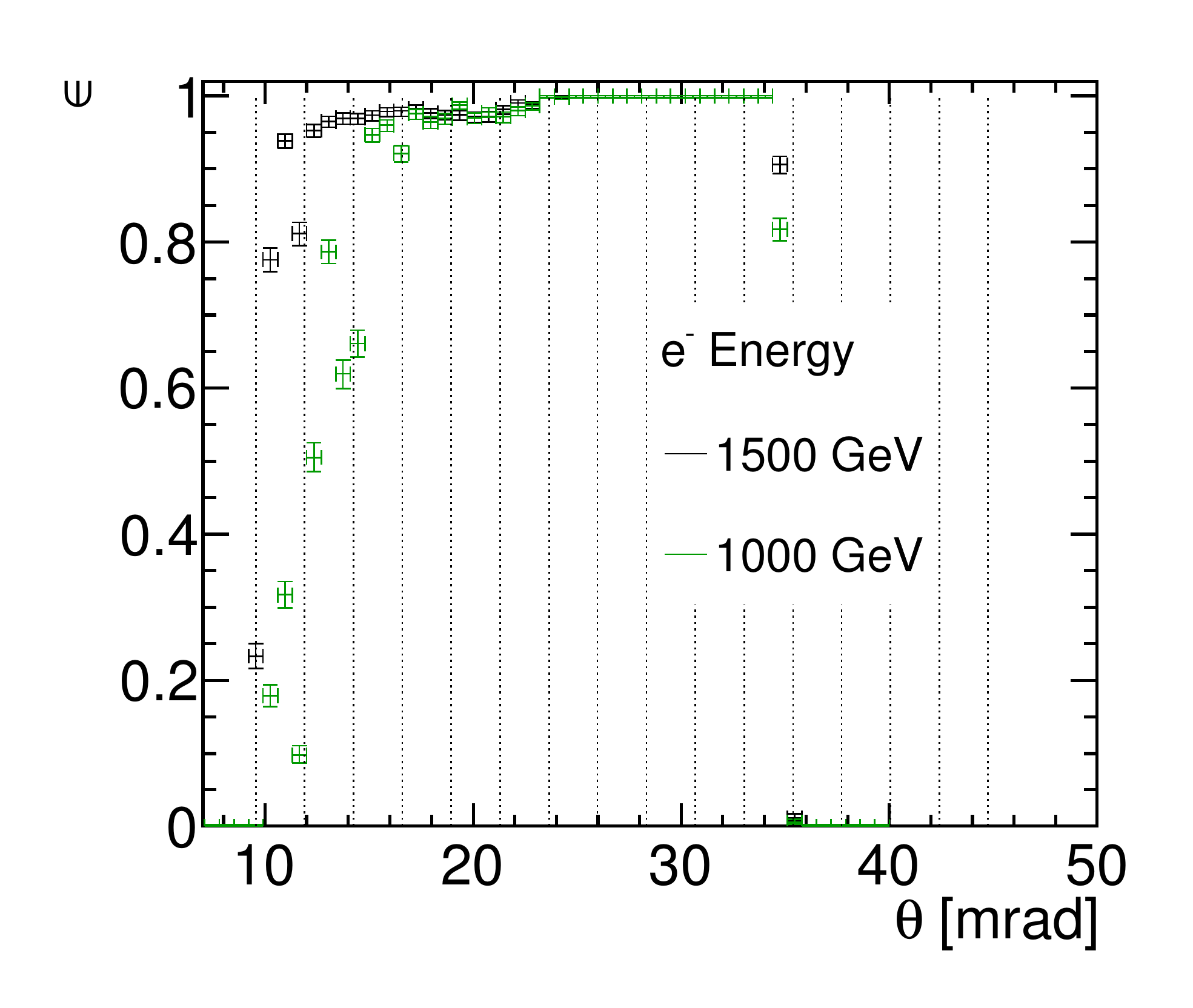}}
\caption{\Subref{fig:efficiency}
The efficiency to detect single high energy electrons
on top of the beamstrahlung background for electron energies of 200 GeV for equal pad sizes (red) and pad sizes growing with the radius (blue). The beamstrahlung is simulated at a centre-of-mass 
energy of 1 TeV.   
\Subref{fig:clic-efficiency}
The efficiency to detect single high energy electrons
on top of the beamstrahlung background for electron energies of 1 and 1.5 TeV as a function of the polar angle for a 3 TeV CLIC~\cite{a_seiler}.
}
\end{figure}

By performing
an appropriate subtraction of the pair deposits
and a shower-finding algorithm which
takes into account the longitudinal shower profile, 
high energy electrons 
can be detected, as shown in Figure~\ref{fig:efficiency}, with high efficiency even in the range of high background.
Particularly in this area the efficiency using a segmentation with pad sizes growing with the radius becomes
superior. 

A similar study is performed for a CLIC detector~\cite{a_seiler}. Even when the background deposition of 40 bunch crossings are
summed up, single high energy electrons will still be detected, as can be seen in Figure~\ref{fig:clic-efficiency}.

\begin{figure}[h]
\begin{center}
\includegraphics[width=0.55\columnwidth,height=0.45\columnwidth]{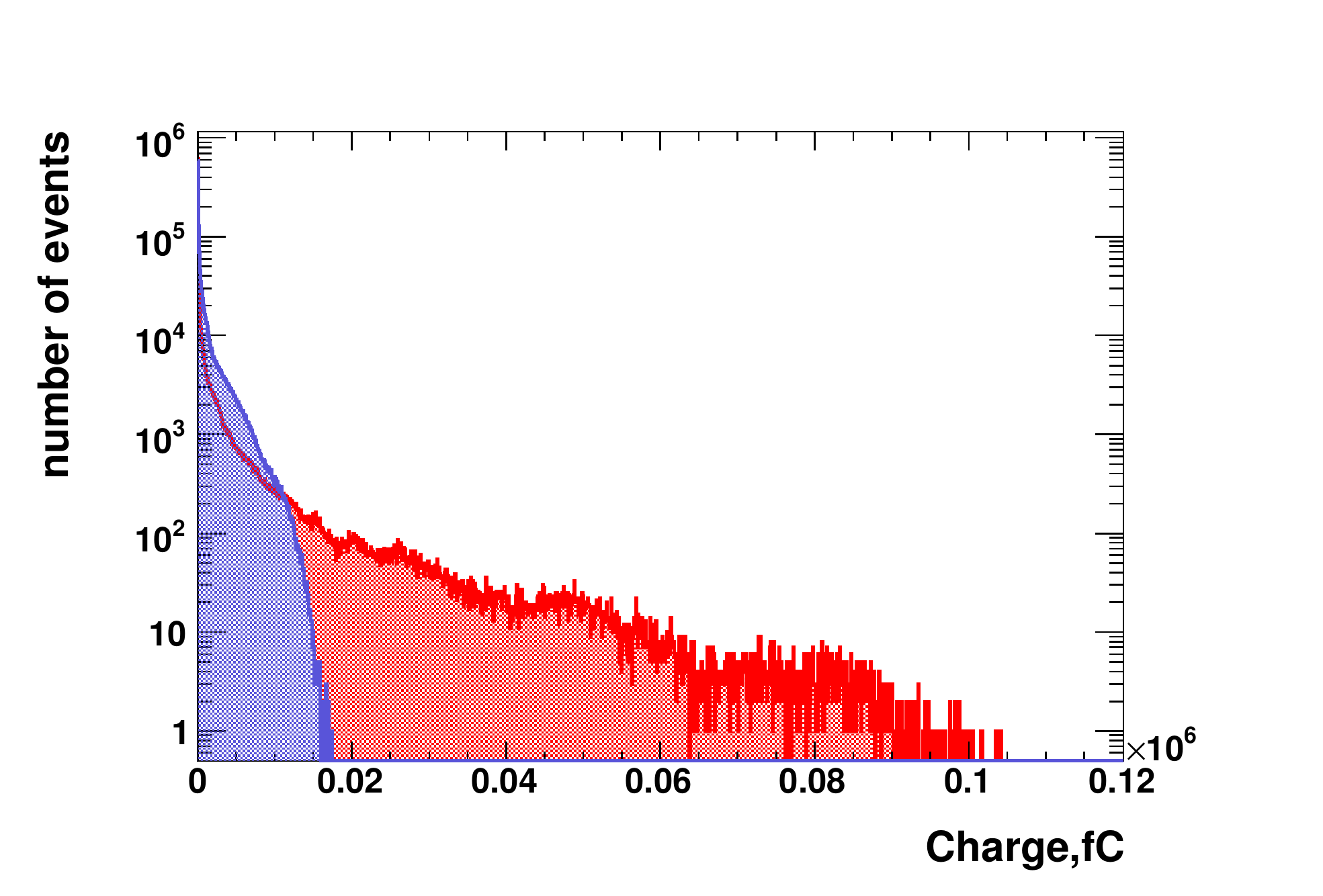}
\end{center}
\caption[]{\label{fig:signal_range}
The signals expected from the depositions after a bunch crossing on the pads
of the BeamCal sensors. The centre-of-mass energy is 1 TeV. The blue histogram results from the pad sizes proportional to the radius while the red - from the equal-pad-size segmentation.
}
\end{figure}

The range of signals expected on the pads of BeamCal at 1 TeV 
centre-of-mass energy
was estimated for polycrystalline diamond and GaAs sensors.
An example for GaAs is shown in Figure~\ref{fig:signal_range}.  
Including the depositions from beamstrahlung, signals up to 10 pC
are expected. Digitising the signals with an ADC with 10 bit resolution
has no impact on the performance of the calorimeter.

\begin{figure}[h]
\subfigure[]{\label{fig:largest_dose}
     \includegraphics[width=0.45\columnwidth,height=6.5cm]{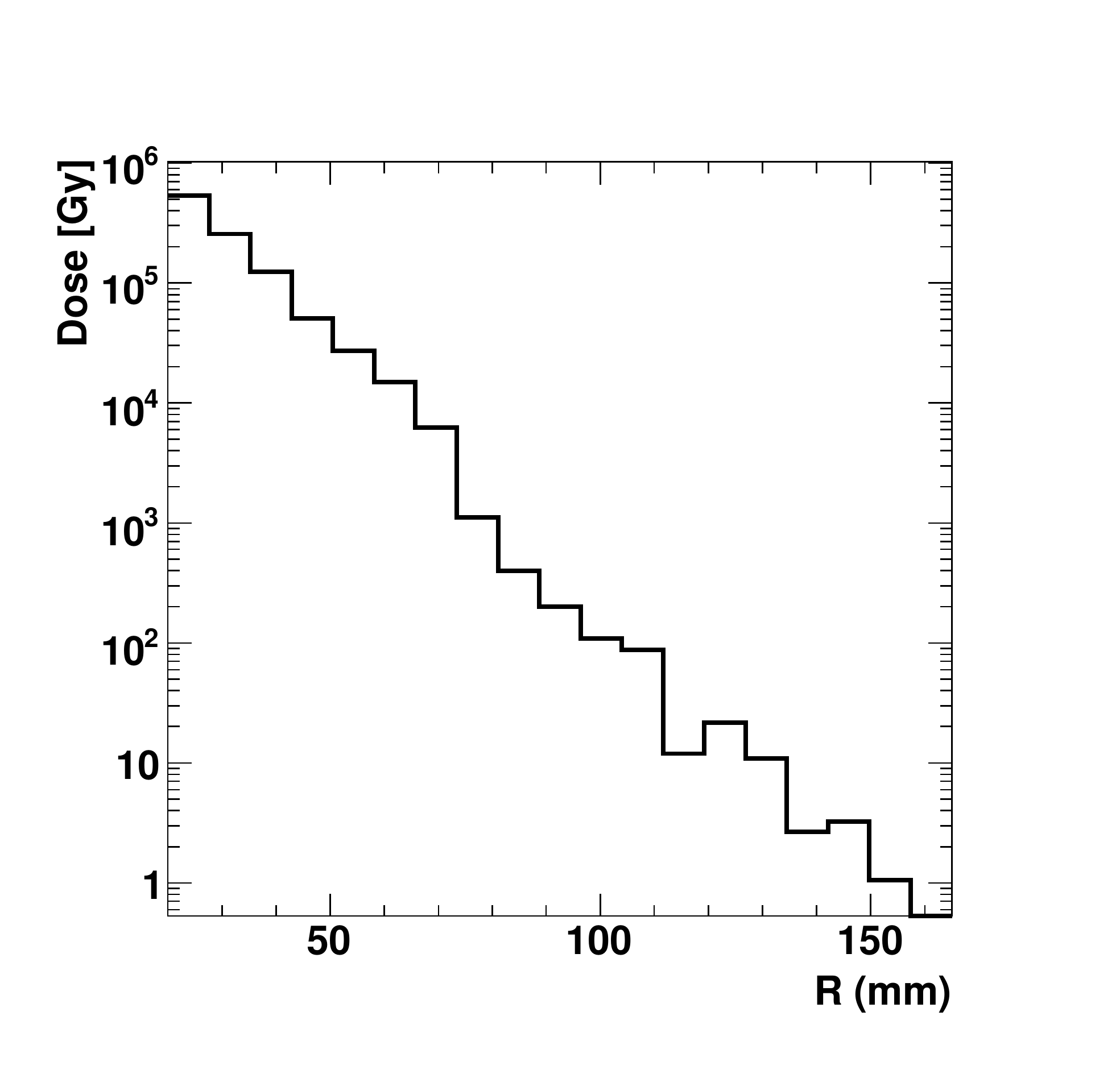}
}
\subfigure[]{
\label{fig:neutron_fluxb}    
\includegraphics[width=0.45\columnwidth,height=6.5cm]{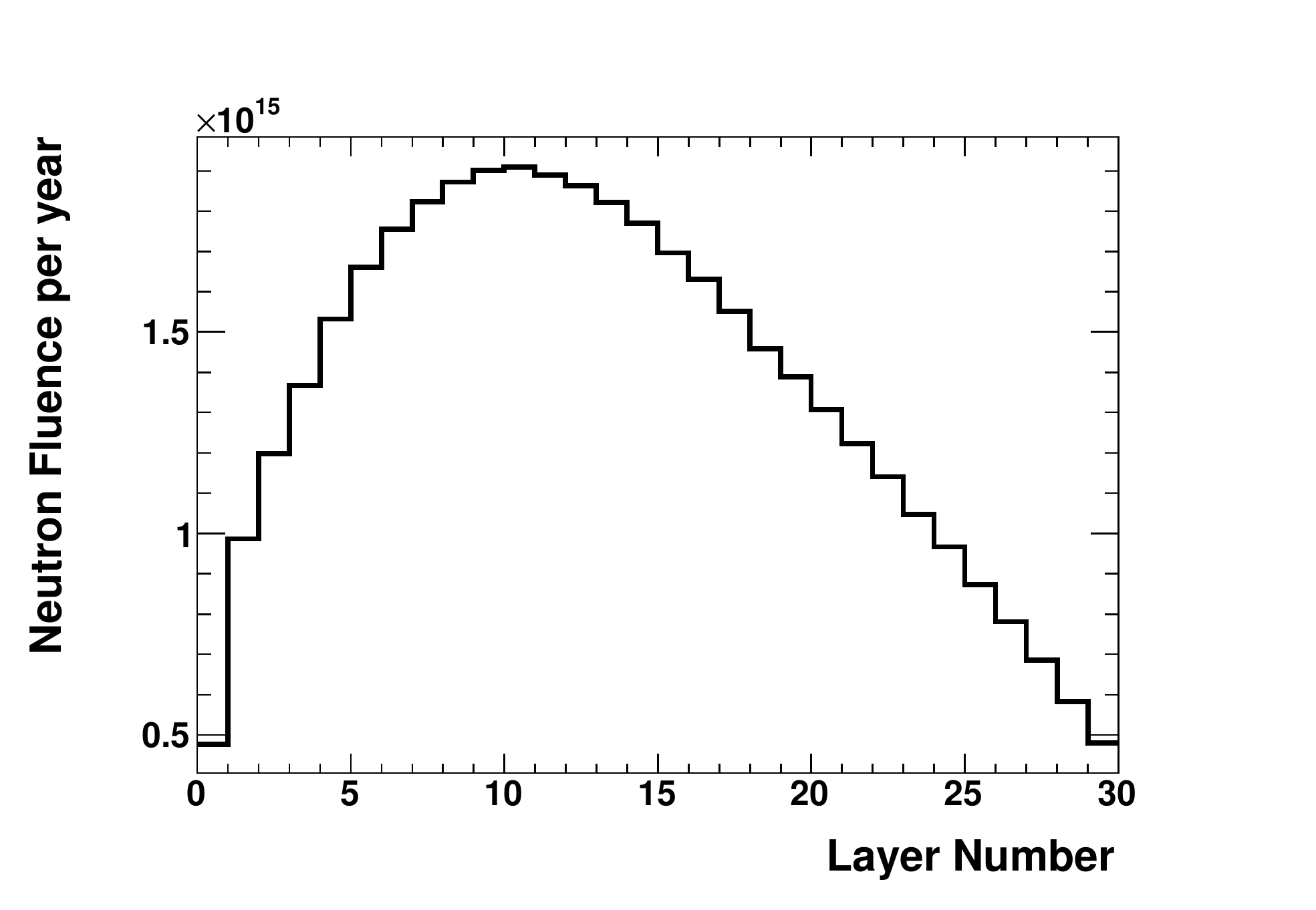}}
\caption{\Subref{fig:largest_dose}
The dose in BeamCal 
sensors per year as a function of the radial distance from the beam.
\Subref{fig:neutron_fluxb} The fluence of neutrons per year inside the sensors of BeamCal
as a function of the sensor layer number using the cascade model of Bertini. A
centre of mass energy of 0.5 TeV 
is assumed}.
\end{figure}
GEANT4 simulations are also used to determine the expected dose
and the neutron fluence in the sensors after one year of operation with nominal
beam parameters.
The dose in a sensor layer at the depths of the shower maximum as a function of 
the radius 
is shown in  Figure~\ref{fig:largest_dose}.
In the innermost ring of the calorimeter a dose of about 0.5 MGy is expected. Since the dose is non-uniformly
distributed as a function of the azimuthal
angle, it approaches 1 MGy per year  in some sensor areas of the inner rings.
A similar study for CLIC conditions at 3 TeV leads to roughly the same dose~\cite{a_seiler}  

\begin{figure}[h]
\begin{center}
\includegraphics[width=0.4\columnwidth]{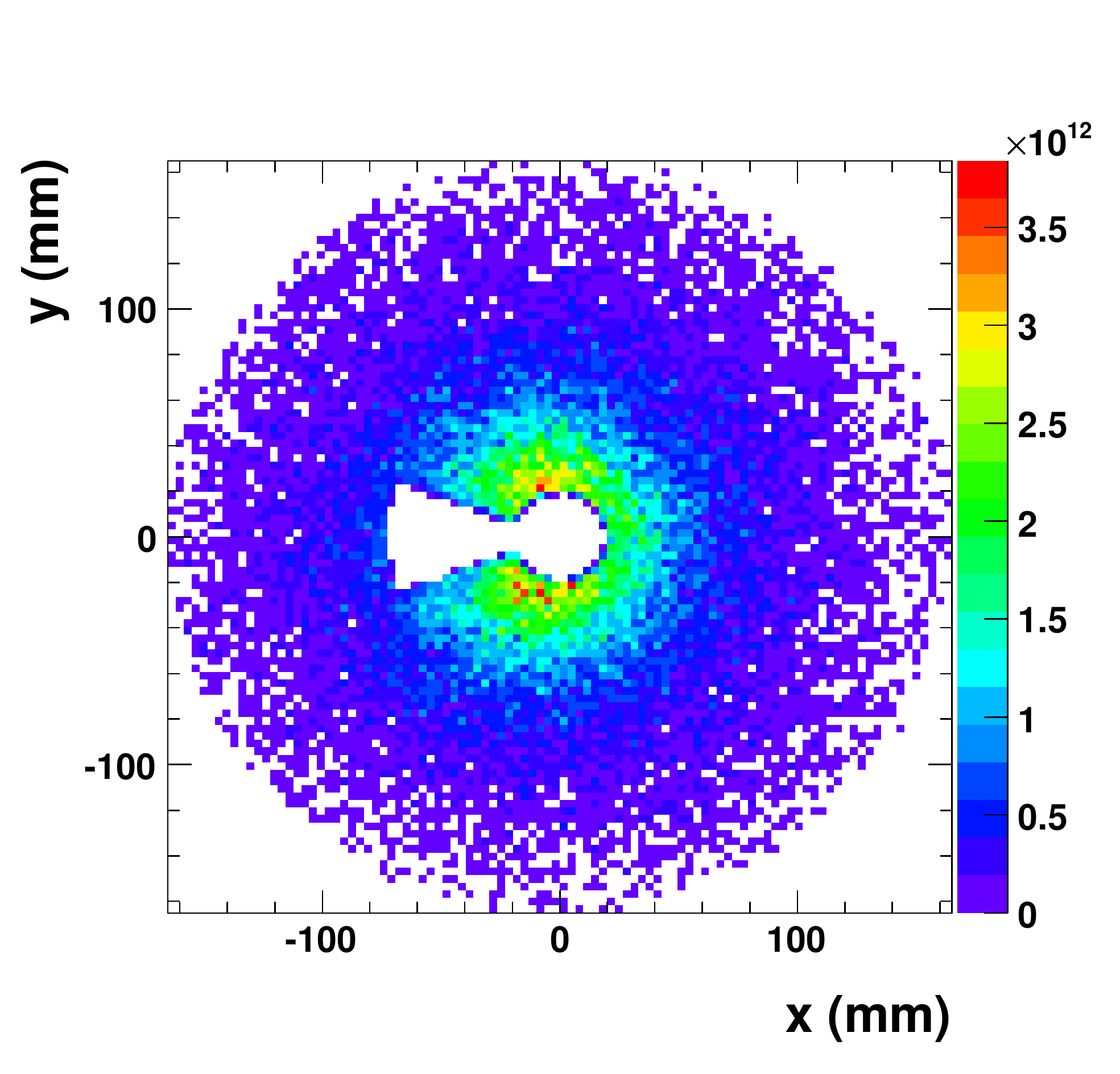}
\caption[]{\label{fig:neutron_flux_rad} 
The fluence of neutrons per $\rm{mm^2}$ and year crossing a sensor of BeamCal
near the shower maximum
using the cascade model of Bertini.
}
\end{center}
\end{figure}
The neutron fluence is estimated by using in GEANT4
the cascade 
model of Bertini~\cite{bertini} (option QGS-BERT-HP).
The fluence per year of running at 
nominal beam parameters is shown in Figure~\ref{fig:neutron_fluxb}
as a function of the sensor layer number. Fluences up to 2$\times$10$^{15}$ per layer are expected
near the shower maximum. 
Other GEANT4 models predict lower neutron fluences, particularly
at low neutron energies~\cite{RomJourPhys}.
The distribution of the fluence of neutrons  
in the sensor 
layer with the maximum fluence is 
shown in Figure~\ref{fig:neutron_flux_rad}.  
With
the cascade model of Bertini,
a neutron fluence of 0.4$\times$ 10$^{12}$ neutrons per mm$^2$ and year  
is expected near the beam-pipe. Albeit this is still an order of magnitude
less than predicted for LHC detectors 
near the beam pipe.

\subsection{Pair monitor simulations} 

Additional and independent information on beam parameters will 
be obtained from the pair monitor~\cite{tauchi1, tauchi2}.
The device will consist of one layer of 
silicon pixel sensors, with pixel size of 400$\times$400 $\mu$m$^2$
just in front of BeamCal,
to
measure the 
number density
distribution of beamstrahlung pairs.
Here
we investigated the sensitivity to the horizontal and vertical 
bunch sizes, $\sigma_{\mathrm{x}}$ and $\sigma_{\mathrm{y}}$, and the ratio 
of the vertical displacement between bunches crossing 
to their vertical size, $\Delta_{\mathrm{y}}$.

\begin{figure}[h]
\begin{center}
\includegraphics*[width=15cm]{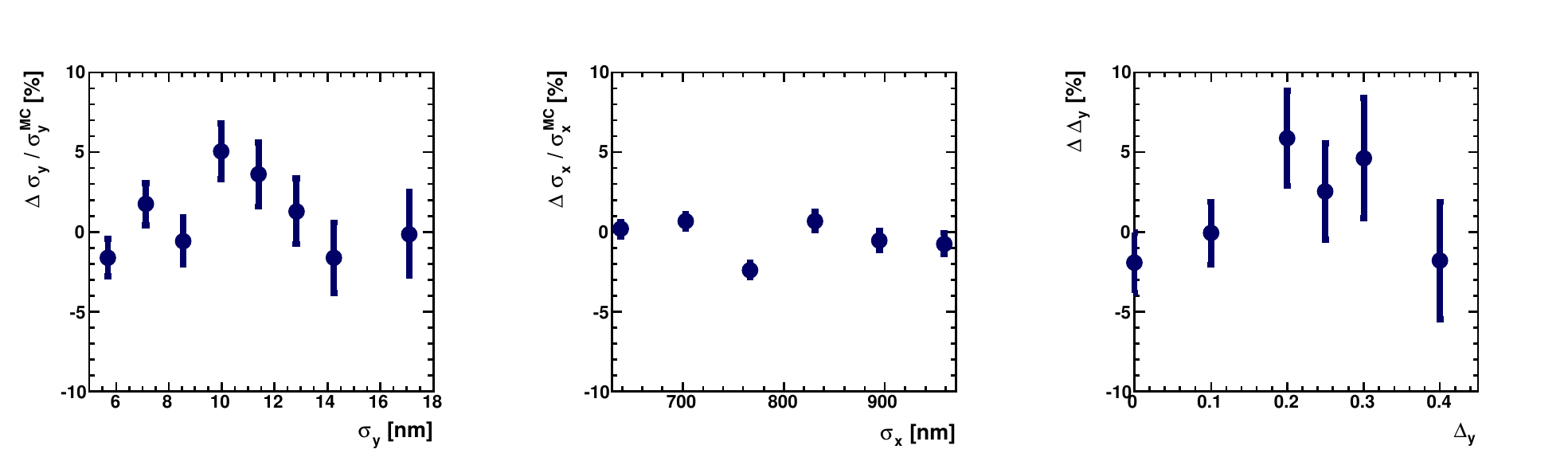}
\end{center}
\caption{\label{fig:reso}
The relative deviations of the vertical, $\sigma_y$, and horizontal, $\sigma_x$, beam sizes, and
the ratio of vertical displacement to the vertical beam size, $\Delta_y$, averaged over 50 bunch crossings 
which are measured by the pair monitor.}
\end{figure}
To reconstruct the beam profile, several observables characterising the
number density of pairs at the front 
face of BeamCal are used~\cite{ito}.
Bunch crossings are simulated for certain ranges of  
$\sigma_{\mathrm{x}}$, $\sigma_{\mathrm{y}}$ and $\Delta_{\mathrm{y}}$, and 
each of these observables is fitted with a second order polynomial.
Then, bunch crossings are generated using a certain set of beam parameters
and $\sigma_{\mathrm{x}}$, $\sigma_{\mathrm{y}}$, and $\Delta_{\mathrm{y}}$ are 
reconstructed with the inverse matrix method. 
Figure \ref{fig:reso} shows the difference between the beam parameters reconstructed and set 
in the simulation divided by the latter, averaged over 50 bunch crossings.     
These quantities are compatible with zero, and the relative uncertainties of the 
vertical and horizontal beam 
sizes and the relative vertical displacement are 10.1\%, 3.2\% and  8.0\%, respectively.

\section{Mechanical concepts}

\begin{figure}[htpb]
\centering
\subfigure[]{\label{fig:beamcal123}
\includegraphics[width=0.4\columnwidth]{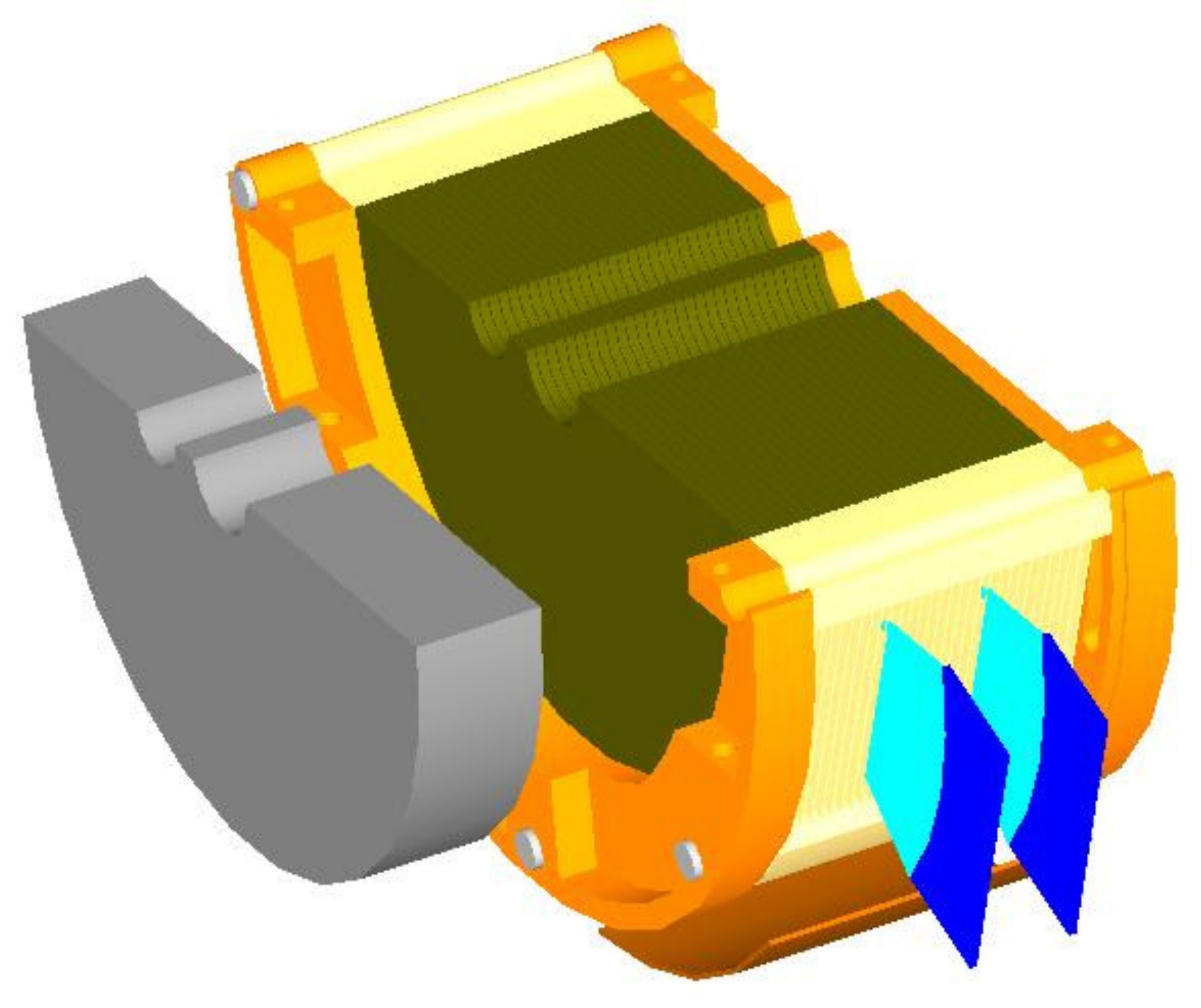}
}
\subfigure[]{\label{fig:beamcal124}
\includegraphics[width=0.4\columnwidth]{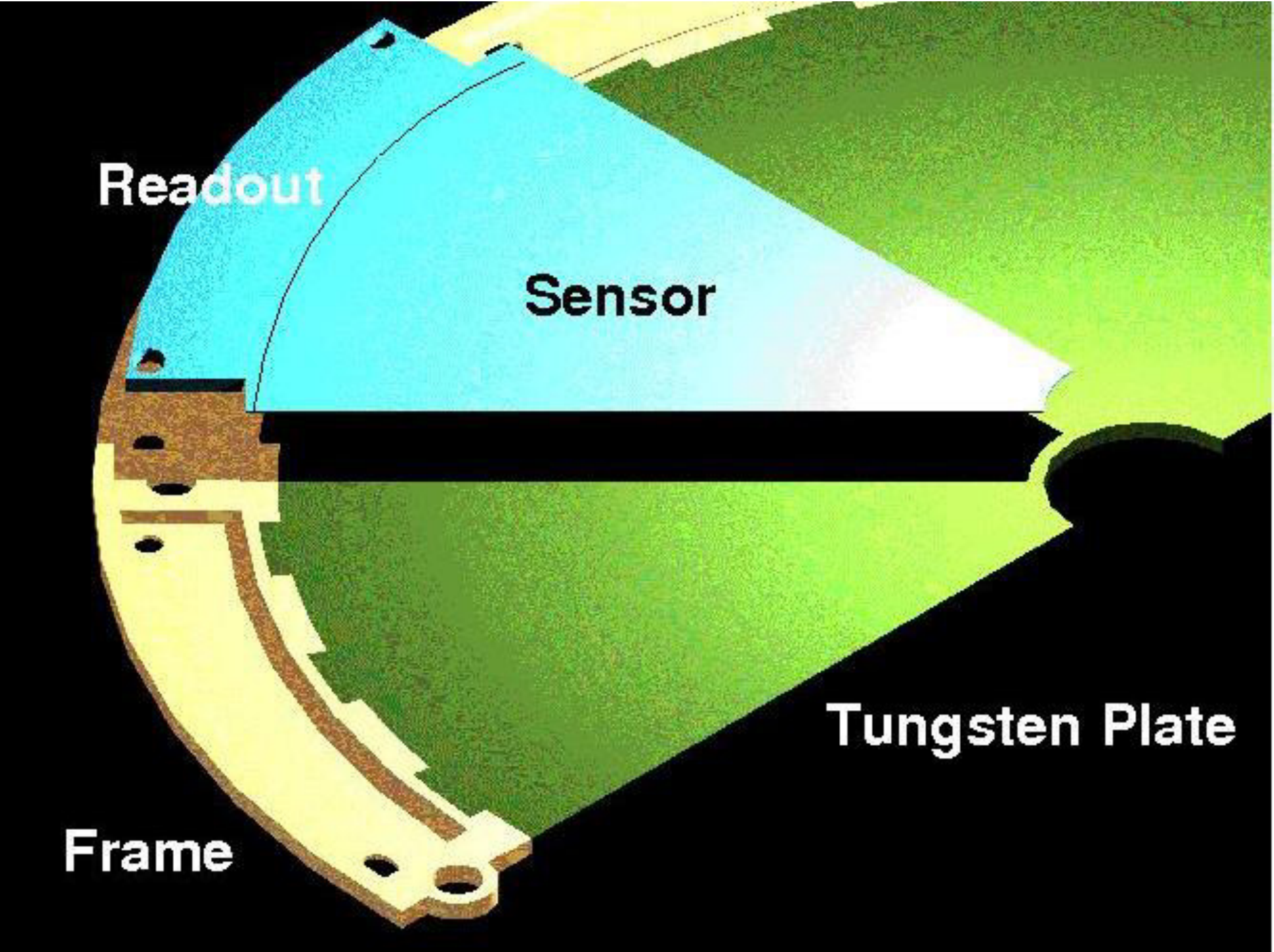}
}
\caption{\Subref{fig:beamcal123} A half-cylinder of BeamCal. The brown block is the 
tungsten absorber structure interspersed with sensor layers. 
The orange structure represents the mechanical frame. The blue segments
at the outer radius indicate the front-end electronics.
In front of the calorimeter a graphite shield, shown in grey, reduces the amount of low energy particles 
back-scattered into the tracking detectors.
\Subref{fig:beamcal124} 
A half-layer of an absorber disk assembled with a sensor sector and the front-end
readout.}
\end{figure}

Mechanical designs of both calorimeters are developed on the basis of the simulation results.
To allow their installation after the beam-pipe is in place, both calorimeters consist of two half-cylinders.
A schematic of a half-cylinder of BeamCal is shown in Figure~\ref{fig:beamcal123}. 
The tungsten absorber disks are embedded in a mechanical frame stabilised by 
steel rods. Each layer is composed of a tungsten half-disc surrounded by a brass 
half-ring as shown in Figure~\ref{fig:beamcal124}. Precise
holes in the brass ring will ensure a position accuracy of better than 100$\mu$m.
The sensors are fixed on the tungsten
and connected via a flexible PCB to the front-end readout.
The distance between two adjacent tungsten plates is kept 
to 1~mm to reach the smallest possible Moli\`{e}re radius.  
Due to the required high radiation tolerance,
GaAs sensors are foreseen. For the innermost part of BeamCal, adjacent to the beam-pipes, also
CVD\footnote{Chemical Vapour Deposition.} diamond is considered.

\begin{figure}
\centerline{\includegraphics[width=0.4\columnwidth]{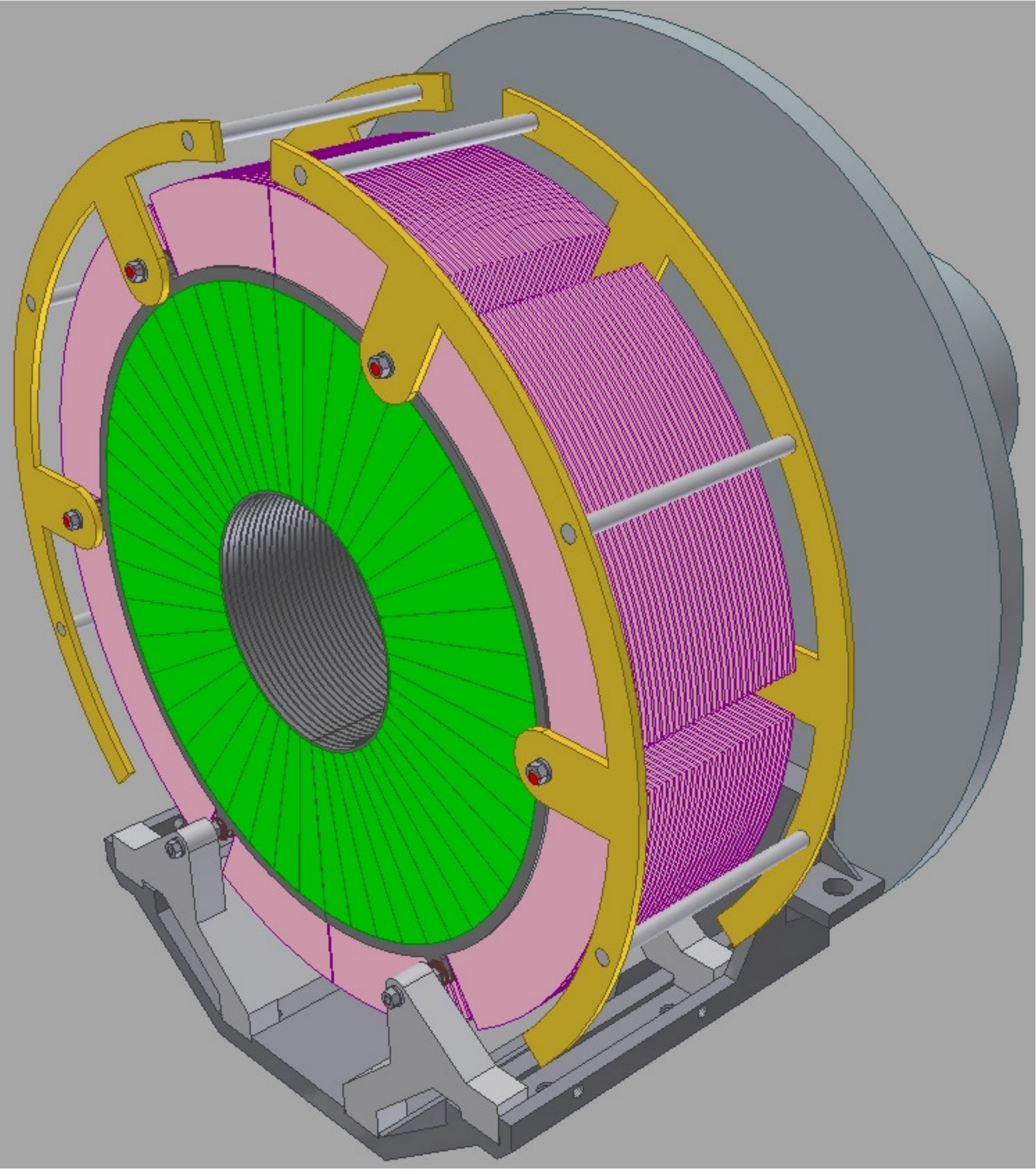}}
\caption{The mechanical structure of LumiCal. Tungsten disks are
precisely positioned using 4 bolts which are stabilised by additional steel 
rings on both sides of the cylinder. 
}\label{fig:lumical}
\end{figure}

The design of LumiCal is similar~\cite{LumiCal_mechanics}. Since it is a precision device,
special care is devoted to the mechanical stability and position control. 
The tungsten half-discs are held by special bolts. 
For a half-barrel structure, as shown in Figure~\ref{fig:lumical}, a finite element 
simulation is performed. The calorimeter weight leads to a maximal vertical displacement
of 20 $\mu$m. For a temperature difference of  1~K
over a disk, the deformation of the shape of the tungsten plate is estimated 
to be 25 $\mu$m. 
To match the 
requirements on the precision of the lower polar angle measurement, the 
sensor positions at the inner acceptance radius must be controlled to better than 40 $\mu$m. 
Other critical quantities are 
the distance between the two calorimeters
and the position of the beam with respect to the
calorimeter axis. The former must be known to about 1 mm and the latter to 500 $\mu$m.
A laser-based position-monitoring system is under development~\cite{laser_alignment}, as described in more  detail
in chapter 7, to control the position
of LumiCal and the sensor planes inside LumiCal
with
$\mu$m precision.

To keep the Moli\`{e}re radius small, the gap for the sensors is 1~mm. 
The signals on the pads of both calorimeters are led by thin copper strips on a 
Kapton foil to 
the front-end electronics positioned at the outer radius of the calorimeter.
Details of the connectivity scheme to be finally
implemented are under investigation.

\section{LumiCal silicon sensor detectors}

\begin{figure}[h]
\begin{center}
\subfigure[]{\label{fig:si-sensor-woj}
\includegraphics[width=.45\textwidth]{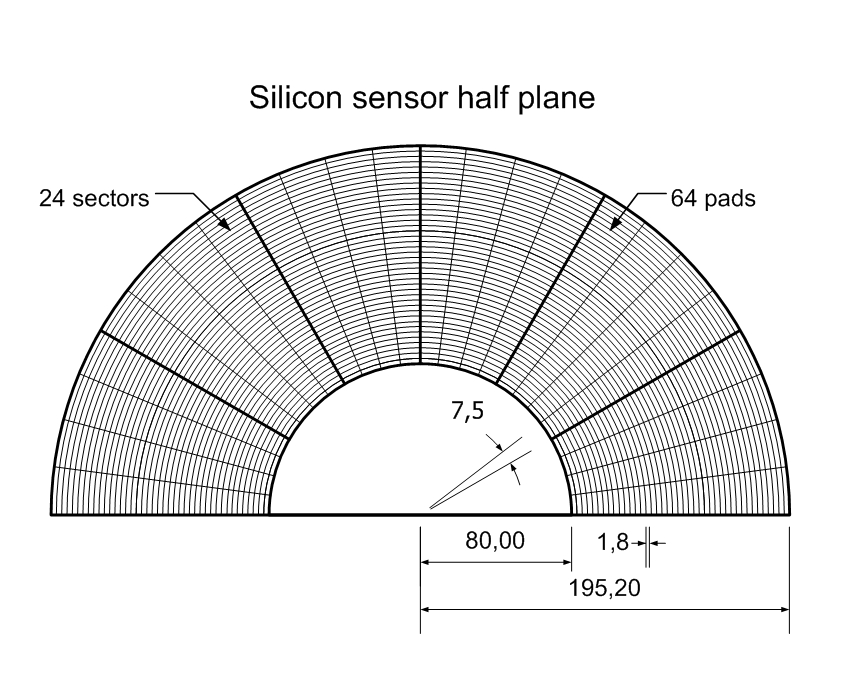}}
\subfigure[]{\label{fig:Gap_detail}
\includegraphics[width=.45\textwidth]{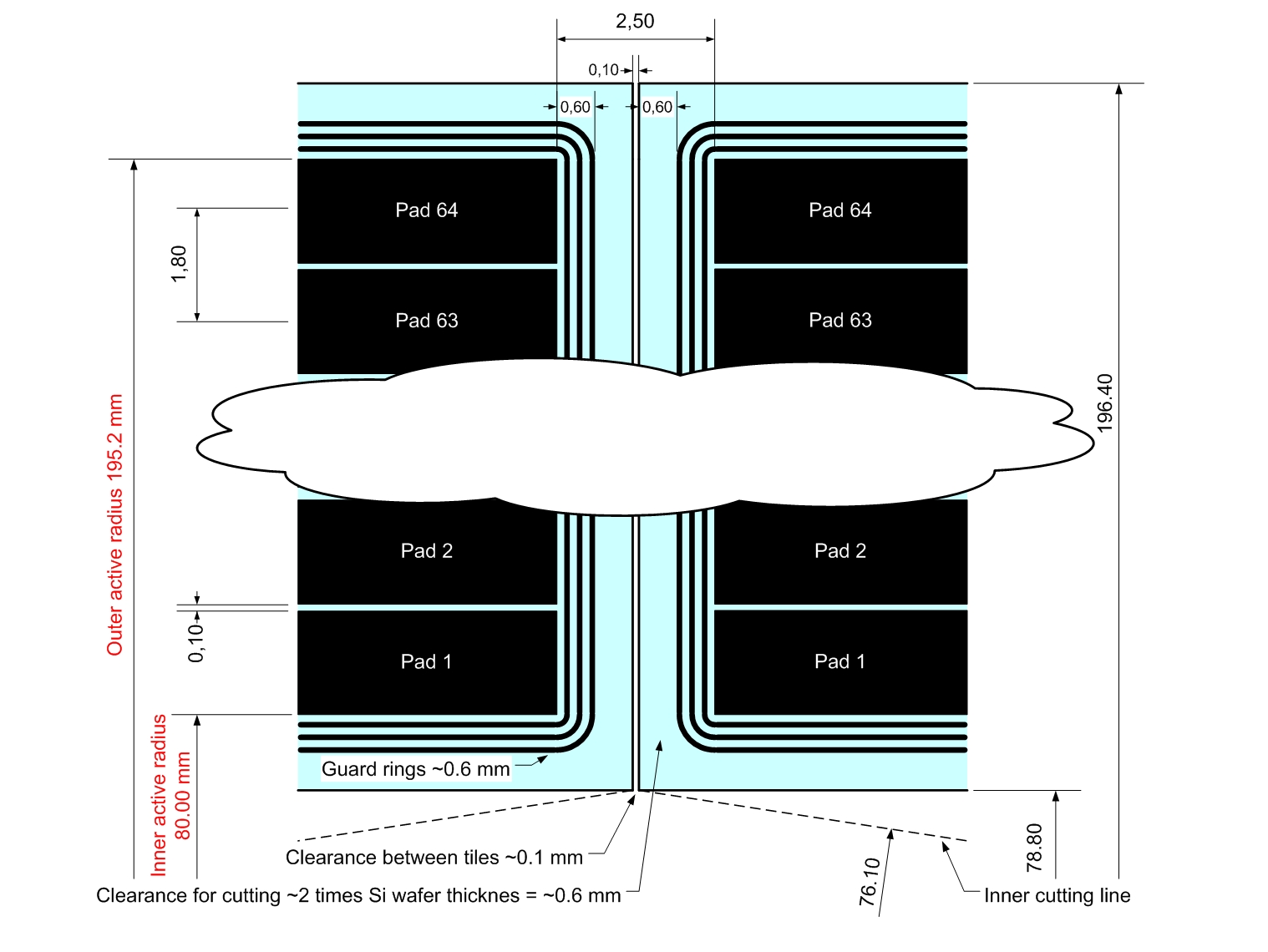}}
\caption{\Subref{fig:si-sensor-woj} A half plane of one of the LumiCal sensors.
 \Subref{fig:Gap_detail} Details of the gap-structure of the LumiCal silicon sensor.
}
\end{center}
\end{figure}

A complete LumiCal calorimeter will require 60 half planes of  silicon sensors (structured in 30 full sensor planes each called a basic module).  Such a half plane is shown in Figure~\ref{fig:si-sensor-woj} and details of the gaps between neighbouring silicon sensors are shown in Figure~\ref{fig:Gap_detail}.

The sensors prototypes were produced using 6-inch wafers by Hamamatsu Photonics with the following design parameters:
\begin{itemize}
\item
n-type silicon, p+ strips, n+ backplane;
\item
crystal  orientation $<100>$;
\item
320 $\mu$m  thickness  $\pm$15 $\mu$m;
\item
pad pitch  1.8 mm;
\item 
pad p+ width 1.6 mm;
\item
pad Al metallization width 1.7 mm;
\item
3 guard rings and an additional wider guard ring.
\end{itemize}
The latter allows for an undistorted electric field in the active area of the sensor.
\begin{figure}[h]
\begin{center}
\subfigure[]{\label{fig:I-V}
\includegraphics[width=.45\textwidth]{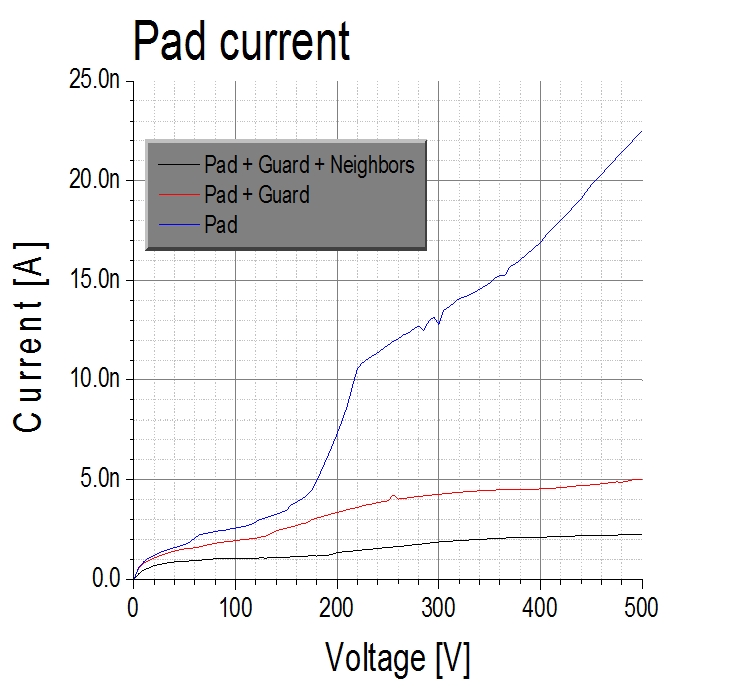}}
\subfigure[]{\label{fig:C-U}
\includegraphics[width=.45\textwidth]{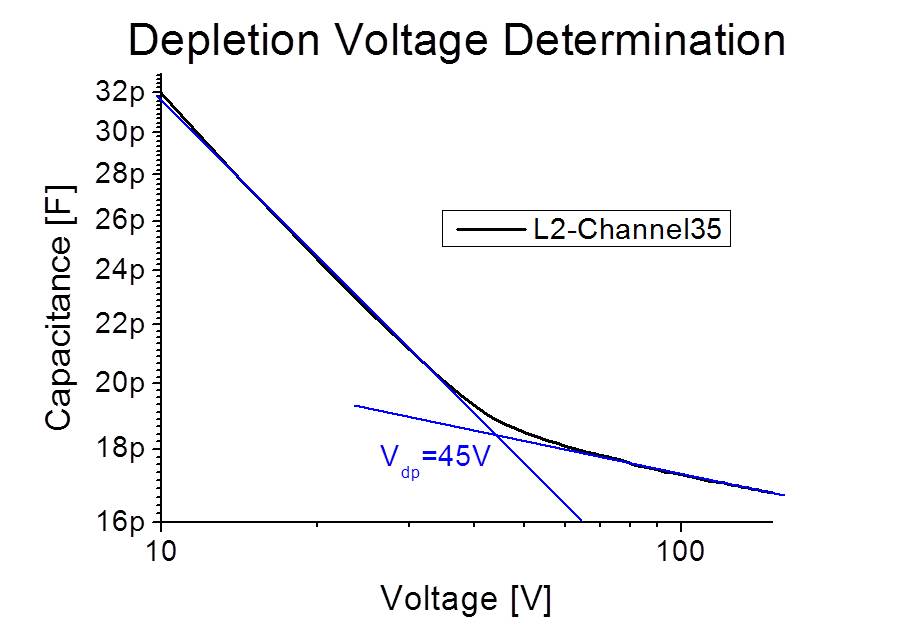}}
\caption{\Subref{fig:I-V} The current $I$ as a function of the voltage $V$ for various measurement setups.
 \Subref{fig:C-U} The capacitance versus voltage of a selected sensor.
}
\end{center}
\end{figure}
Delivered sensors were qualified by an optical inspection and current vs voltage as well as capacitance vs voltage measurements of individual pads. All measured parameters meet the design values.
Figure~\ref{fig:I-V} shows a typical current measurement of a single pad. Measured without neighbouring pads the current goes to higher values due to a non-homogenious electric field inside the bulk of the sensor. With all surrounding structures connected the current is significantly lower and stable up to 500V.
The capacitance vs voltage measurements were used to determine the full depletion voltage of the sensors. Figure~\ref{fig:C-U} shows such a measurment. The extrapolated full depletion voltage was found to be less than 50 volts for all measured pads.

The sensor qualification demonstrated a very good sensor quality.

\section{Sensors for BeamCal}
\subsection{Sensor requirements}
 BeamCal sensor pads adjacent to the beam-pipe are irradiated with a dose of a MGy per year of operation due to the depositions
 of low energy pairs from beamstrahlung. Therefore, the sensors for BeamCal must provide
radiation hardness to operate in these conditions. 

To avoid uninstrumented regions and due to limited space, no cooling is foreseen. The device will 
be operated at room temperature; this constraint rules out the use of currently available silicon detectors.
Two alternative sensor thechnologies were investigated, semi-insulating Gallium Arsenide and diamond. The detectors based on these 
materials function as a solid-state ionisation chamber. In both cases the detector does not have a p-n junction and the applied electric 
field only serves to provide charge carrier drift. The physcial properties of the material compared to silicon are given in 
Table~\ref{tab:TheMaterialProperties}. 

\begin{table}[htb]
    \centering
        \begin{tabular}{|c||c|c|c|}\hline
   & Silicon & Diamond & Gallium Arsenide\\
  \hline
Density, [g/cm$^{3}]$ &  2.33 & 3.52 & 5.32\\
Dielectric constant & 11.9 & 5.7 & 12.9\\
Band gap, [eV] & 1.12 & 5.45 & 1.42\\
e$^{-}$-h pair creation energy, [eV] & 3.6 & 13 & 4.3\\ 
Most probable energy  & & & \\
deposition per 300 $\mu$m & & & \\
 for 10 MeV e$^{-}$, [keV] & 82 & 130 & 167\\\hline

\end{tabular}
    \caption{Relevant Material properties of Silicon, Diamond and Gallium Arsenide.}
    \label{tab:TheMaterialProperties}
\end{table}

\begin{figure}[htpb]
  \centering 
  \subfigure[]{
	\label{fig:gaas} 
	\includegraphics[width=0.45\textwidth,angle=90]{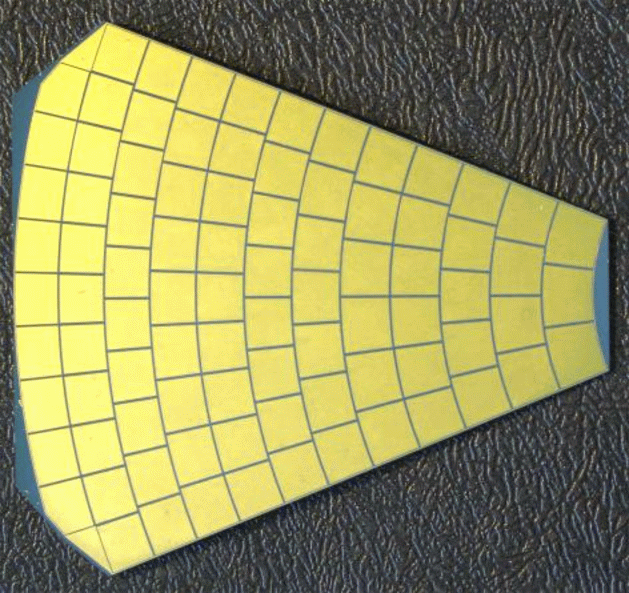}}
\subfigure[]{
     \label{fig:diam}
      \includegraphics[width=0.45\textwidth]{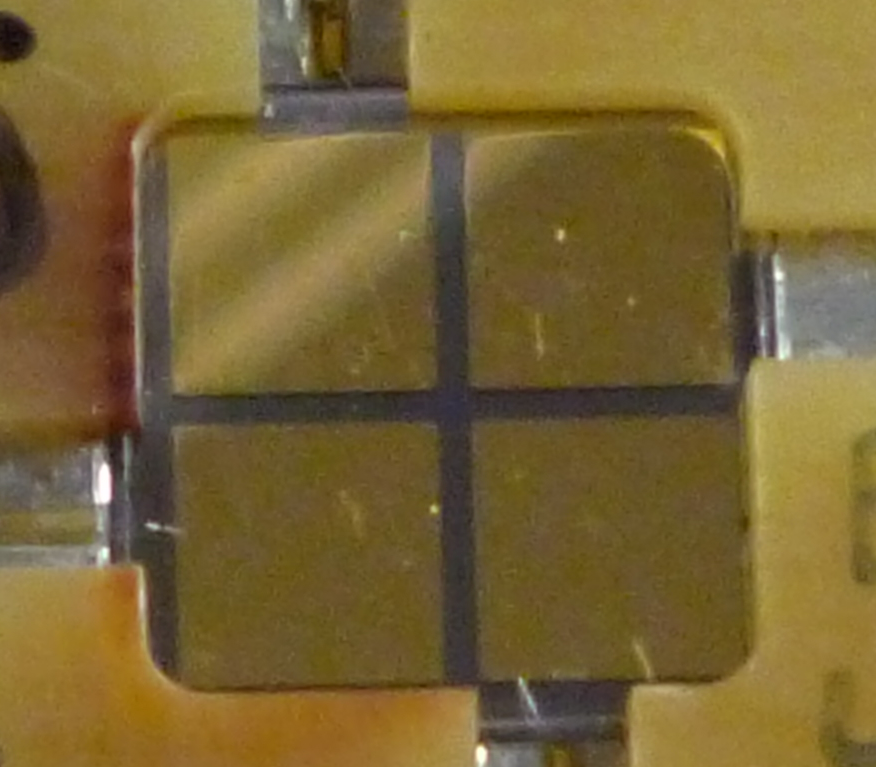}}
   \caption{
\Subref{fig:gaas} A prototype of a GaAs sensor sector for BeamCal. A gallium arsenide plate of 500~$\mu$m thickness has a shape of a 45$^{\circ}$ sector.  The metallisation layer is subdivided into 12 ring segments with innermost radius 20~mm and outermost radius 84~mm. Each ring segment is divided into pads of approximately 5$\times$5~mm$^2$ size.
\Subref{fig:diam} A polycrystalline CVD diamond sensor is 10$\times$10~mm$^2$ mm in size. The thickness of the sensor is 500~$\mu$m. The metallisation pattern defines 4 pads of 4$\times$4~mm$^2$ size}
  \end{figure}

\subsection{Gallium Arsenide sensors}

Large area semi-insulating GaAs sensors,
as shown in Figure ~\ref{fig:gaas}, were obtained from the Tomsk State University. They are produced using 
the liquid encapsulated Czochralski method and are doped with tin or tellur as shallow donors 
and compensated with chromium as a deep acceptor. 
This results in a semi-insulating GaAs material with a resistivity of about 10$^7$ $\Omega$m.

\begin{figure}[htpb]
  \centering 
  \subfigure[]{
  \label{fig:gaas_cce} 
      \includegraphics[width=0.45\textwidth]{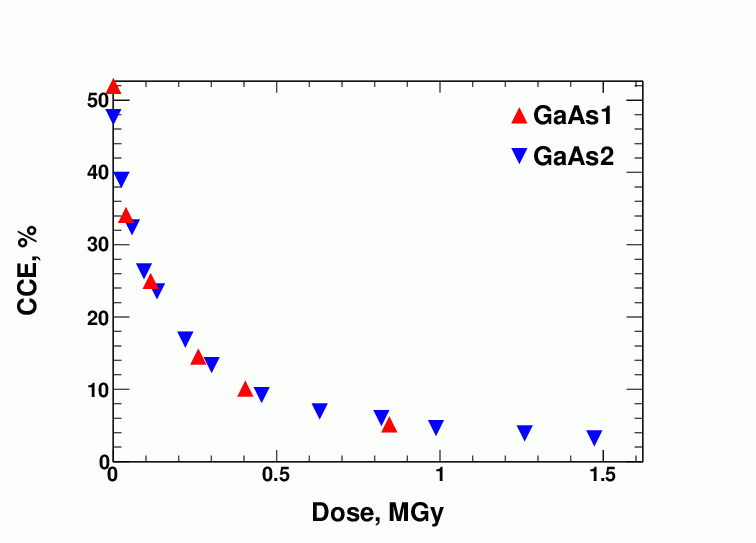}} 
  \subfigure[]{
     \label{fig:gaas_iv}
      \includegraphics[width=0.45\textwidth]{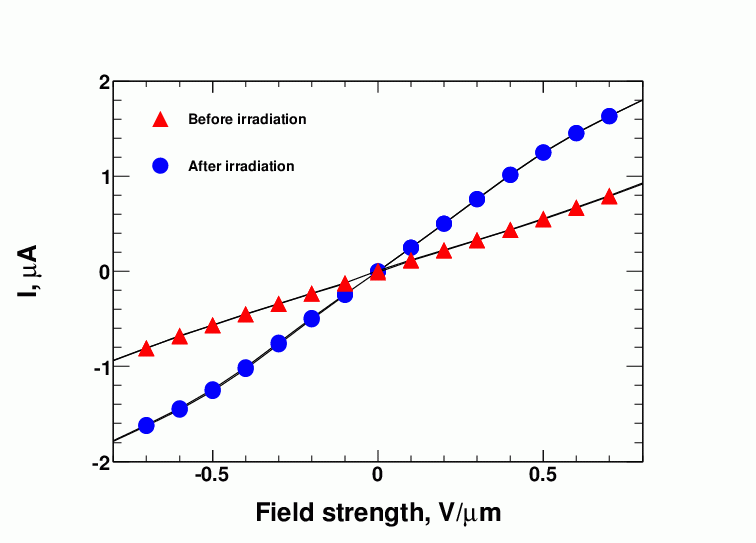}}
   \caption{
\Subref{fig:gaas_cce} The CCE as a function of the absorbed dose for the GaAs sensors.
\Subref{fig:gaas_iv} The dependence of the leakage current on the bias voltage 
for a single pad of the GaAs sensor before and after irradiation to a dose of 1.5 MGy.}
  \end{figure}

A study was performed to determine the radiation hardness of GaAs sensors~\cite{jinst_gaas}. 
Several sensors were irradiated in a 10 MeV electron beam. The maximum accumulated dose was 1.5 MGy. 
The charge collection efficiency, CCE, was measured as a function of the absorbed dose. The saturated value of the CCE reaches 
about 50\% for unirradiated samples. 
For irradiated samples the CCE is reduced to about 10\% of the initial value after an absorbed dose of 1 MGy, 
as can be seen in Figure~\ref{fig:gaas_cce}). The signals of minimum-ionising particles are still separated from the noise after a dose of 1 MGy, allowing for calibration of the gain of the sensor pads using e.g. muons from beam halo.

The dark current was measured to be about 0.45~$\mu$A per pad for a field strength of 
0.4~V/$\mu$m for the unirradiated sample and about 1~$\mu$A per pad after an absorbed dose of 1.5 MGy, as shown in Figure.~\ref{fig:gaas_iv}.    

\subsubsection{CVD diamond sensors}

An example of a polycrystalline chemical vapuor deposited, pCVD, diamond sensor is shown in Figure~\ref{fig:diam}. 
The radiation hardness and linearity of response were investigated for pCVD diamond 
sensors in test beams~\cite{ieee2,ieee3}.

\begin{figure}[h]
  \centering 
  \subfigure[]{
  \label{fig:diam_cce} 
      \includegraphics[width=0.45\textwidth]{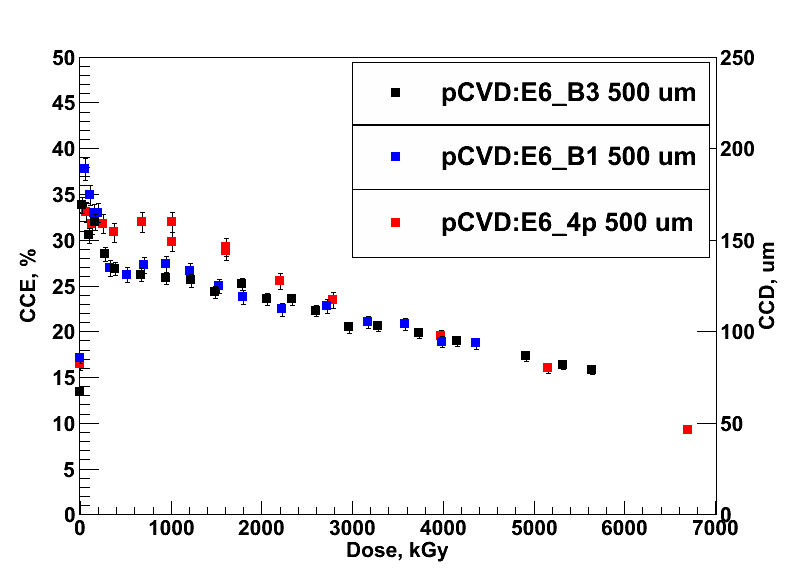}} 
  \subfigure[]{
     \label{fig:diam_linear}
      \includegraphics[width=0.45\textwidth]{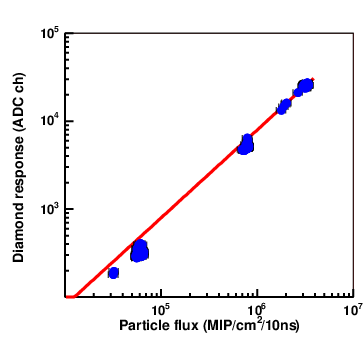}}
   \caption{
\Subref{fig:diam_cce} The CCE as a function of the absorbed dose for the pCVD diamond sensors.
\Subref{fig:diam_linear} The response of polycrystalline diamond sensors as a function
of the particle flux in units of the number of particles per 10 ns.}
  \end{figure}

\begin{figure}[h]
  \centering
  \includegraphics[width=0.45\textwidth]{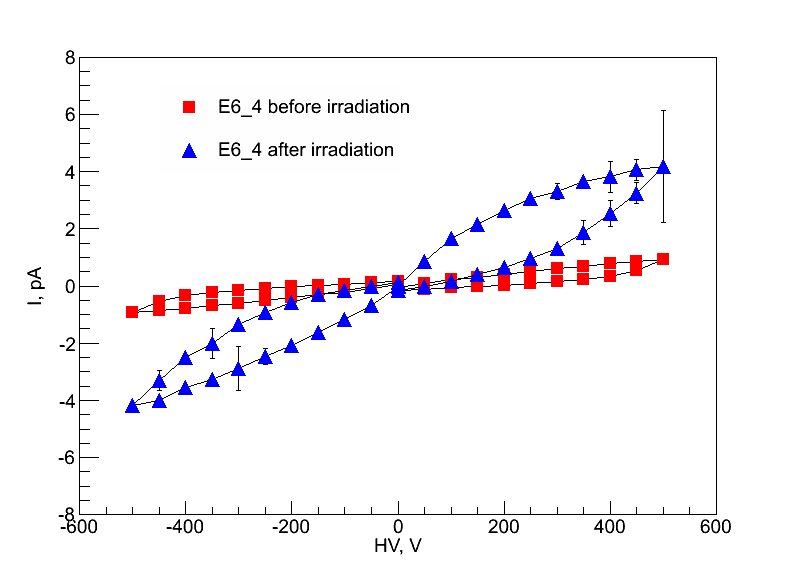}
  \caption{The dependence of the leakage current on the bias voltage 
for a pCVD diamond sensor before and after irradiation to an absorbed dose of 7 MGy.} \label{fig:ivdiam}
\end{figure}

The CCE of the sensors was about 15\% before irradiation, as can be seen in 
 Figure~\ref{fig:diam_cce}. 
After irradiation with a dose of about 10 Gy, the CCE value increased to 
about 35\%. This is known as a priming effect. After irradiation to an absorbed dose of 
7 MGy the CCE of the sensor dropped to approximately the initial value of the CCE or to 
about 30\% of the maximum value. The behavior is similar for different sensors. 
The leakage current, shown in Figure ~\ref{fig:ivdiam} as a function on the applied voltage, 
is less than 1 pA at 500 V for an unirradiated sample, and rises to about 4 pA 
after an absorbed dose of 7 MGy.

The linearity of the response of pCVD diamond sensors up to a particle 
flux of 5$\times$10$^6$ particles in 10 ns is 
demonstrated in Figure~\ref{fig:diam_linear}. The red line is a fit 
including also the response for a single minimum ionising particle.
Very good linearity over more than 6 orders of magnitude is found. 

Since large area CVD diamond sensors are very expensive, they may be 
used only at the innermost part of BeamCal, where the radiation level is highest.

\newpage

\section{ASICs for FCAL detectors}

\subsection{Development of BeamCal ASICs}
\label{ASICs_beamcal}

Regarding the development of BeamCal readout, progress in the following topics 
is reported: 
the development of a design-oriented noise analysis technique for charge amplifiers; 
the development of a mathematical framework for a design-oriented analysis of discrete-time 
filters in the discrete-time domain; and the design and implementation of a 10-bit fully-differential 
successive approximation register (SAR) ADC with configurable integral non-linearity (INL) to cancel 
out the charge amplifier non-linearity, along with the implementation of customized metal-oxide-metal (MOM) capacitors.

\subsubsection{Charge Sense Amplifier Design}
Typically, a low noise front-end design is achieved by following a simple recipe for the input 
transistor of the charge-sense amplifier (CSA): maximum available current, optimal capacitance 
matching at the input node, and minimum-length input device. These guidelines, obtained from 
analysis on simple transistor noise models and neglecting flicker noise, produce acceptable but 
sub-optimal results. We have developed a new technique for noise analysis in CSAs, which allows 
to find the optimal operation point of the input device for maximum resolution, and also provides 
a deeper insight on the noise limits mechanism from a more design-oriented point of view  ~\cite{angel1},{angel2}.

\subsubsection{Discrete-Time Pulse-Shaper}
Discrete-time filters represent a promising solution for pulse-processing in high energy physics 
experiments due to their flexibility and their capability to synthesize weighting functions with 
virtually any shape. One of the major concerns when designing one of these filters is to calculate 
the filter parameters that maximize the signal-to-noise ratio. In the context of the design of a 
switched-capacitor pulse-shaper for the BeamCal IC, we have developed a mathematical framework for 
a design-oriented analysis of discrete-time filters in the discrete-time domain. By using this 
analysis, a closed-form expression for the front-end noise can be obtained, suitable for computer 
automatic evaluation and optimization procedures~\cite{angel3}.

\subsubsection{Analog-to-Digital Converter}

\begin{figure}[h]
\centerline{\includegraphics[width=2.5in]{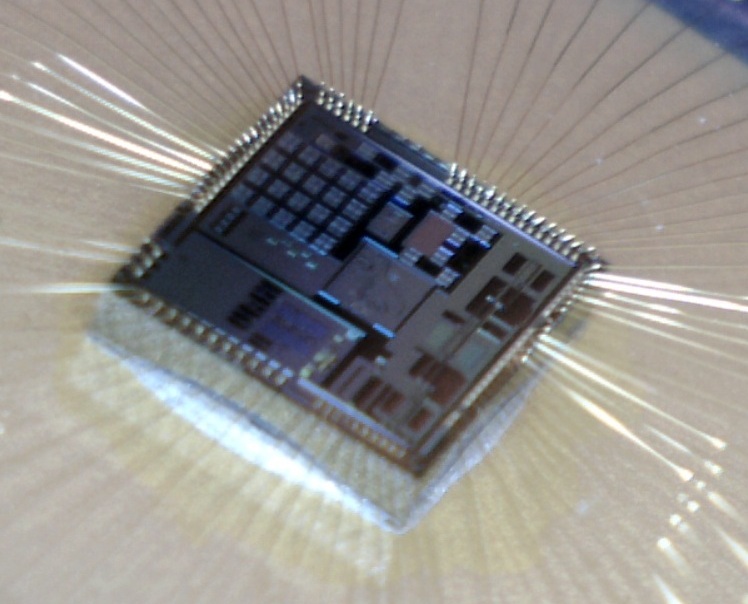}}
\caption{Fabricated IC to test different capacitors structures for the ADC.}
\label{chip}
\end{figure}

A 10-bit fully-differential successive approximation register (SAR) ADC with configurable 
integral non-linearity (INL) has been designed and implemented in a 0.18-$\mu$m technology. 
Figure~\ref{chip} shows the fabricated IC, which contains, among other circuits, three ADCs using different capacitor structures.

\begin{figure}[h]
\centerline{\includegraphics[width=4in]{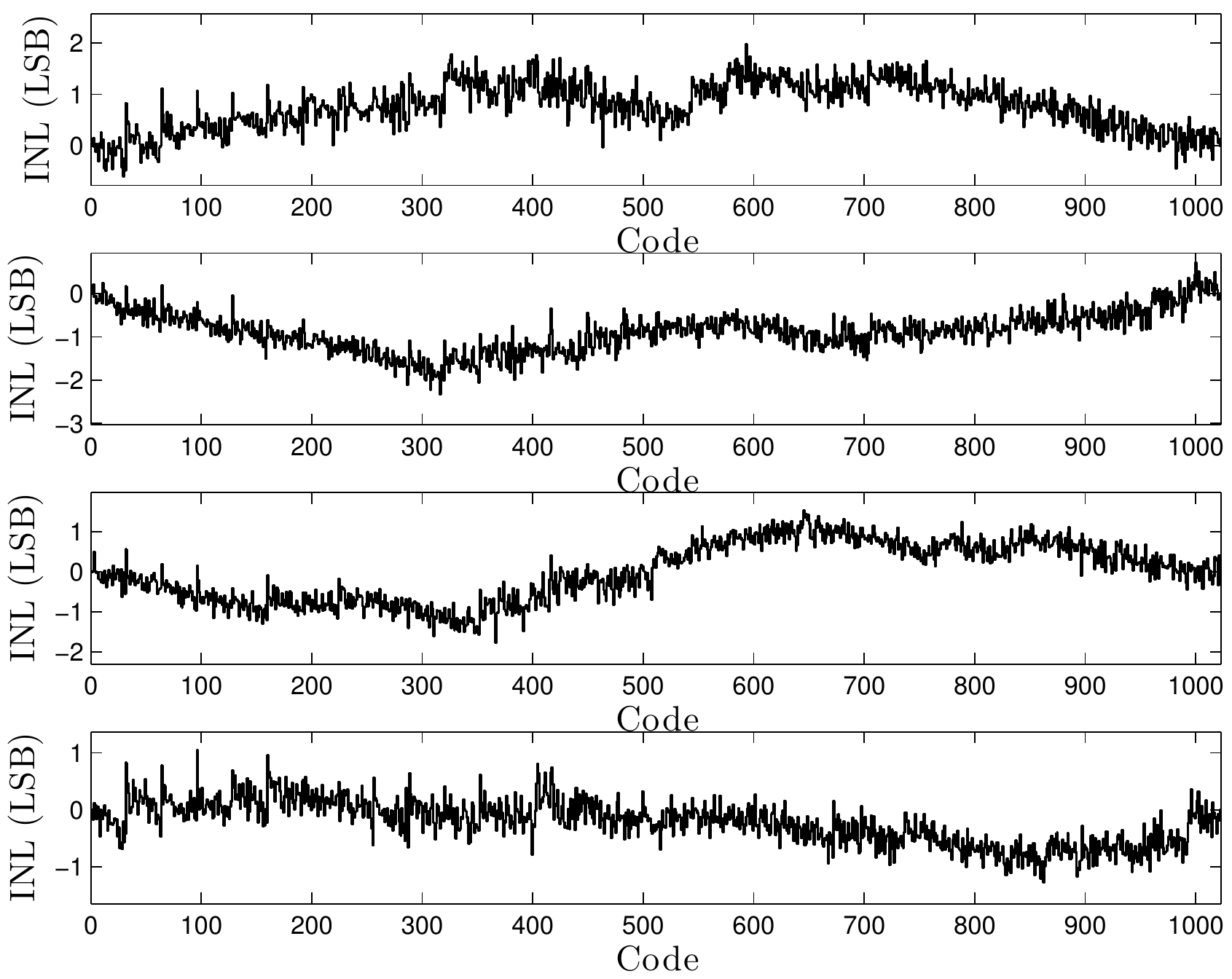}}
\caption{ADC non-linearity resulting from different INL shaper inputs.}
\label{inls}
\end{figure}

The ADCs comprise a charge-redistribution switched-capacitor digital-to-analog 
converter (DAC) network, the SAR logic and a voltage comparator preceded by an amplifier. 
All digital circuits in the ADCs are based on standard CMOS logic gates. The DAC capacitor 
array is divided into two sections: a binary-weighted one and a thermometer-coded one. 
Since the thermometer-coded capacitor array improves the ADC DNL, small capacitors (from 2 to 5~fF) 
were used to reduce the converter input capacitance and power consumption. The capacitors, 
implemented through a parallel connection of unit capacitors scattered over a 32x32 array, 
are subject to radial effects such as the copper dishing during the chemical and mechanical 
planarization (CMP), so the ADC INL is dependent on the order in which these capacitors are 
connected during the conversion process. The SAR logic contains a block called INL shaper, w
which consists of a combinational circuit that defines the order in which the DAC array 
capacitors are connected throughout a conversion, so the ADC INL can be manipulated and used 
to correct or cancel out the non-linearity of the CSA. As depicted in Figure~\ref{inls}, the 
results obtained so far show that the ADC non-linearity can be effectively manipulated by 
changing the order in which the capacitors are connected during a conversion.

\subsubsection{Current and Future Work}

Currently, the complete characterization of the fabricated ADCs is being performed. Also, the design 
of the CSA and the SC pulse-shaper using the techniques described above 
has already been started. The production of a single channel prototype of the BeamCal IC is planned for September, 2013.

\subsection{Development of LumiCal ASICs}
\label{ASICs_lumical}

The architecture of the LumiCal readout electronics depends on several assumptions 
and requirements concerning the luminosity detector~\cite{VFCAL_memo}. 
The readout should work in two modes: the physics mode and the calibration mode.
In the physics mode (low gain), the detector should be sensitive to electromagnetic showers
resulting in high energy deposition and the front-end electronics should process signals up to about 6-10~pC per channel.  
In the calibration mode (high gain), it should detect signals from relativistic muons, i.e.
should be able to register minimum ionizing particles (MIPs). 
The proposed sensor geometry (sensor and fanout) results in a wide range 5~--~50~pF of capacitive load connected to a single front-end channel. 
Because of high expected pad occupancy, the front-end 
should be fast enough to resolve signals from the subsequent beam bunches which 
are separated in time by about 350~ns.
The simulations of LumiCal indicate that the reconstruction procedure needs 
about 10-bit precision on the measurement of the deposited energy, which means that a 10-bit Analog--to--Digital Converter (ADC) is needed in the readout system.
Severe requirements set on readout electronics power dissipation
may be strongly relaxed if switching off the power between bunch trains 
is implemented. This is feasible since in the ILC experiment, after
each 1~ms bunch train there will be about 200~ms pause~\cite{beam}.
\begin{figure}[htb!]
\centerline{\includegraphics[width=1\columnwidth]{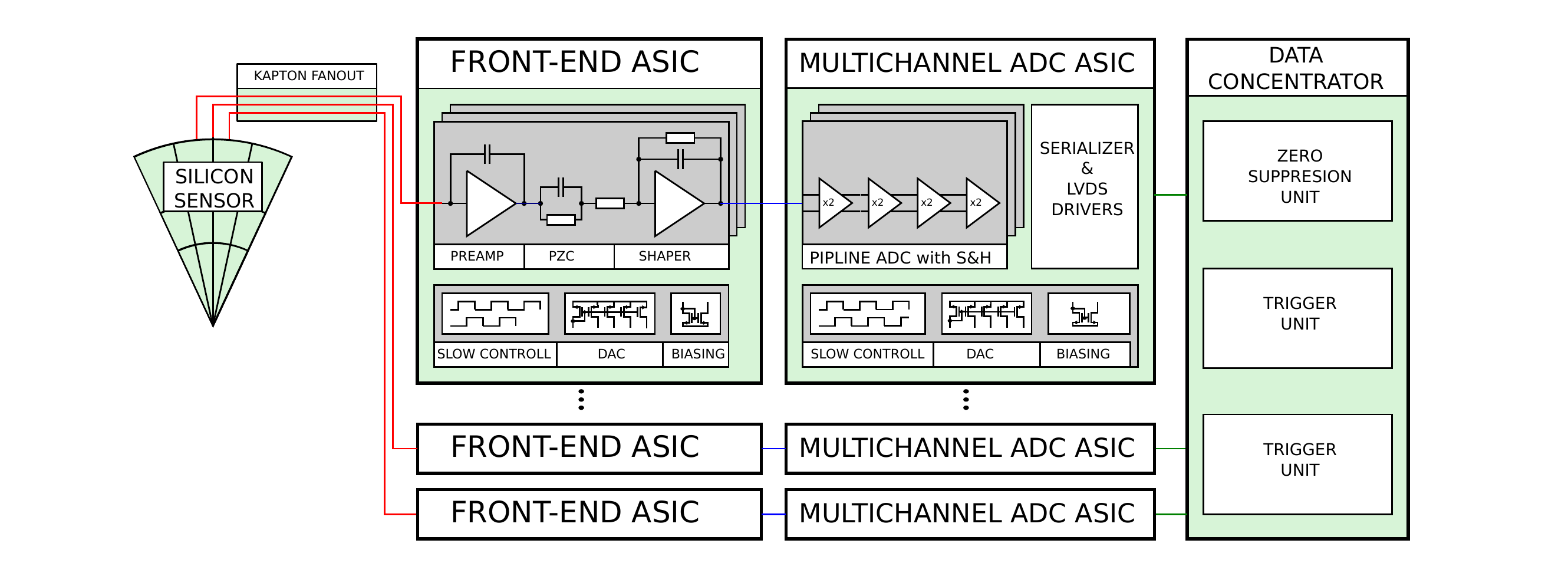}}
\caption{Block diagram of the LumiCal full readout chain.}\label{fig:fcal_lumical_system}
\end{figure}
From the above specifications, a general concept of the LumiCal readout chain 
is outlined, as shown in Figure~\ref{fig:fcal_lumical_system}. The main blocks of the signal processing chain are radiation sensor, front-end electronics, ADC conversion and data 
concentrator.

\subsubsection {Front-end electronics }
\label{fcal_lumical_fe}

\begin{figure}[h!]
  \begin{minipage}[t]{0.5\linewidth}
    \centering
    \centerline{\includegraphics[width=1\columnwidth]{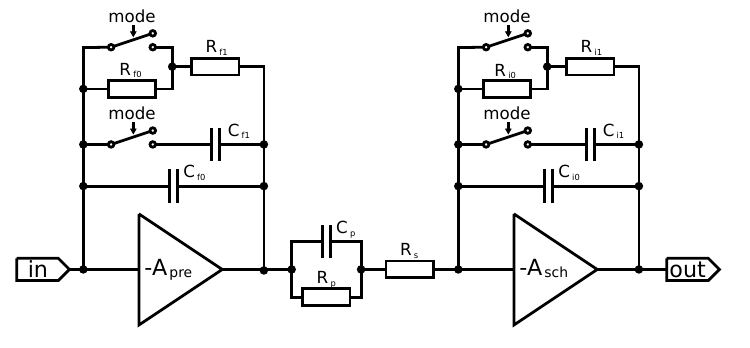}}
    \caption{Block diagram of front-end channel.}
    \label{fig:frontend}
  \end{minipage}
  \hspace*{0.02\linewidth}
  \begin{minipage}[t]{0.5\linewidth}
    \centering
    \centering \includegraphics[width=0.8\columnwidth]{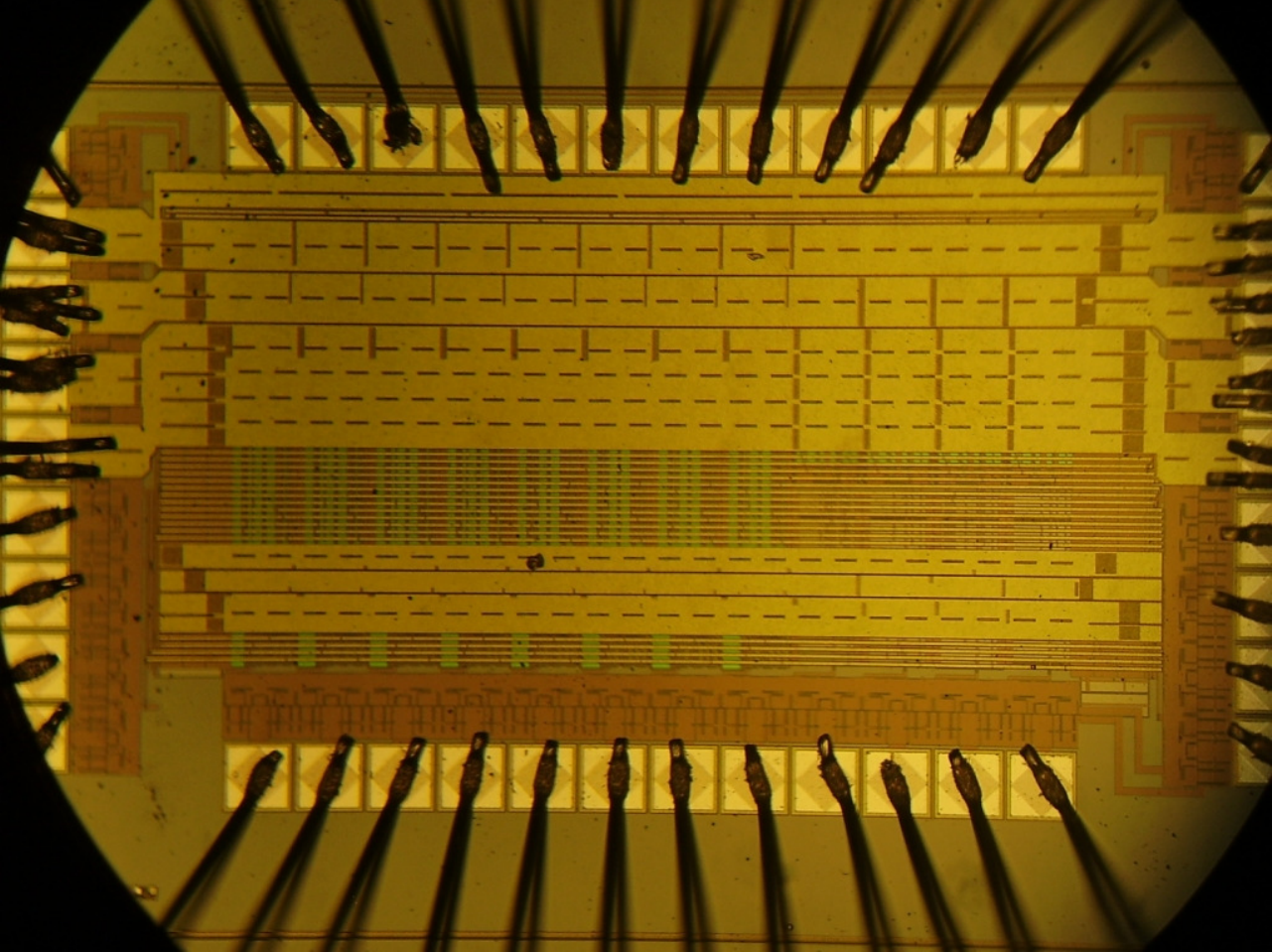}
     \caption{Micrograph of front-end ASIC.}\label{fig:frontend_photo}
  \end{minipage}
\end{figure}

To fulfill the above requirements, the front-end 
architecture~\cite{pzc,lumi_frontend} 
comprising a charge sensitive amplifier, a pole--zero cancellation circuit (PZC) and a shaper were chosen, as shown in Figure~\ref{fig:frontend}.
In order to cope with large charges in the physics mode and the small ones 
in the calibration mode a variable gain in both the charge amplifier and 
the shaper is implemented. The ``mode'' switch in Figure~\ref{fig:frontend} changes 
effective values of the feedback circuit components $R_f$, $C_f$, $R_i$, $C_i$ 
and so changes the transimpedance gain of the front-end.

Setting properly the PZC parameters ($C_f R_f = C_p R_p$) and equalizing 
shaping time constants ($C_i R_i = C_p (R_p || R_s)$) one obtains the first 
order shaping, equivalent to a CR--RC filter, with a peaking time $T_{peak}=C_iR_i$, which was set close to 60~ns.

The prototype ASIC, containing 8 front-end channels, was designed and fabricated in 0.35~$\mu$m, four--metal, two--poly CMOS technology. 
The area occupied by a single channel is $630\mu$m $\times$ 100$\mu$m. 
The photograph of the prototype glued and bonded on the PCB is shown Figure \ref{fig:frontend_photo}. 
After checking the basic front-end functionality  
systematic measurements of the essential parameters like gain, noise, 
high count rate performance and crosstalk,
were done, confirming the expected features of the ASIC. 
The detailed description and results of the measurements have been published elsewhere~\cite{lumi_frontend}.

\subsubsection {Multichannel ADC }
\label{fcal_lumical_adc}
To apply the analog--to--digital conversion in each front-end channel, a dedicated low power, small area, multichannel ADC is needed.
For the ILC detector a sampling rate of about 3~MS/s will be sufficient, but since at the design time the configuration of one ADC per eight channels was also considered, the solution with a general purpose variable sampling rate ADC with scalable power consumption was chosen. 
\begin{figure} 
 \begin{minipage}[t]{0.5\linewidth}
    %\centering
    \includegraphics[width=1\columnwidth]{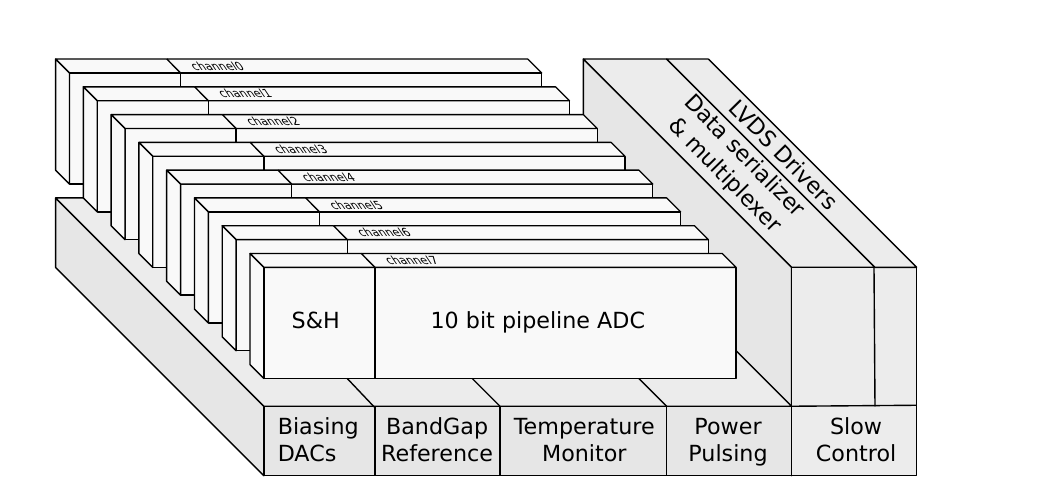}
    \caption{Multichannel ADC block diagram.}
    \label{fig:multi_adc_diagram}
  \end{minipage}
  \hspace*{0.02\linewidth}
  \begin{minipage}[t]{0.5\linewidth}
    %\centering
    \includegraphics[width=0.5\columnwidth,angle=90]{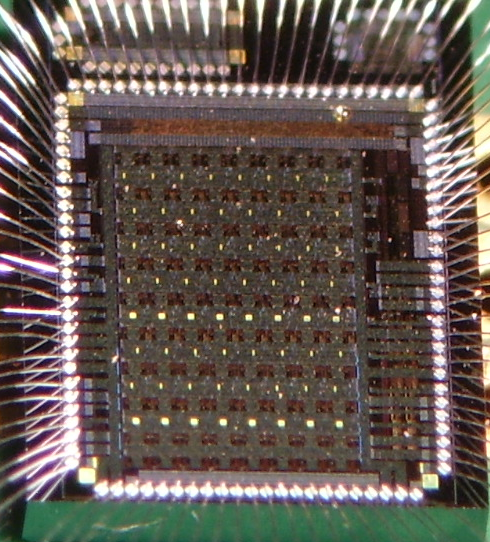}
    \caption{Micrograph of ADC ASIC.}
    \label{fig:adc_photo}
  \end{minipage}
\end{figure}

The block diagram of the designed multichannel digitizer ASIC~\cite{lumi_adc} is shown in Figure~\ref{fig:multi_adc_diagram}. 
It comprises eight 10-bit power and frequency (up to 24MSps) scalable pipeline ADCs, a configurable digital serializer circuit, 
fast power-scalable Low Voltage Differential Signaling (LVDS) I/O circuits, a set of Digital-to-Analog converters (DACs) for automatic internal current and voltage control, 
a precise band-gap voltage reference and a temperature sensor. The only external signals needed, apart from a power supply, are reference voltages to set the range of ADC input signal.
The developed digitizer comprises also the power pulsing functionality. In the current version this feature is implemented in all ADC channels, while the band-gap, temperature sensor and DACs are working continuously. About 10 ADC clock periods are needed to restart the correct ADC operation.  

The prototype ASIC was fabricated in 0.35~$\mu$m, four-metal two-poly CMOS
the analog and digital peripheral circuits are on the ASIC sides.  
After positive tests of the ASIC functionality, the
complete measurements of ADC static parameters, dynamic parameters, power scaling, crosstalk, and others, were performed, 
showing an effective resolution ENOB=9.7. The detailed results have been published elsewhere~\cite{lumi_adc}.

\subsubsection {New developments in 130 nm CMOS technology}
\begin{figure}[h!]
\centering
\includegraphics[width=0.4\columnwidth,angle=90]{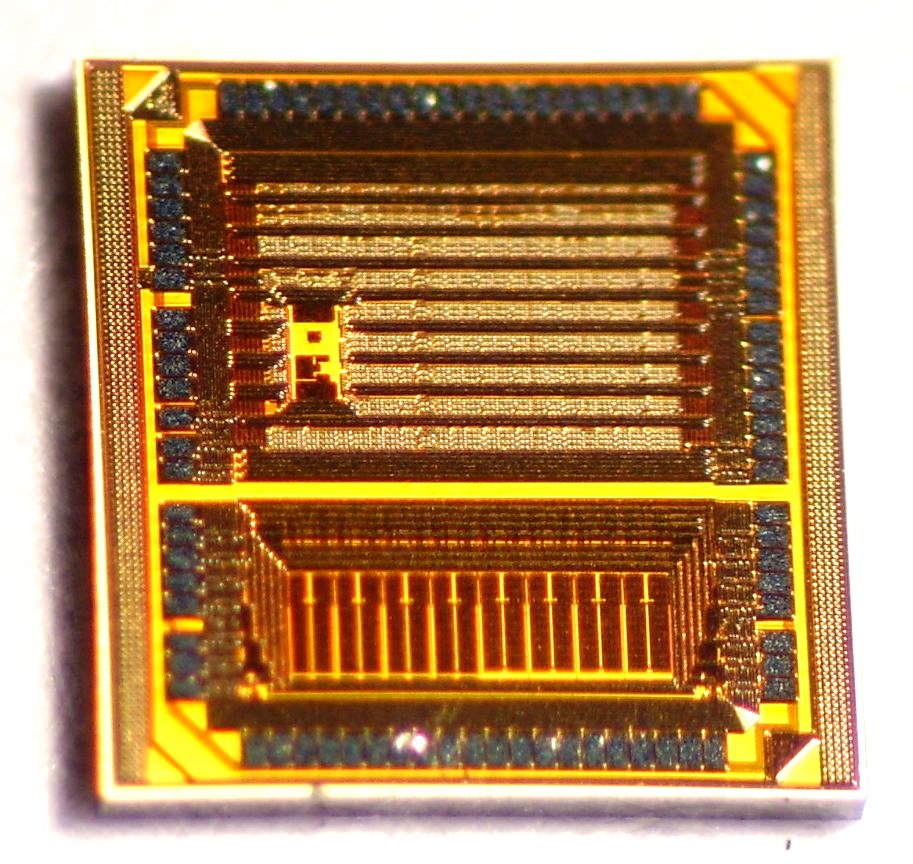}
\caption{Photograph of ADC prototype in CMOS 130 nm technology.}
\label{fig:photo_adc_ibm_10b}
\end{figure}

The described prototype system containing multichannel front-end and ADC ASICs, 
developed in 0.35~$\mu$m CMOS technology, has been already used successfully in test-beam measurements, 
serving for the readout of both the LumiCal and the Beamcal prototype detectors.  
In view of the final LumiCal readout, having in mind the importance of extremely low power 
consumption and taking into account possible radiation hardness issues, a new development of the front-end and ADC ASICs has been started in deep submicron 130~nm CMOS technology.
The overall readout architecture has not been changed and comprises the front-end and ADC in each readout channel. 
For the front-end the goal is to decrease the power consumption per channel by about a factor of 5, and achieve 
a peak power of about 1-2~mW. The architecture of the ADC has been changed from pipeline to successive approximation (SAR) 
in order to reduce the power consumption by more than one order of magnitude.
First prototypes of both front-end and ADC ASICs have been already designed. The ADC has been fabricated and is presently under tests. The photograph of a bare chip is shown in Figure~\ref{fig:photo_adc_ibm_10b}. First measurements have shown good functionality of SAR ADC and 
more detailed measurements are in progress.

\subsection{Integration and DAQ}

On the basis of the developed front-end and ADC ASICs in 0.35~$\mu$m AMS CMOS technology,  the prototype module of the LumiCal readout system was developed and built~\cite{fcal_readout_board}.
The block diagram of the developed system reflects the architecture
shown in Figure~\ref{fig:fcal_lumical_system}.
In order to increase the system flexibility and to allow the operation with different sensors, the signal is sent to the front-end electronics through a multi-way connector.
The signal is then amplified and shaped in the front-end ASIC, and
digitized in the multichannel ADC ASIC, continuously sampling the front-end output.
There are four pairs of front-end and ADC ASICs, 8~channels each, giving 32 channels in total. 
Power pulsing was implemented to the ASICs on the readout module. For the ADC it was done simply by sending a control signal, while for the front-end, additional external switches disconnecting the bias currents were used. 
To allow precise analysis of detector data, 
i.e. the reconstruction of event amplitude and time (time reconstruction is needed in the test-beams), and to perform pile-up studies, special attention was given to ensure high enough ADC sampling rate and very high internal data throughput between the ADC and the FPGA. The signal is sampled with 20~MSps rate, and digitized with 10-bit resolution, resulting in the raw data stream of about 6.4~Gbps.
The digitized data stream is continuously recorded in a buffer inside the Field Programmable Gate Array (FPGA).
When a trigger condition occurs, the acquisition is interrupted and the micro-controller firmware builds an event packet and transmits it to a host PC.

\begin{figure}[h!]
\centering
\includegraphics[width=0.4\columnwidth,angle=90]{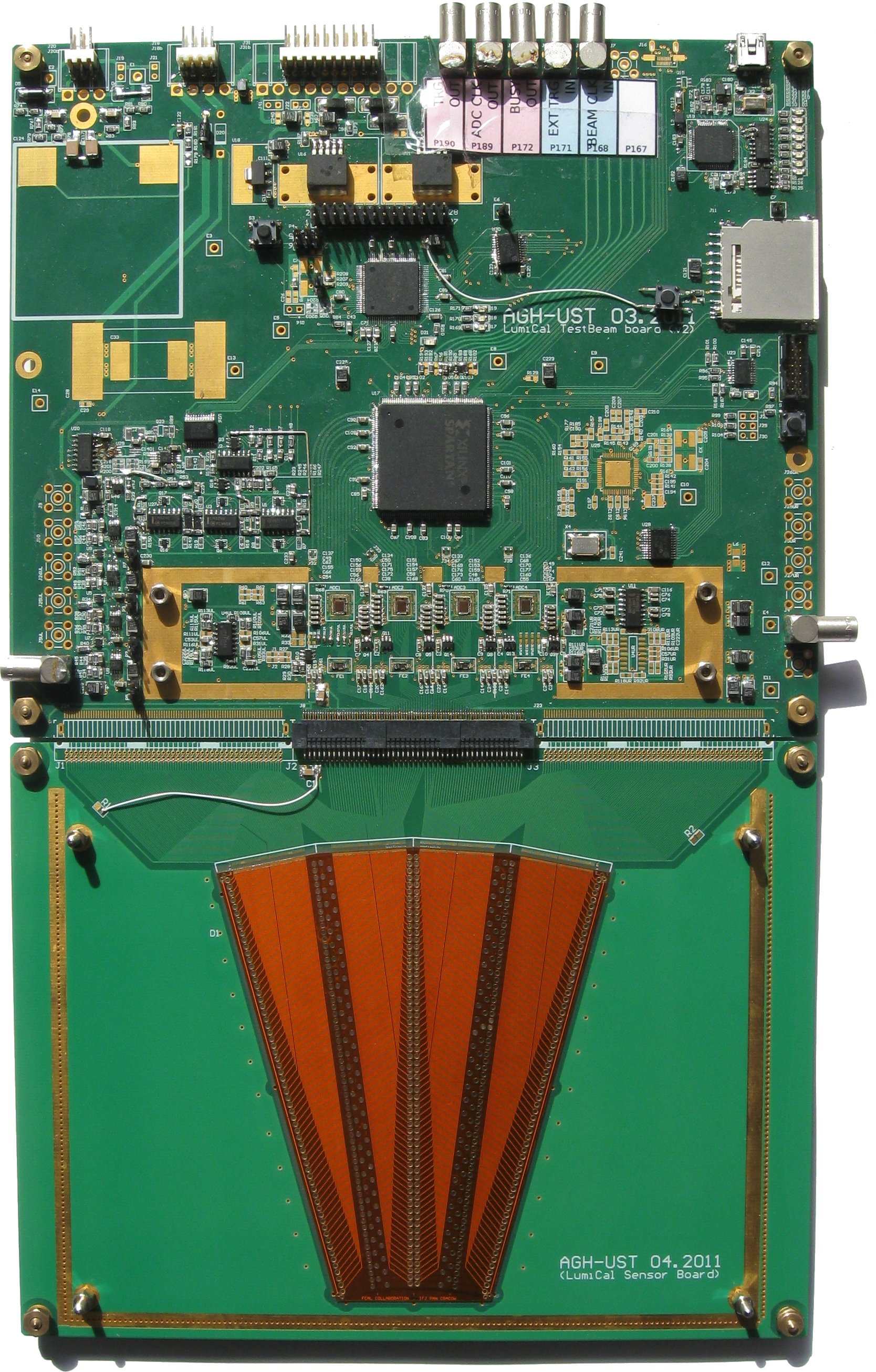}
\caption{Photograph of LumiCal readout module with sensor connected.}
\label{fig:fcal_lumical_module_photo}
\end{figure}

The prototype system was fabricated and assembled on a 6-layer Printed Circuit Board~(PCB). 
The photograph of assembled readout system, with the LumiCal sensor connected, is shown in Figure~\ref{fig:fcal_lumical_module_photo}.
To verify and to quantify the multichannel readout performance, various laboratory measurements of different system sections  and the complete readout chain were done.
For the analog part, the measurements of input dynamic range, gain, and noise and their uniformity were performed. 
For the digital part, the data transmission rate and the event trigger rate in different readout conditions were measured. 
The performance of the complete readout system in self-triggering mode was verified with cosmic rays. Additionally, 
the measurements of power consumption and thermal system behavior in the power pulsing mode were done. All the performed measurements are described and published in~\cite{fcal_readout_board}.  

The developed readout system may be used both with the LumiCal and the BeamCal sensor boards. It has been successfully used in last years to collect data during the LumiCal and BeamCal test-beams.

\newpage
\section{Laser alignment}

The control of displacements of the LumiCal during  
data taking  is necessary to ensure the required accuracy in luminosity measurements. 
For this purpose an alignment
system  is designed which will measure the absolute distances between both 
calorimeters as well as the relative position of each of them to a reference frame using the precisely positioned QDO magnets.
The required precision is better than  500 $\mu$m in beam direction over the distance of 5m,  
and better than 1 mm for transversal displacements.  
In addition, the inner radius of the silicon sensors layers has to be known with an accuracy better than 40 $\mu$m.
A system to monitor the position over short distances has been developed and studied in a test-bench~\cite{lc_p7,lc_p8}.

The midterm goal now is to build a prototype of a laser alignment system (LAS) 
to be used to monitor the positions of sensors and the calorimeter in the test-beam
and to integrate the measurements in the DAQ.

It combines two methods;
\begin{itemize}
\item an infra-red collimated beam from a diode laser and semi-transparent sensors, referred hereafter as TDS;
\item a tunable laser with interferometers and femtowatt photo-detector
 will be used to build a Frequency Scanning Interferometry (FSI) system 
to allow for a measurement of the absolute distances between calorimeters.
\end{itemize}
For a first prototype of the TDS, components of a system in the ZEUS experiment \cite{lc_p9 } will be used. 
The semi-transparent amorphous sensors, DPSD-516    
have light transmission above 85\% for a laser beam with $\lambda >$ 780 nm in an active  
area of 5 $\times$ 5 mm$^2$, allowing several sensors to be used  in one laser beam. 
Each sensor contains 16 horizontal 
and 16 vertical strips. Their signals are  used  for the determination of the mean position
of the sensor in the  X and Y  directions. 
The expected precision  of the position measurements is about 10 $\mu$m.
The  TSD system will be used for displacement monitoring of the calorimeter over short distances 
and the monitoring of displacements of the sensor layers.
The laser power will be about 5 mW.
Data are read from special PC boards   
 via a VM-USB VME controller with USB2 interface.  

\begin{figure}[h]
\begin{center}
\begin{minipage} [h] {7.cm}
\includegraphics[width=60mm]{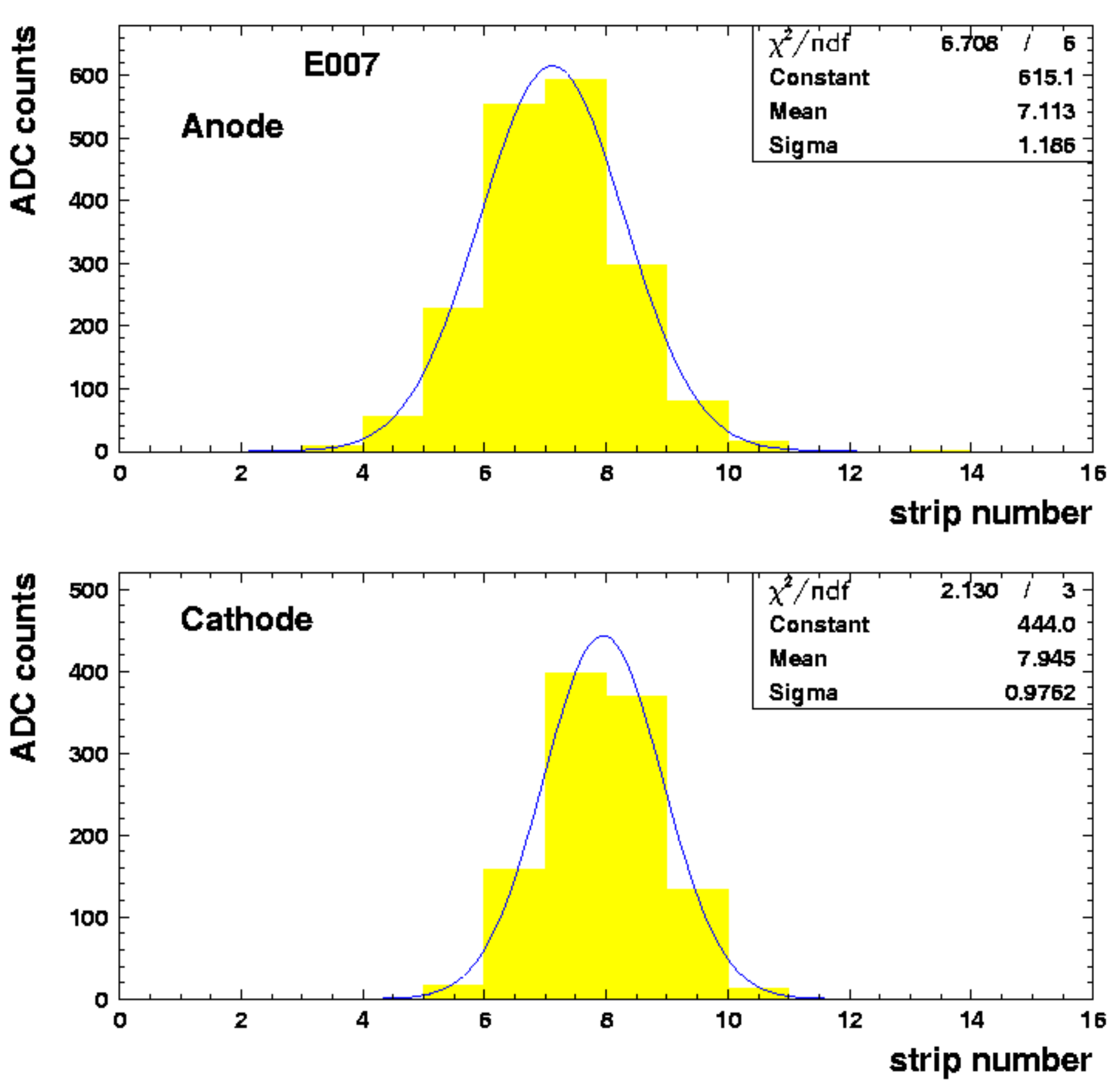}
\end{minipage}
\begin{minipage}[h] {7.cm}
\includegraphics[width=68mm]{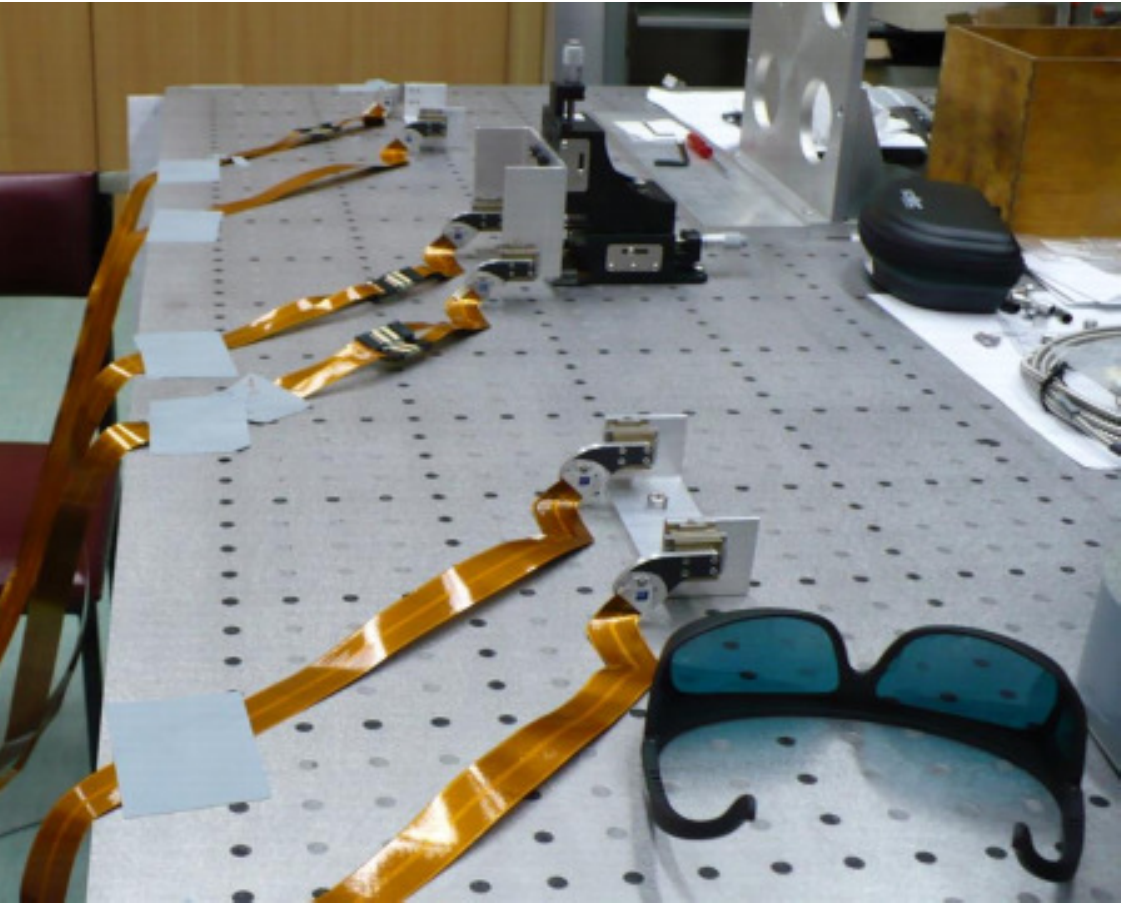}
\end{minipage}
\caption{Left: A typical  signal distribution  for a semitransparent sensor. 
Right:  A photograph of the test setup  which contains  six  semi-transparent sensors.}
\label{cr5}
\end{center}
\end{figure}

The functionality of the sensors has been tested on a probe-station using a laser beam.
Figure \ref{cr5} (left) shows as an example the signal distribution from X (cathode) and Y (anode) strips for one of the 
sensors.  
In Figure \ref{cr5} (right) the prototype of the TSD with  six semi-transparent sensors 
placed on the  optical table is shown.
Four of these  sensors were used to obtain the reference line
 whereas  the other two 
were mounted on a movable table. 
Moving these two sensors in steps of 100$\mu$m,  
data for the displacement measurement were taken.
By comparison with the precisely-known step-width, a measurement precision of about 10$\mu$m was obtained.

In the second method an interferometric length measurement, 
FSI~\cite{lc_p10, lc_p11, lc_p12, lc_p13} will be used. 
The  laser beams  from a tunable laser are split and sent to
a reference interferometer of known length  and to one of unknown length. 
From the measured phase shift in the interferometers the unknown length is measured. 

\begin {figure}[th]
\begin{center}
\includegraphics[width=140mm,height=98mm]{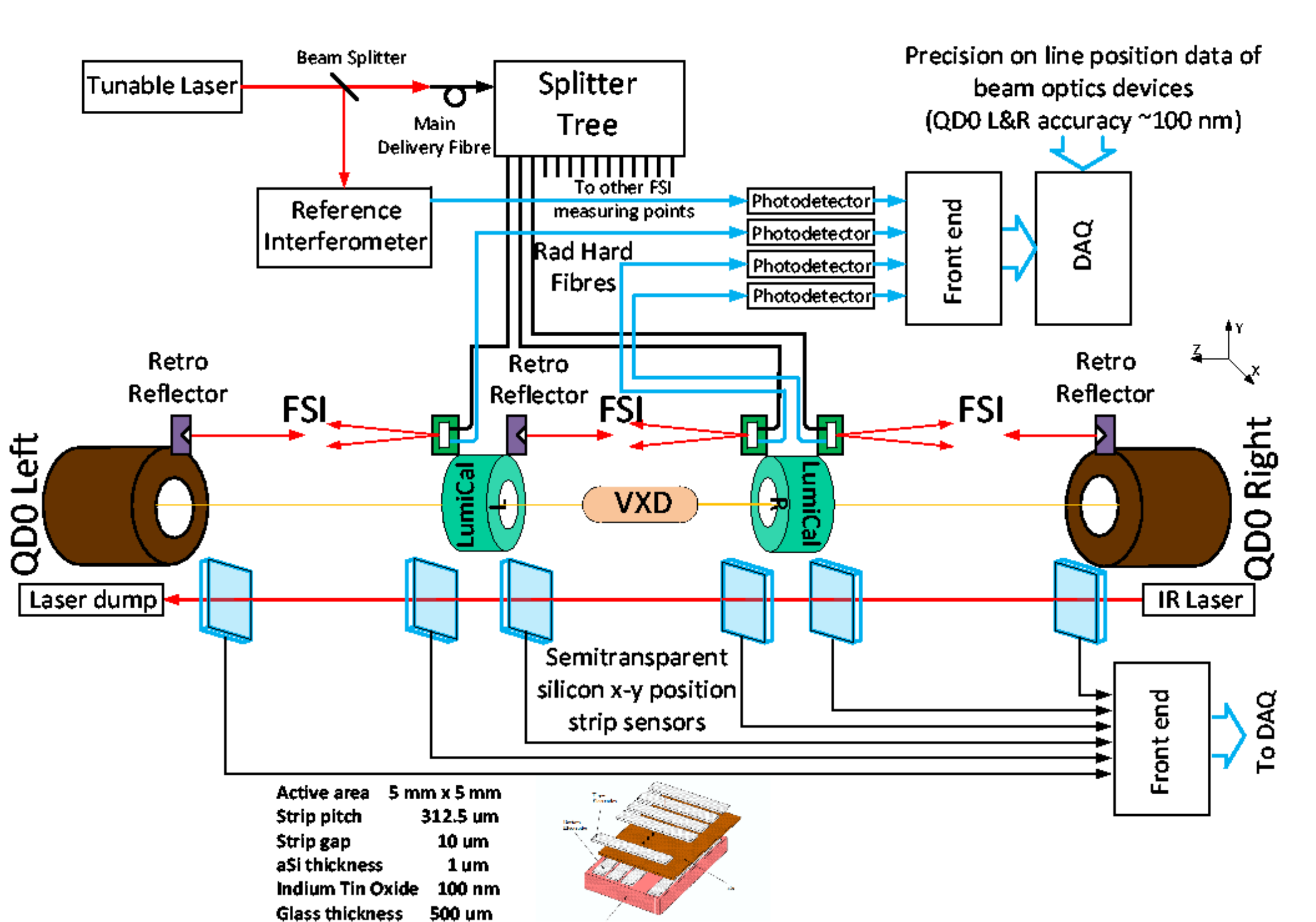}
\caption{Laser alignment  system proposed for the LumiCal detector.}
\label{cr2}
\end{center}
\end{figure}

Such an  FSI will allow an
absolute distance measurement between both LumiCal calorimeters and a monitoring of the displacement of each of the calorimeters relative
to a reference frame given by the QD0 magnet. 
The expected accuracy in the distance measurements will be a few $\mu$m over 5 m.
A concept is under development for an FSI system
integrated in a detector at ILC or CLIC. A  possible scheme is shown in 
Figure \ref{cr2}.

\section{Testbeam results}
During 2010 and 2011 the FCAL collaboration performed three beam tests.
These were the first tests of the LumiCal silicon- and the BeamCal GaAs-sensors prototypes equipped with a full readout chain. The readout chain included sensors, fan-outs, dedicated front-end electronics and, during the 2011 beam-tests, also the newly-developed 10-bit pipeline ADCs was included. 
The sensors performance results, shown briefly here, includes spectrum and correlation analyses, combinations between sensor information and position-reconstruction and the development of electromagnetic showers in the Tungsten absorber.

\subsection{Tests beam setup}
\begin{figure}[!htb]
\begin{center}
\includegraphics[width=0.5\textwidth]{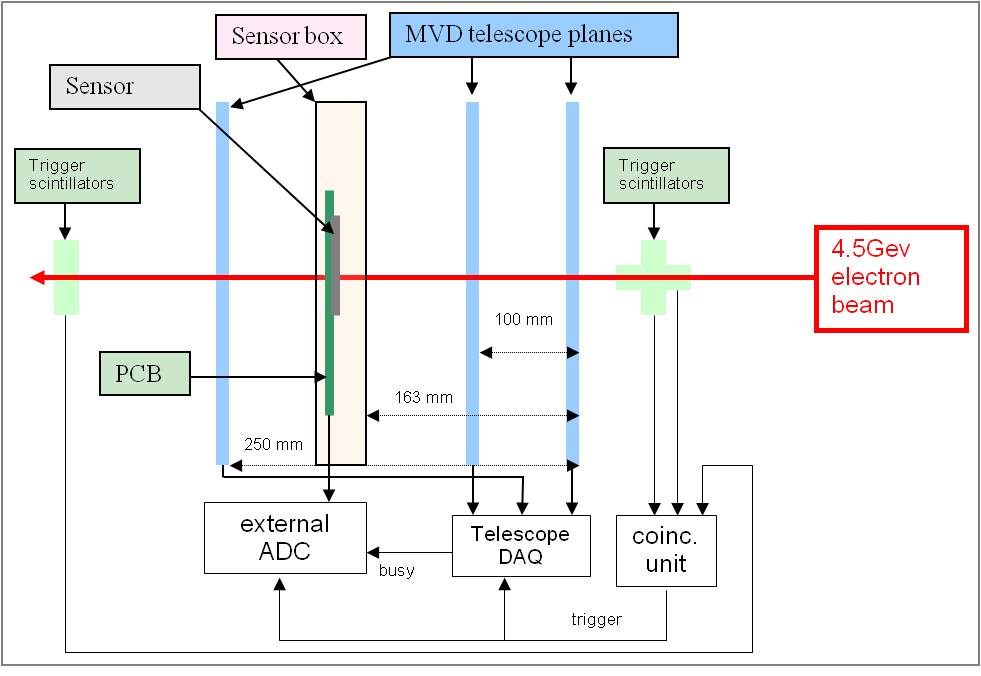}
\end{center}
\caption{Scheme of the beam tests set-up.}
\label{fig:setup}
\end{figure}

The beam tests were performed at the DESY-II ring with secondary electrons at a beam energy of 4.5~GeV.
A scheme of the set-up is shown in Figure~\ref{fig:setup}.
%\itamar_change\
The particles cross three planes of a silicon-strip telescope (MVD), equipped with horizontally
and vertically oriented silicon strip sensors each.
The horizontal and vertical beam size was about 5x5 $mm^2$.
In the 2010 tests the sensor was mounted between the second and third MVD plane inside a shielded PCB box. In 2011 the sensor was installed downstream of the three MVD planes.
The sensor was mounted on a remotely movable X-Y table. Special runs were used to align the MVD planes. A precision of better than 10 $\mu$m was reached.
Scintillation counters upstream and downstream of the setup are used in coincidence as trigger counters. 

\subsection{BeamCal}
\subsubsection{2010 beam test}

\begin{figure}[!htb]
\centering
\subfigure[]{\label{fig:fig:beamcal_2010_signal_a}\includegraphics[width=0.48\textwidth]{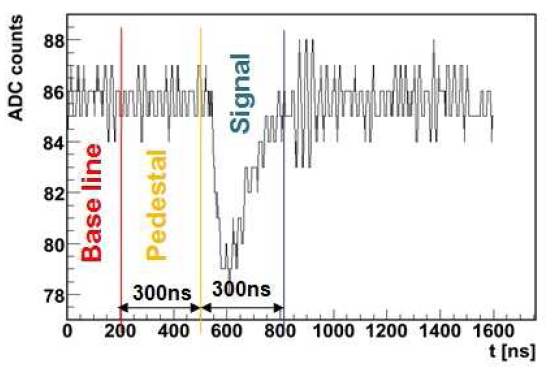}}
\subfigure[]{\label{fig:fig:beamcal_2010_signal_b}\includegraphics[width=0.48\textwidth]{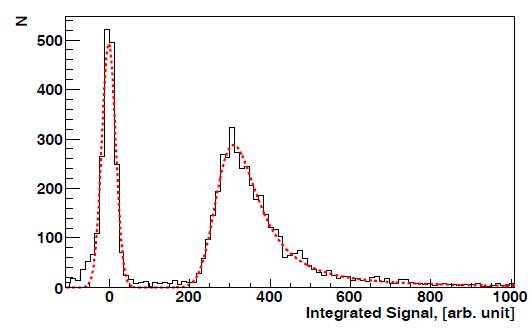}}
\caption{ (a) A digitized signal of one channel recorded with a 500 Ms/s ADC. (b)  The spectrum of triggered events. The signal is integrated over the wavefront in (a). }
\label{fig:beamcal_2010_signal}
\end{figure}

A GaAs sensor module was tested in 2010. Two areas of 8 pads each were connected
to via the fan-out to front-end ASICs and a stand-alone ADC was used. Several million triggers were recorded.
An example of a signal recorded by the 8-bit flash-ADC (CAEN
V1721) is shown in Figure~\ref{fig:beamcal_2010_signal} together with a spectrum of the signal size obtained as the integrated amplitude waveform over time. 
Each trigger was analysed using several time windows of the recorded wave-form.
From the first 100 ADC values the baseline is determined.
In the following two time windows of 300 ns each the charge is integrated, first to measure the pedestal, and then to record the signal. 
The spectrum is well described by a Gaussian for the pedestal and a Landau distribution convoluted with
a Gaussian for the signal. From the most probable value of the signal and the width of
the pedestal distribution signal-to-noise ratios of better than 20 are measured for all 16
investigated GaAs pads.

The signal size determined from the fit is also used to determine the charge collection
efficiency, CCE, as the ratio of the detected charge to the charge expected to be released by
the crossing particle. The latter is obtained by a Monte-Carlo simulation to estimate the
energy loss of a 4.5 GeV electron in the sensor and using the average energy to create an
electron-hole pair in GaAs. To measure the detected charge the readout-chain is calibrated
by the injection of a known charge. The CCE obtained for the saturated signals is 42\%. The
CCE measurement is compatible with lab measurements of GaAs sensors.
When checking CCE as function of bias voltage, it is found that above 60 V the CCE remains constant.

\begin{figure}[!htb]
\centering
\includegraphics[width=1\textwidth]{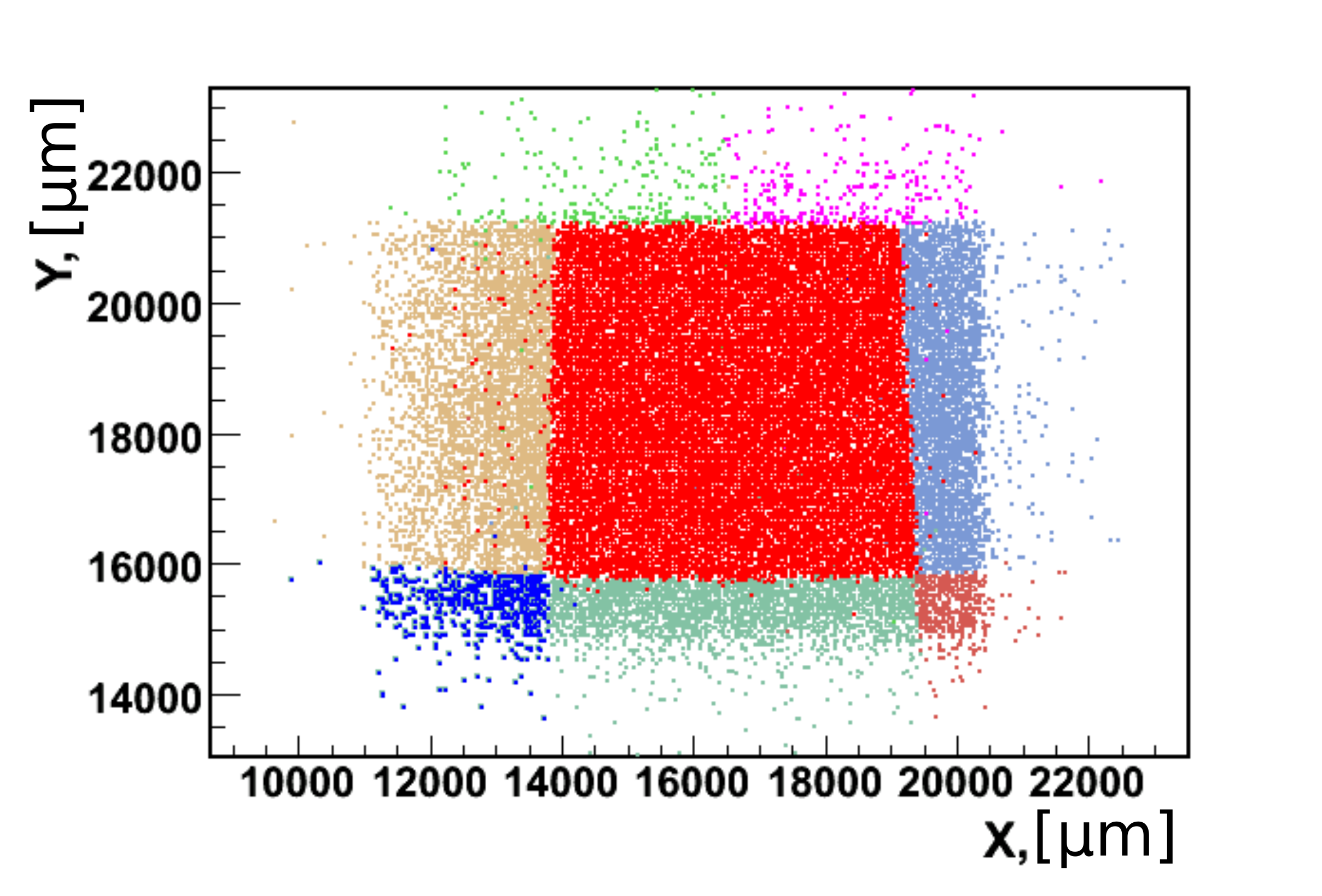}
\caption{Pads struture of the GaAs sensor reonstruted from telesope data.}
\label{fig:beamcal_hitmap}
\end{figure}

In Figure~\ref{fig:beamcal_hitmap} the impact points predicted by the MVD telescope are plotted over the area of approximately one pad on the GaAs sensor
The different colors, characterising the pads, are assigned to hits for which the signal in the
pad is above a certain threshold. The pad structure of the GaAs BeamCal sensor is accurately reproduced in the plot.

\begin{figure}[!htb]
\centering
\includegraphics[width=1\textwidth]{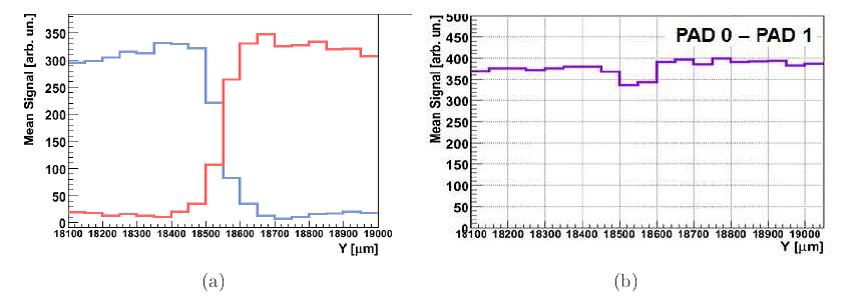}
\caption{(a) The signal integral mean value as a function of the hit position on the border
between two pads. (b) The signal integral sum of two pads as a function of hit position on the border between two pads.}
\label{fig:beacal_pad_gaps}
\end{figure}

Signal spectra have been compared for different areas on the pad and were found to be identical, i.e. the most probable values of the signal peaks agree within their measurement uncertainties.
High statistics runs allowed a scan over the gap of 200 $\mu$m in between pads. For events
with impact points near the pad boundary or in the gap, sharing of the signal between the
adjacent pads is observed. The sum of the signals of two adjacent pads is shown in
Figure~\ref{fig:beacal_pad_gaps} as a function of the impact point position in the coordinate perpendicular to the
gap. The signal sum drops by about 10\% in the region of the gap.

\subsubsection{2011 beam tests}
\begin{figure}[!htb]
\centering
\includegraphics[width=1\textwidth]{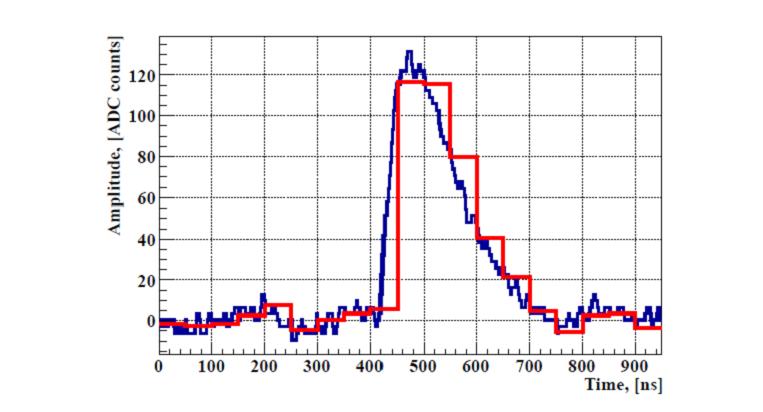}
\caption{An example of the signal digitized by the external ADC (blue) and the ADC ASIC (red).}
\label{fig:beamcal_signals_2}
\end{figure}

\begin{figure}[!htb]
\centering
\includegraphics[width=1\textwidth]{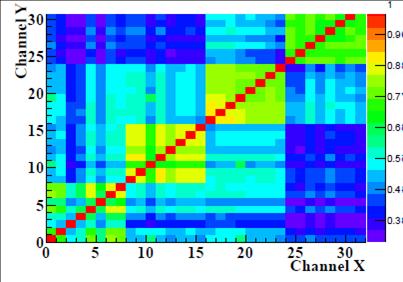}
\caption{Correlation coefficients between the ADC sample values of 32 channels.}
\label{fig:beacal_correletion32}
\end{figure}

For the 2011 tests the BeamCal sensor had two readout systems, one with the internal 10-bit pipe-line ADC and one using an external ADC as in 2010.
Signals from the two different ADCs are compared  in Figure~\ref{fig:beamcal_signals_2}.
The shape of the digitized  signals is very similar, given the different sampling time.
Amplitude and integral of the signals were calculated with the same procedure as described in the previous section.
The 2011 test differ for the BeamCal sensor from the 2010 test in the amount of Common Mode Noise (CMN) measured.
Due to the CMN, a high level of correlation between channels, shown in Figure~\ref{fig:beacal_correletion32}, was found.  
A method to correct for the common mode noise is developed and applied off-line to the data.

\begin{figure}[!htb]
\centering
\includegraphics[width=1\textwidth]{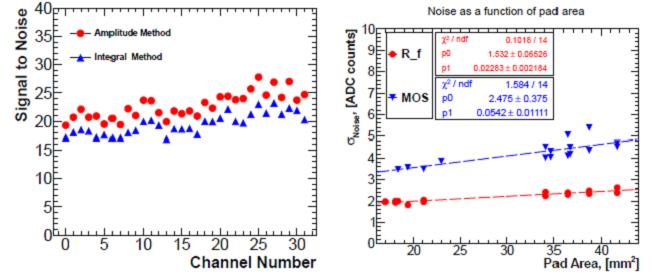}
\caption{(a) The S/N of all measured channels for the amplitude method (red dot) and
integral method (blue triangles). (b) The pedestal standard deviation as a function of the pad area.}
\label{fig:beacal_snr}
\end{figure}

The gain in the preamplifier and shaper can be switched between "Low" and "High"; altogether four combinations are possible.
The mode "Low Low" will be  used as physics mode, i.e. for electromagnetic showers with high energy deposition in the sensor.
The mode "High High" is intended for the measurements of MIPs for calibration and alignment.
Data were taken for all 4 combinations of gains, but only "High High" and "High Low" modes are sensitive to MIP signals and can be used to measure the S/N.
The S/N results obtained are shown in Figure~\ref{fig:beacal_snr} using the integral over the signal or the peak value of the amplitude. 
With the amplitude method a S/N of more than 20 was obtained.
The variation between channels is less than 16\%. 

In both test beam measurements the MIP position is shown to be stable for all measured channels.
Figure~\ref{fig:beacal_snr}(a) shows a slight increase of the S/N with channel number.
In Figure~\ref{fig:beacal_snr}(b) is shown that the noise linearly depends on the pad size. 
This is consistent with the expectation that the noise depends on the pad capacitance, thus the pad area.

\begin{figure}[!htb]
\centering
\includegraphics[width=1\textwidth]{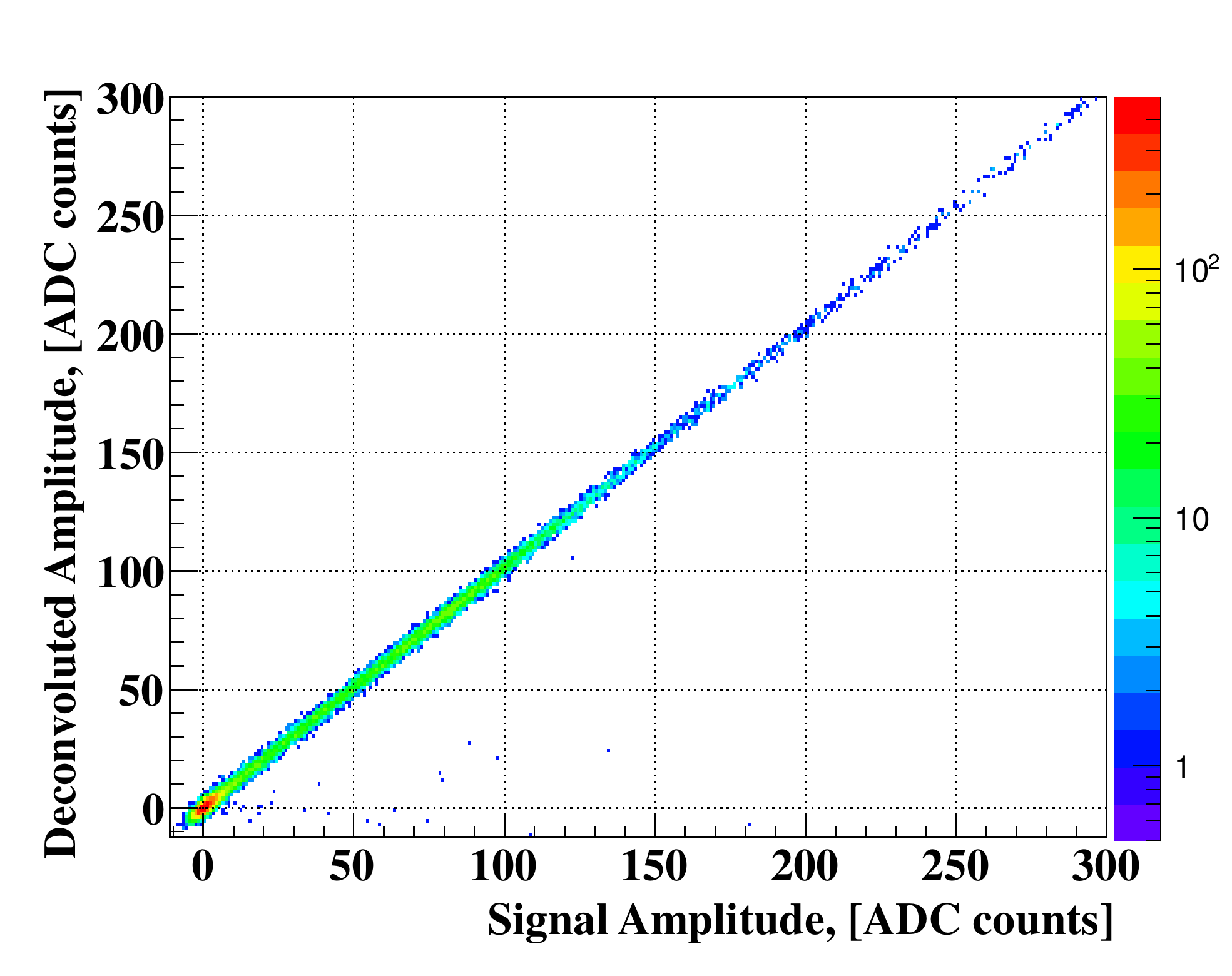}
\caption{The deconvoluted amplitude as a function of the signal amplitude for the synchronous read out mode.}
\label{fig:beacal_dcomvolotion}
\end{figure}

During the 2011 beam tests data were taken in both the synchronous (ILC like) and the asynchronous (beam test) mode.
To analyze data from the synchronous mode a deconvolution method ~\cite{Kulis_phd} was applied and its performance studied.
The resulting comparison between the signal amplitude from the ADC and the output of the deconvolution method is shown in Figure~\ref{fig:beacal_dcomvolotion}.

\begin{figure}[!htb]
\centering
\includegraphics[width=1\textwidth]{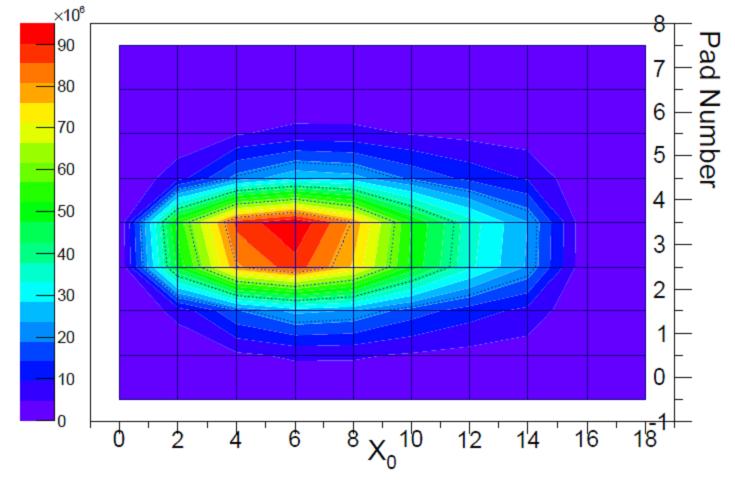}
\caption{The two dimensional shower profile using data taken with the "High High" readout
mode. }
\label{fig:beamcal_elec_shower}
\end{figure}

Also in 2011, e.m. shower data was taken using tungsten absorber plates of different thicknesses
installed upstream of the GaAs sensor plane. This served to test the performance of the readout when many particles are crossing the sensor.
As an example, a 2D profile of particles in the shower is shown in Figure~\ref{fig:beamcal_elec_shower}.

\subsection{LumiCal}
\subsubsection{2010 beam test}

\begin{figure}[!htb]
\centering
\subfigure[]{\label{fig:8_channel}\includegraphics[width=0.4\textwidth]{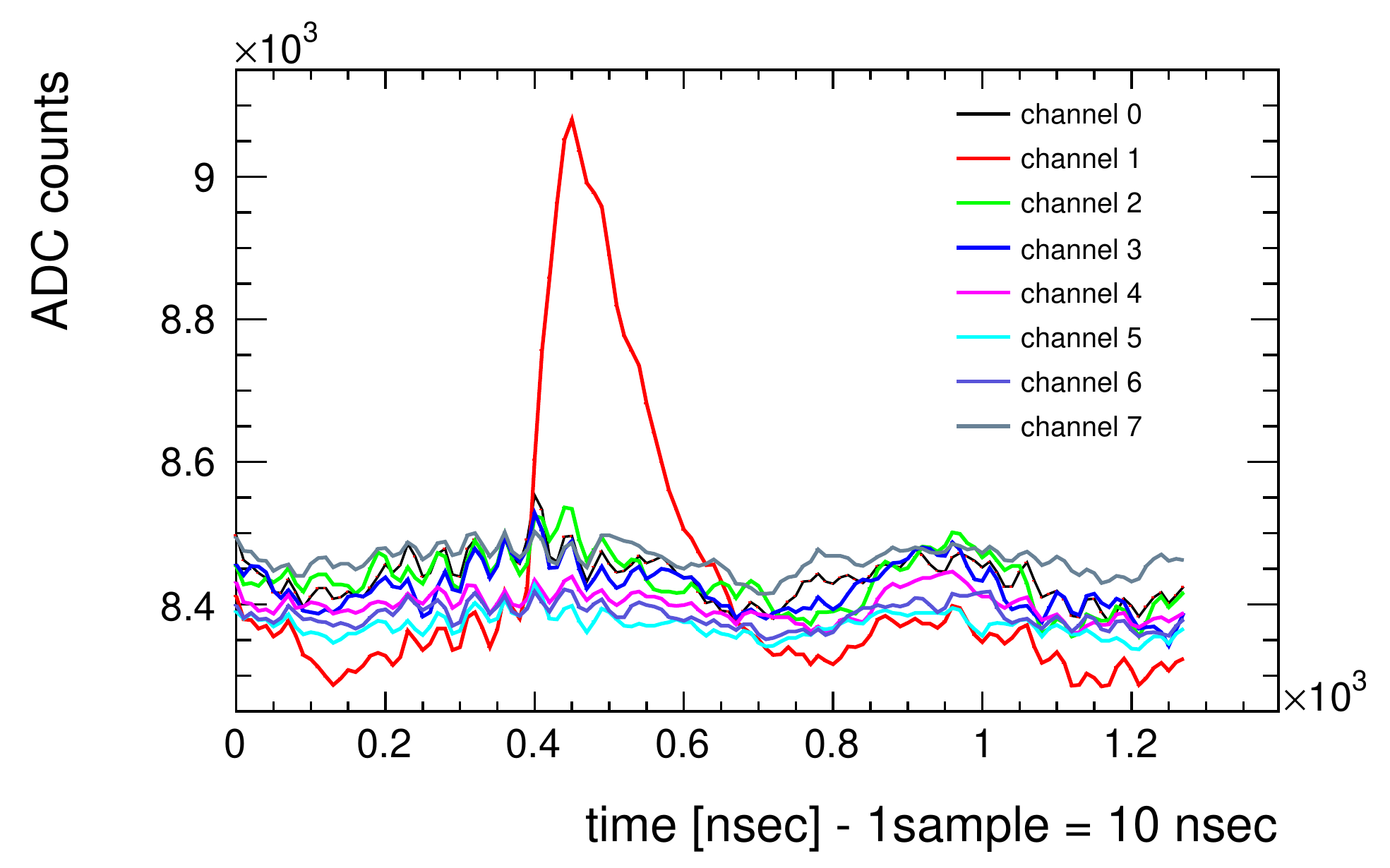}}
\subfigure[]{\label{fig:cross_talk_mesurement_C}\includegraphics[width=0.4\textwidth]{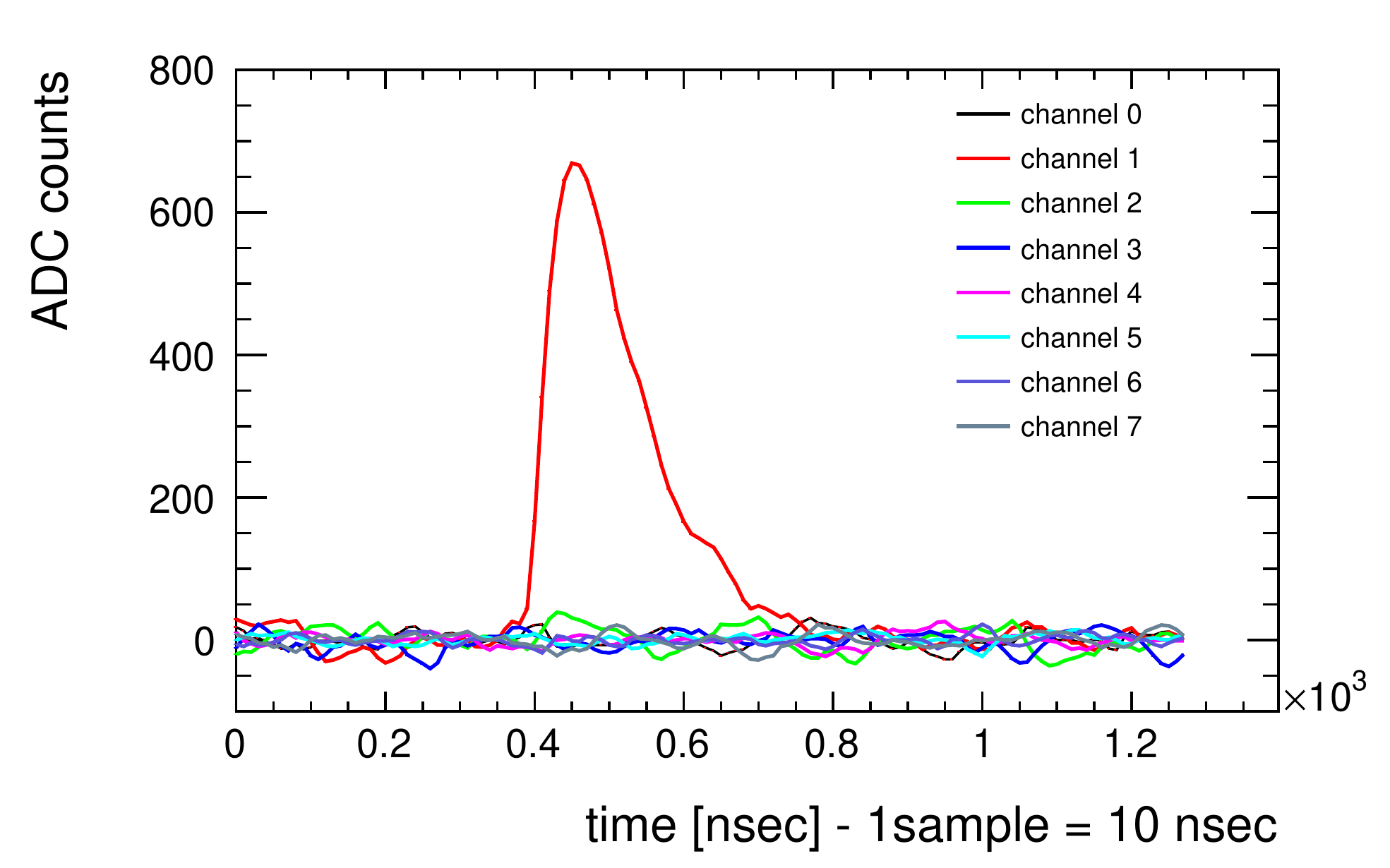}}
\caption{The digitized waveforms of 8 channels, before (a) and after (b) CMN subtracted. }
\label{fig:cross_talk_mesurement}
\end{figure}

In 2010, several million triggers have been recorded for two areas of 8 pads each on a sensor equipped with ASICs via a Kapton fan-out.
%\itamar_change\
In the ASICs 8 channels are implemented, four of them have been designed with a passive resistor as a feedback, while the other four channels use a MOS transistor in the feedback loop.
The analog signals were digitised by an external 14-bit ADC (CAEN V1724) with 100 MSp/s.
The digitized waveforms of 8 channels read out in parallel are shown in Figure~\ref{fig:8_channel}~\cite{Itamar} . 
The Common Mode Noise, CMN, is determined and subtracted from the raw data, taking into account the different gain in the channels (there is a difference of a factor of 2 in gain between the two technologies used).

\begin{figure}[!htb]
\centering
\includegraphics[width=1\textwidth]{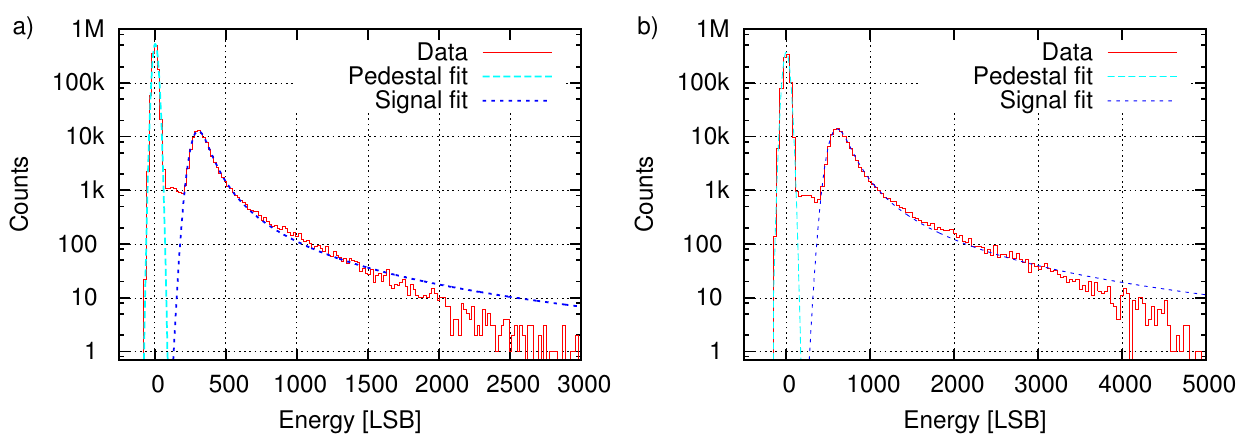}
\caption{Energy deposition spectrum in a) A5 pad (passive feedback) and b) A3 pad (active feedback).}
\label{fig:spectra}
\end{figure}

The pulse height spectrum, shown in Figure~\ref{fig:spectra} for both technologies, is fitted well by a convolution of a
Landau distribution and a Gaussian. Using the most probable value and the width of the
pedestal, signal-to-noise values between 16 and 22 are estimated before the CMN subtraction, and between 28 and 33 after it.
%\itamar_change\

\begin{figure}[!htb]
\centering
\includegraphics[width=1\textwidth]{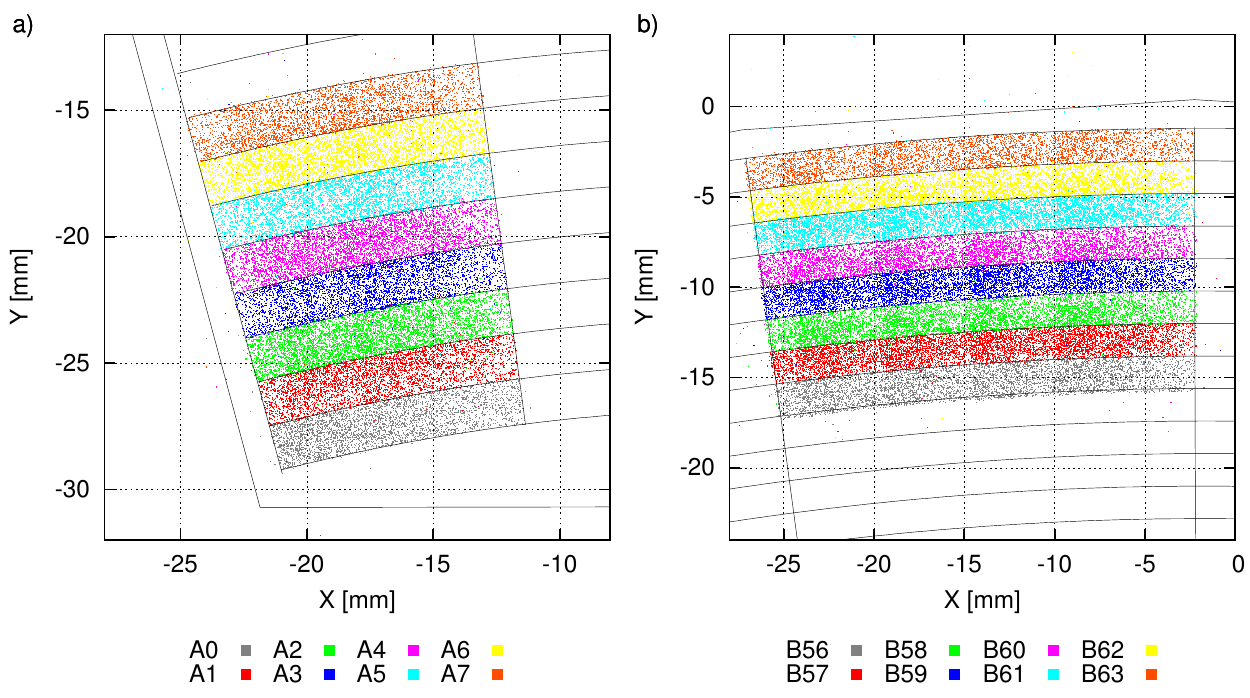}%\itamar_change\
\caption{Reconstructed position of impact point on sensor with color assigned correspond to the pad with hit above threshold registered in the LumiCal: a) (smaller) bottom pads and b) (wider) top pads area.}%\itamar_change\
\label{fig:position}
\end{figure}

\begin{figure}[!htb]
\centering
\includegraphics[width=1\textwidth]{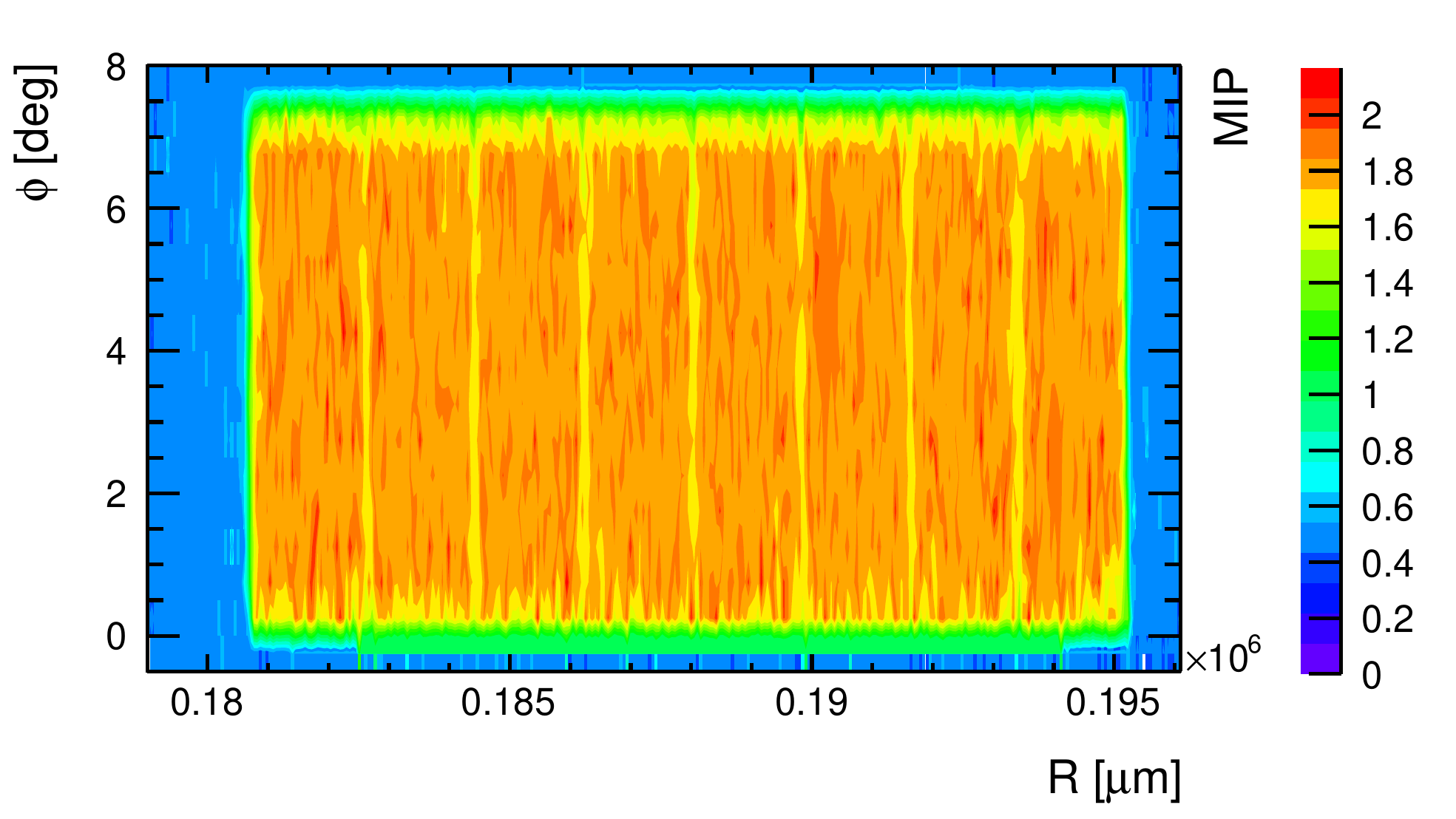}
\caption{Signal size as a function of hit position on the sensor. A signal drop of  about 10\% in between pads is observed.}
\label{fig:pad_gap}
\end{figure}

\begin{figure}[!htb]
\centering
\includegraphics[width=1\textwidth]{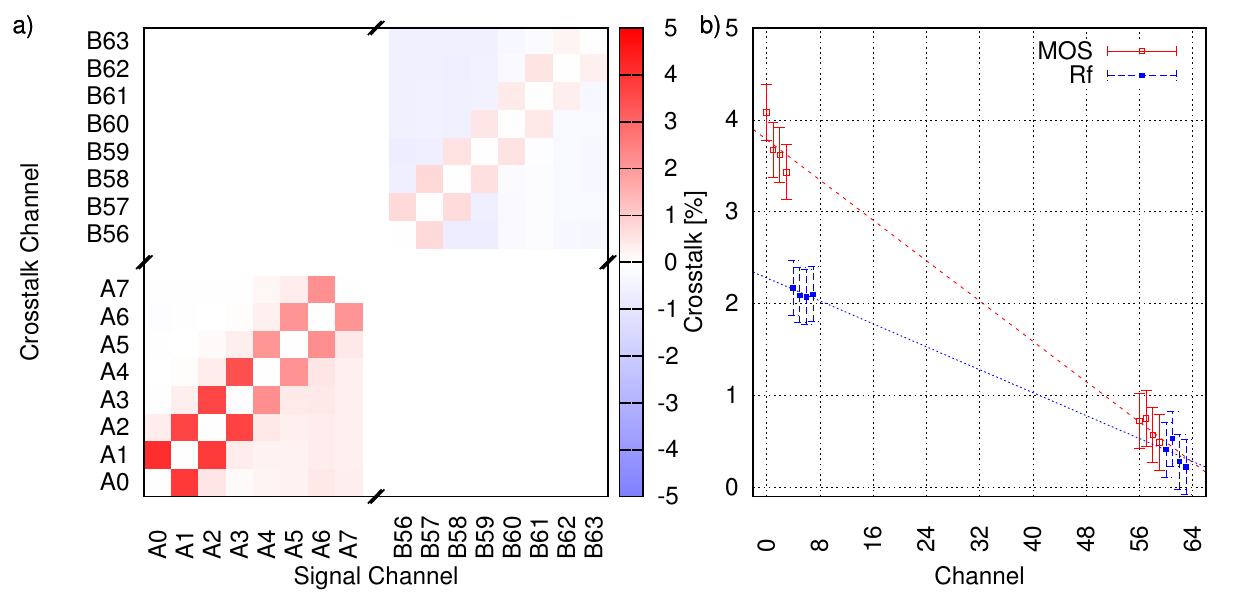}%\itamar_change\
\caption{a) Measured crosstalk coefficients for groups of eight pads. b) Measured crosstalk coefficients
to the closest neighbor as a function of pad number.}%\itamar_change\
\label{fig:cross_talk}
\end{figure}

Using a track fit from the MVD telescope the particle impact position on the sensor is predicted.
The distribution of hits is shown in Figure~\ref{fig:position}. The colors, characterising the pads, are assigned to each point in case the signal in the
pad is above a certain threshold. The pad structure for the LumiCal sensor is correctly reproduced.
%\itamar_change\
In Figure~\ref{fig:pad_gap} the average signal size is shown as a function of the position. In between the
pads a drop of the signal size of about 10\% is observed in a range of 200 $\mu$m near the 100
$\mu$m gap between pads. 
On the pads the signal variation is due to the limited statistics.
The relation between signal amplitudes on neighboring pads was studied to evaluate the crosstalk level between pads.
Only events identified to lie further than 200 $\mu$m from the pad border were used, to minimize the effect of
charge sharing. One can see the positive correlation between these amplitudes, that can be
interpreted as a crosstalk.
For the crosstalk measurement the common mode subtraction procedure was not applied. This
avoids artificially introduced biases; however, this comes at the cost of broader distributions,
and therefore less precise results.
The crosstalk coefficients between all channels are presented in Figure~\ref{fig:cross_talk}.
Significant crosstalk can be observed only between the closest neighbors. 
The
highest crosstalk is observed for pads with the longest fanout attached to it. Moreover, there is a
significant difference between channels with different feedback type, which can be explained by
the differences in the input impedance.

%\itamar_change\

%\itamar_change\

\subsubsection{2011 beam tests}

In a second test-beam period in August 2011 a complete LumiCal sensor and electronics configuration was tested. The 32 largest area pads, in the uppermost part of the sensor, were connected to the electronics board.
Signals were amplified and shaped using front-end ASICs and, for the first time, the digitized in pipeline 10-bit ADC ASICs. 
The complete set-up was found to be fully operational at the testbeam. Data were collected in two modes of operation: synchronous (ILC mode) and
asynchronous (test beam mode).

To recover information on the signal waveform a deconvolution method was applied~\cite{Kulis_phd} .
Using the deconvoluted amplitude waveform the signal size is determined.
A similar analysis to the 2010 test was performed,  and a slightly worse signal to noise performance was observed, with values between 11 and 25 before the CMN subtraction, and between 19 and 22 after it.

\begin{figure}[!htb]
\includegraphics[width=0.9\textwidth]{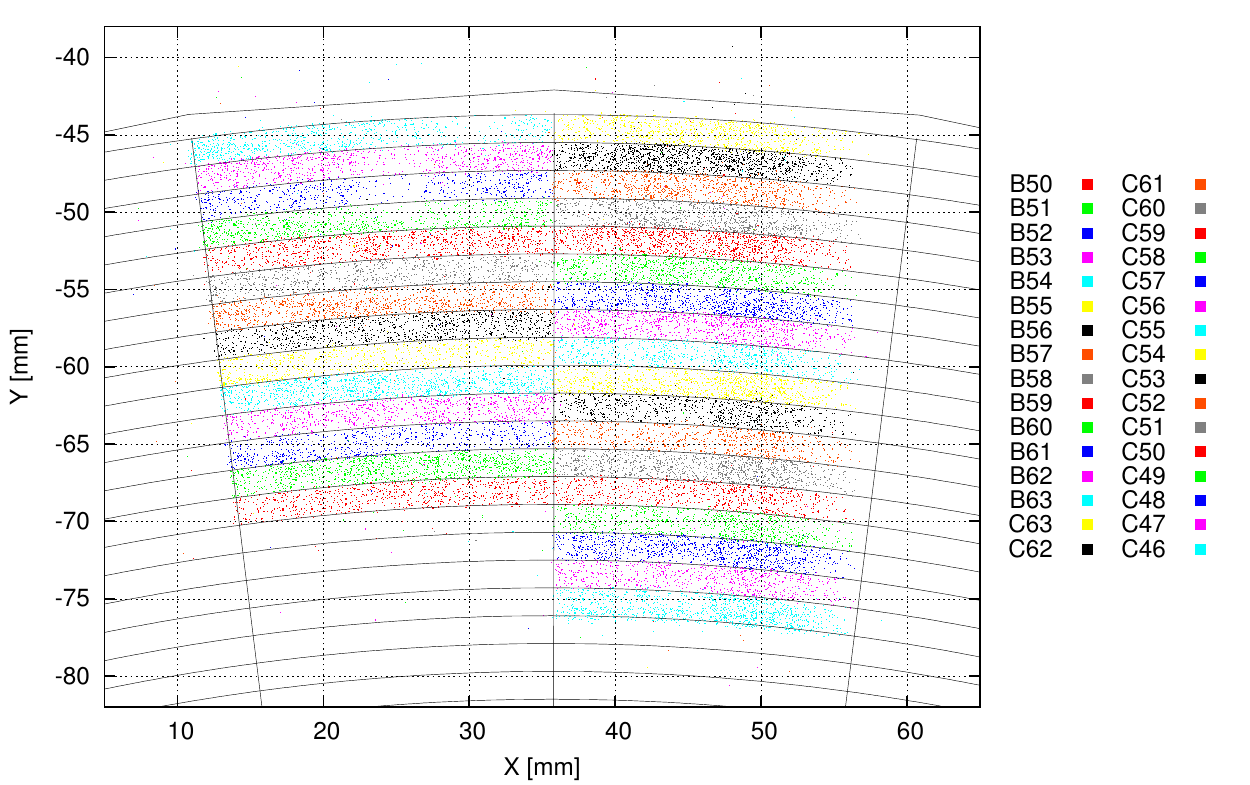}%\itamar_change\
\caption{Reconstructed position of impact point on sensor with color assigned correspond to the pad with hit above threshold registered in the LumiCal at the 2011 beam test.}%\itamar_change\
\label{fig:position_2011}
\end{figure}

A full beam scan of the connected area of the sensor was performed by moving the sensor mounted on the X-Y table.
Signals from the sensor plane combined with reconstructed particle positions obtained from the telescope reveal the pad structure of the LumiCal sensor, as shown in Figure~\ref{fig:position_2011}. 
The lack of points close to the edge of the pads on the right in Figure~\ref{fig:position_2011} is related to the limited range of the X-Y table used.

Several data sets were taken with tungsten absorbers of different thickness in front of the sensor, to study the electromagnetic shower development.
The electron beam was directed to the center of the sensor plane, and one or several 3.5 mm thick tungsten plates, corresponding to one radiation length each,
were positioned in front of the sensor.
Because of the stochastic character of the development of an electromagnetic shower, and due to the limited number of sensor layers, only statistical parameters can be analyzed.
The geometry and material of the experimental setup was approximated in the MC simulations, using GEANT4.
The electron beam was modeled as a collection of electrons with parallel tracks, collimated to obtain a square beam profile similar to the one in the testbeam.

\begin{figure}[!htb]
\includegraphics[width=1\textwidth]{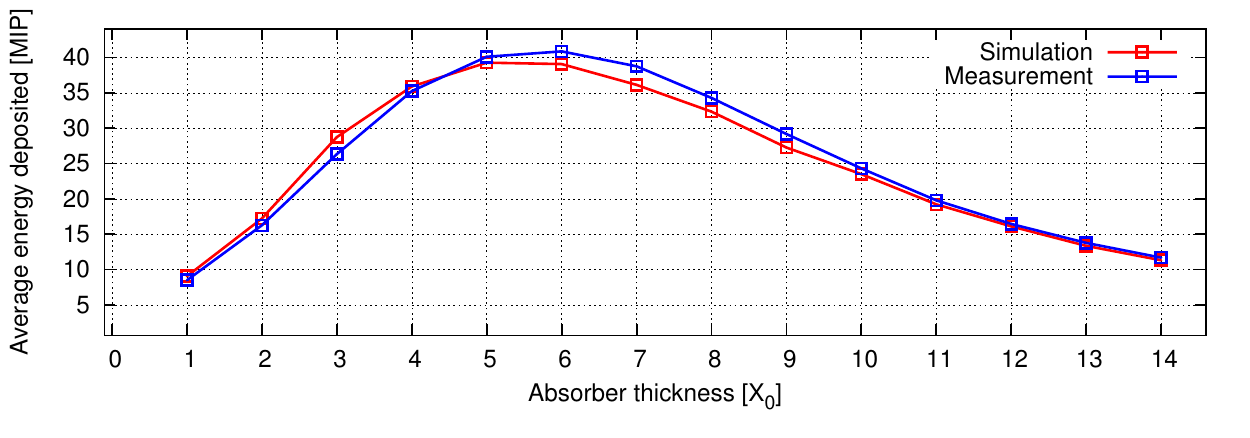}%\itamar_change\
\caption{Average energy deposited in the instrumented area as a function of the tungsten absorber
thickness expressed in radiation lengths.}%\itamar_change\
\label{fig:absorber}
\end{figure}

Spectra of the total energy deposited in the sensor at different phases of shower development were studied and compared to the MC simulation.
The average energy deposited in the instrumented area as a function of the tungsten
thickness is shown in Figure~\ref{fig:absorber}.The measured shower maximum is observed after six radiation lengths, and the agreement with MC is found to be reasonable.
Details of the shower development will be addressed
in future studies of the FCAL collaboration, when a LumiCal detector prototype with several sensor planes is available.

\section{Luminosity Spectrum}
\label{sec:lumiSpecReco}

Small, nanometre-sized beams are necessary to reach the required luminosity at
future linear colliders. Together with the high energy, the small beams cause
large electromagnetic fields during the bunch crossing. These intense fields at
the interaction point squeeze the beams. This so called \emph{pinch effect}
increases the instantaneous luminosity. However, the deflection of the particles
also leads to the emission of Beamstrahlung photons -- which reduce the nominal
energy of colliding particles -- and produces collisions below the nominal
centre-of-mass
energy~\cite{chen_beam,Chen:242895,Schroeder:216344,schulte1996}. The resulting
distribution of centre-of-mass energies is the \emph{luminosity spectrum}.  

A good knowledge of the luminosity spectrum is mandatory for many measurements
at future \epem colliders. As the beam-parameters determining the luminosity
spectrum cannot be measured precisely, the luminosity spectrum has to be
measured through a gauge process with the detector. The measured distributions,
used to reconstruct the spectrum, depend on Initial-State Radiation (ISR),
cross-section, and Final-State Radiation (FSR). To extract the basic luminosity
spectrum a parametric model of the luminosity spectrum
\lumispec{\Eele,\Epos; \parset{0}} is created. \Eele and \Epos are the energies
of the electron and positron prior to the Initial-State Radiation and scattering
event, and \parset{0} is the set of parameters of the parametrisation.  With the
model a reweighting technique is used to extract the luminosity spectrum from
measured Bhabha event observables. Details in the reconstruction of the ILC and
CLIC luminosity spectrum can be found in the
references~\cite{sailer,LCD-2011-040}.

For the linear collider design studies, the beam-beam effects are simulated with
the \guineapig program~\cite{schulte1996}. It is a Monte Carlo program, which
uses a distribution of beam-particles as the input to calculate the fields and
deflection during the collisions.

\subsection{Luminosity Spectrum Reconstruction with a Reweighting Fit}
\label{sec:LSRreweighting}
The luminosity spectrum can be reconstructed with a reweighting fit, which
allows for the efficient reconstruction of an underlying distribution. A
measured distribution $f(O_{1},O_{2},\ldots)$ of observables $O_{i}$ can be
approximately written as
\begin{multline}
  \label{eq:rewfit2}
  f(O_{1},O_{2},\ldots) \approx\\
  \sigma(O_{1},O_{2},\ldots;\Eele,\Epos) \times \lumispec{\Eele,\Epos} \conv
  \ISR{\Eele,\Epos} \conv
  \FSR{O_{1},O_{2},\ldots} \conv \Det{O_{1},O_{2},\ldots},
\end{multline}
where D are the detector resolutions of the respective
observable, and $\sigma$ is the transfer-function from the initial electron and positron
energies to the observables, for example the differential Bhabha
cross-section. 

Three observables, which can be extracted from the final-state electrons, are
used for the reconstruction of the luminosity spectrum: the relative
centre-of-mass energy calculated from the polar angles of the outgoing electrons
${\rootsaco}/{\rootsnom}$ and the final-state energy of the electron and
positron. The relative centre-of-mass energy reconstructed from the
acollinearity of the final-state electron and positron
is~\cite{sailer,moenig:DiffLumi}

\begin{equation}
  \label{eq:acollinearity}
  \frac{\rootsaco}{\rootsnom} = \sqrt{\frac{\sin(\theta_{\ele})+\sin(\theta_{\pos})+\sin(\theta_{\ele}+\theta_{\pos})}{\sin(\theta_{\ele})+\sin(\theta_{\pos})-\sin(\theta_{\ele}+\theta_{\pos})}},
\end{equation}
where \rootsnom is the nominal centre-of-mass energy, and where $\theta_\ele$ is
the polar angle of the electron and $\theta_\pos$ that of the positron. The
polar angles of the electron and positron can be measured very precisely (to the
order of a few tens of microradians), so
that the distribution of the relative centre-of-mass energy is least affected
by detector resolutions. The relative centre-of-mass energy therefore gives a
precise handle on the energy spectrum, while the particle energies help to
disentangle which of the particles lost energy prior to the collision. 

For the reweighting fit: first, events \varEv are generated according to the
parametrised luminosity spectrum
$\lumispec{\Eele^{\varEv},\Epos^{\varEv}; \parset{0}}$ with the initial set of parameter
values $\parset{0}$; the Bhabha scattering is simulated with a Monte Carlo
generator, for example \bhwide~\cite{BHWIDE}; and the detector
resolutions are modelled with a \geant-based detector simulation. The actual fit
is a \chisquare-minimisation between the histograms of the observed distribution
$f_{\mathrm{Data}}$ and the simulated distribution $f_{\mathrm{Monte
    Carlo}}$. For each iteration, the weight of all events $w^{\varEv}$ in the simulated
sample is varied based on the parametrisation of the luminosity spectrum
\lumispecNoArg, with parameter values $\parset{N}$ varied by the minimiser
\begin{equation}\label{eq:eventweight}
    w^{\varEv} = \frac{\lumispec{\Eele^{\varEv},\Epos^{\varEv}; \parset{N}}}{\lumispec{\Eele^{\varEv},\Epos^{\varEv}; \parset{0}}}.
\end{equation}

Because the simulation of the events only has to be done once and the
calculation of the event probability based on the parametrisation is very fast,
the reweighting fit allows for an efficient reconstruction of the luminosity
spectrum. It does, however, require a reliable parametric description of the
spectrum.

Of all the current linear collider designs, the 3~TeV CLIC offers the most
challenging luminosity spectrum. Recent effort has focused on the reconstruction
of the 3~TeV CLIC spectrum~\cite{LCD-2011-040}, but the reweighting procedure was
previously applied at the ILC case with some adaptions to the
parametrisation~\cite{sailer}.

\begin{figure}[tbp]
  \centering
   \includegraphics[width=0.49\textwidth]{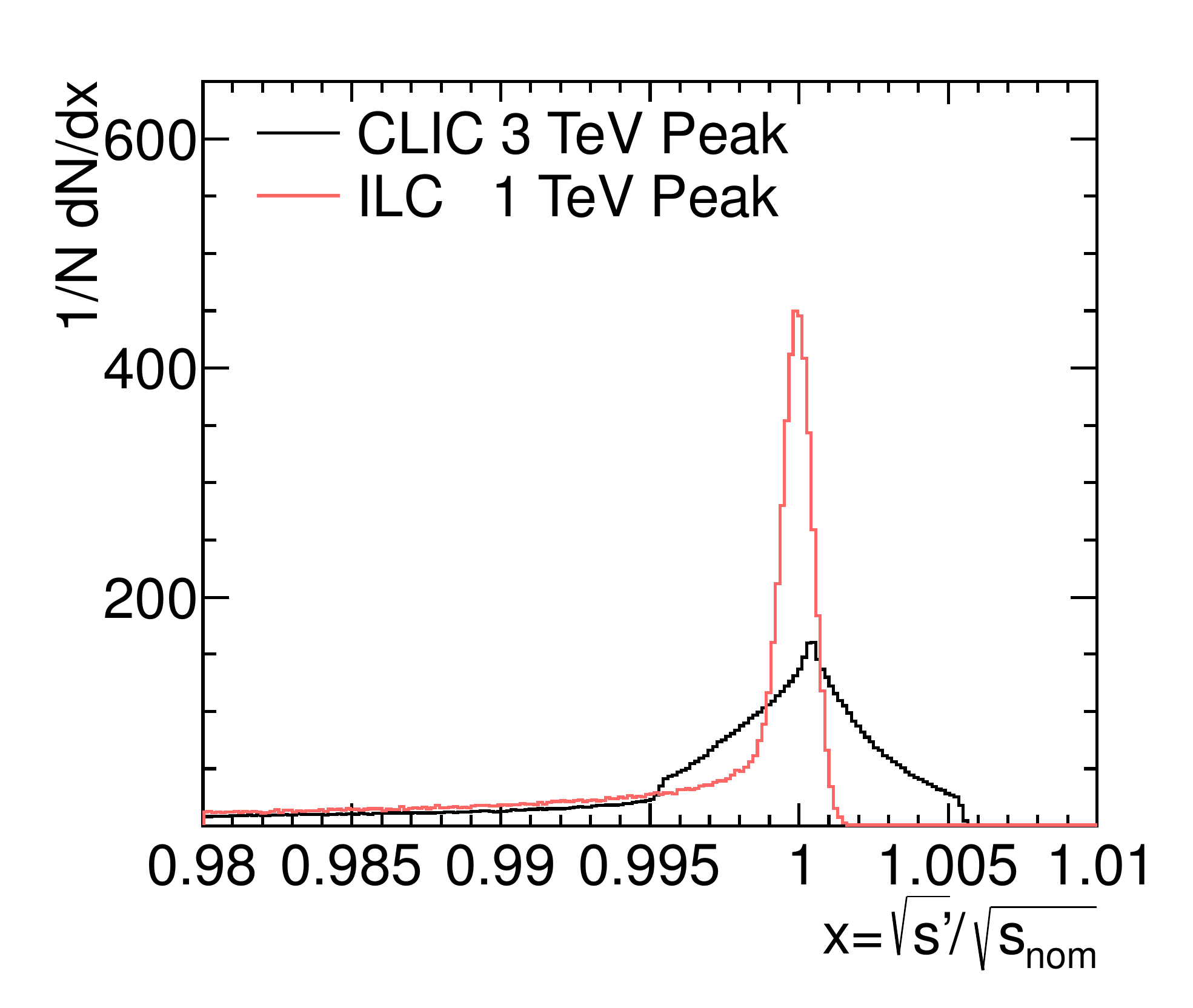}
 \includegraphics[width=0.49\textwidth]{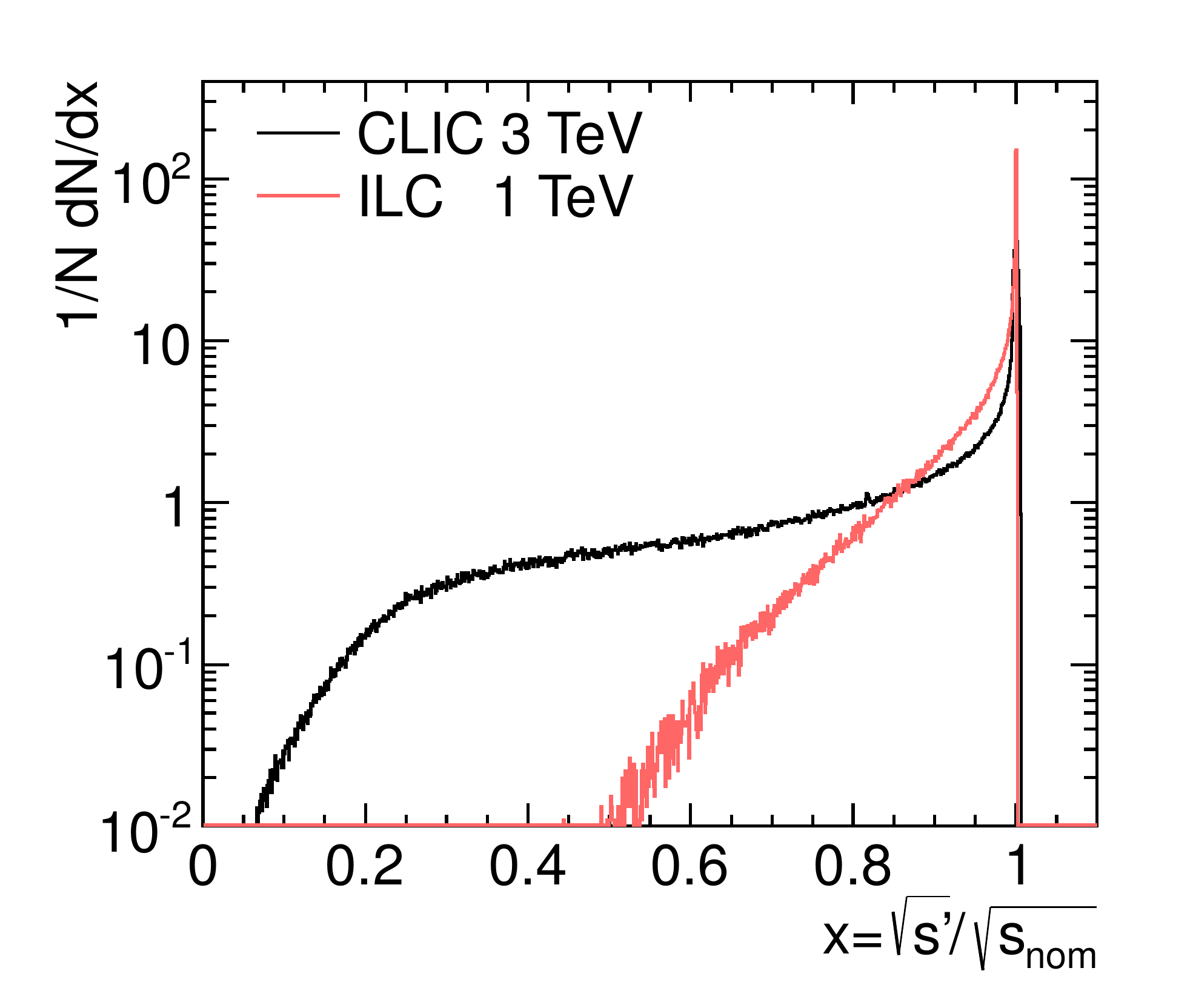}
  \caption{Left: Peak of the luminosity spectrum of 3~TeV CLIC and 1~TeV ILC.
 Right: Luminosity spectrum for 3~TeV CLIC, and 1~TeV ILC.}
  \label{fig:LSRlumiSpectrum}
\end{figure}

One challenge for the modelling of the spectrum is the non-Gaussian beam-energy
spread. The non-Gaussian beam-energy spread also results in a non-Gaussian shape
of the luminosity spectrum, shown in Figure~\ref{fig:LSRlumiSpectrum} (right).

The second challenge at CLIC is the long tail of centre-of-mass energies. As can
be seen in Figure~\ref{fig:LSRlumiSpectrum}, the tail of centre-of-mass energies
extends almost to zero for the 3~TeV CLIC case, for the 1~TeV ILC the luminosity
spectrum ends around 550~GeV. 
The details of the parametrisation for the 3~TeV
CLIC can be found in the reference~\cite{LCD-2011-040}.

\begin{figure}[tbp]
  \centering
  \includegraphics[width=0.49\textwidth]{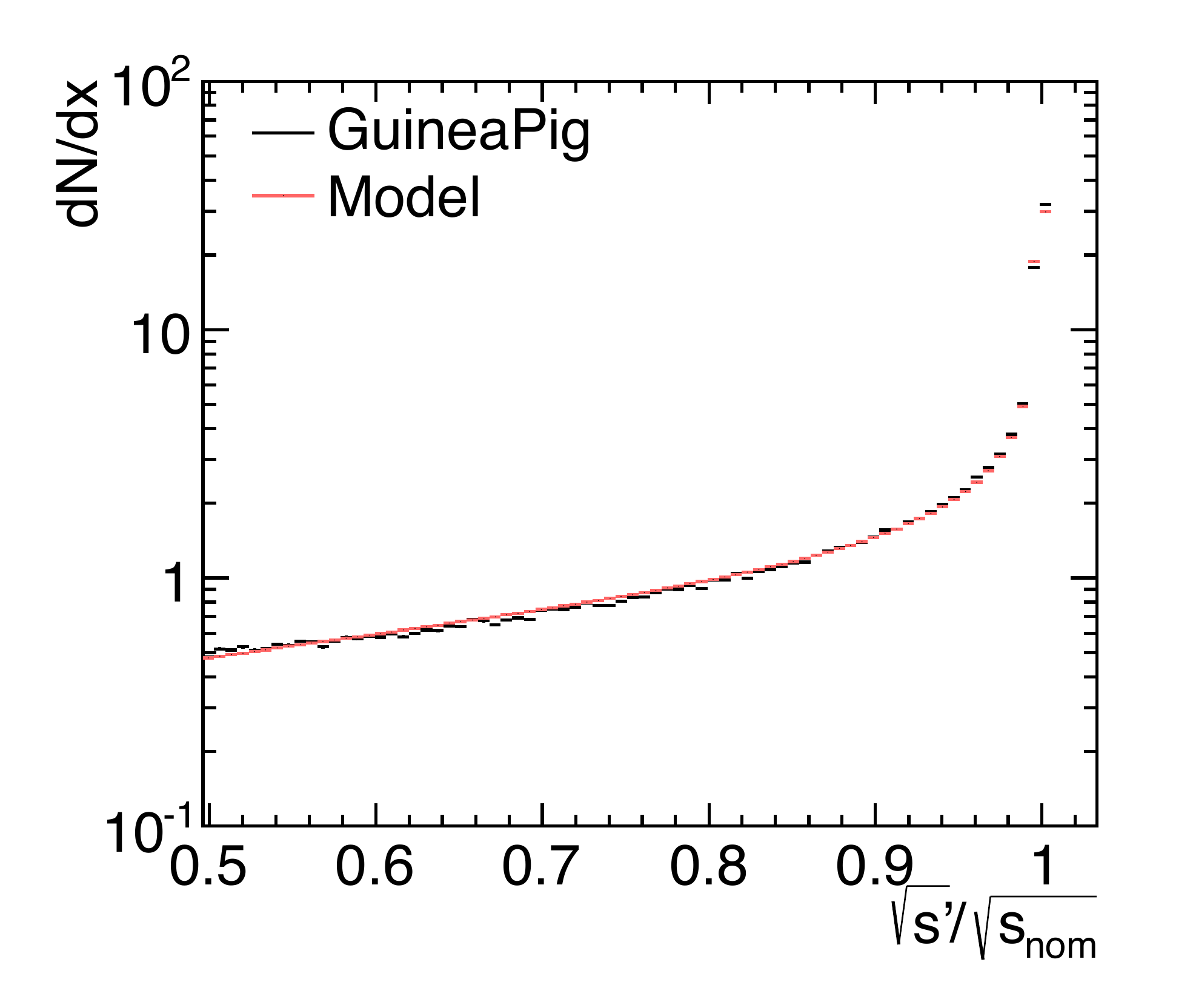}
  \includegraphics[width=0.49\textwidth]{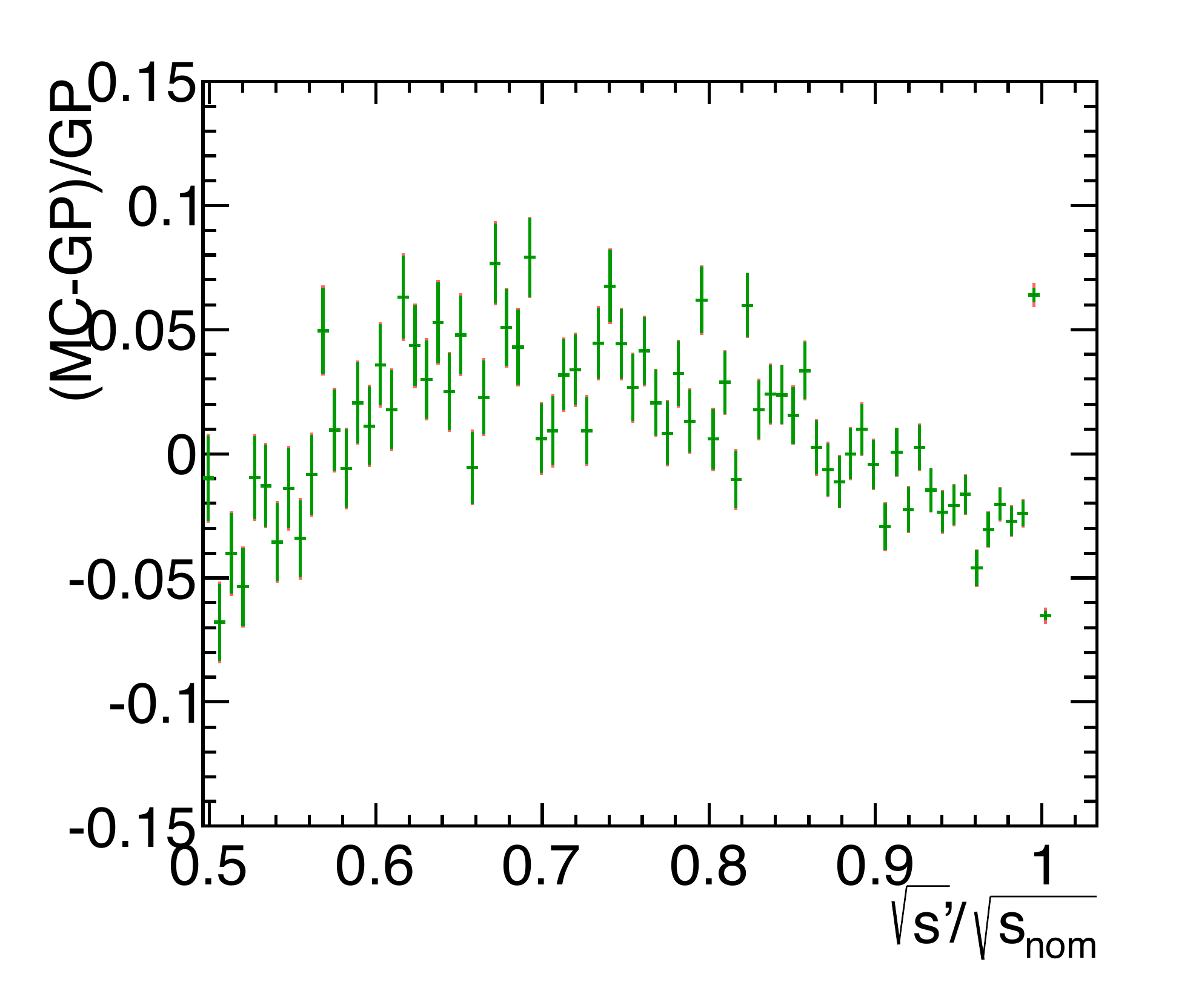}
  \caption{Left: \guineapig and reconstructed Spectrum of the 3~TeV CLIC. Right:
    Relative difference between the reconstructed spectrum and \guineapig. Error
    bars are the statistical uncertainty from the \guineapig sample.}
\label{fig:LSRrecoSpec}
\end{figure}

Figure~\ref{fig:LSRrecoSpec} shows the \guineapig and reconstructed spectrum of
3~TeV CLIC, and the relative difference between the two spectra. The
cross-section, Initial-State Radiation, Final-State Radiation, and detector
resolutions were applied on the luminosity spectra. The simulated data set
consisted of 2~million Bhabha events which correspond to about 400~\fbinv. For
the Monte Carlo sample, 10~million events are used in the fit. The largest
difference is observed close to the nominal centre-of-mass energy, because of
the difficulty in modelling the beam-energy spread at CLIC. Over the rest of the
range, the difference is smaller than 5\% and overlapping within the statistical
uncertainty. It now has to be studied, whether the precision of this
reconstruction is sufficient for the ambitious goals of the linear collider
physics programme. 
Improvements of the model can be envisaged, but would likely entail increasing its complexity
and thus the number of parameters. This may result in a disproportionate increase in 
the computing resources needed. The use of observables different from the Bhabha
scattering ones used so far may also be considered.

\section{Background and systematics of the luminosity measurement} 

The most precise method for a luminosity measurement at linear colliders is to count Bhabha-scattering events in the very forward region.
Such events are identified by coincident detection of showers, in a given (high) energy range, in the fiducial volume in both halves of the luminometer (LumiCal). The luminometer at a future linear collider will be designed as a pair of finely granulated calorimeters of high energy and polar angle resolution in the very forward region \cite{ieee1}, allowing full reconstruction of the four-vectors of the detected particles. In the ILC case the fiducial volume of the LumiCal covers the angular range between 41 and 67 mrad. 

At LEP this technique allowed reaching sub-permille precision \cite{Opal00}. A number of systematic effects present a challenge to reach permille precision at future linear colliders. The two dominant systematic effects, the beam-beam effects and the physics background, are briefly discussed here. The uncertainties from all known sources are listed at the end of this section.

\subsection{Beam-beam effects}

At future linear colliders the center-of-mass (CM) energy will be 3 to 30 times higher than at LEP, and the instantaneous luminosity up to three orders of magnitude higher \cite{CLIC, ILC}. In such conditions, the colliding bunches induce a strong focusing effect upon each other \cite{Yok91}. This, in turn, causes emission of intense \emph{Beamstrahlung}. The energy loss on Beamstrahlung creates a long tail towards the lower energies in the luminosity spectrum. The movement of the CM frame due to forward-backward asymmetry of the Beamstrahlung reduces the effective angular acceptance of the LumiCal and creates an overall \emph{angular counting loss} of the order of 10\%. Besides that, the electromagnetic deflection (EMD) of the final electrons in the field of the opposite bunch induces an additional counting loss in the order of 1-2 permille \cite{Luk12b}.

Since high energy photons are indistinguishable from electrons in the calorimeter, the final-state radiation is added to the electrons detected in the LumiCal. The resulting four-vectors allow reconstruction of the kinematics of the hard-collision frame, i.e. the CM energy $E_{CM}$, the velocity of the collision frame $\beta_{coll}$ and the scattering angle $\theta_{coll}$. The reconstructed $\beta_{coll}$ and $\theta_{coll}$ can be used to correct the reduction of the angular acceptance on an event-by-event basis \cite{Luk12, Luk12b}. 

\begin{figure}
\centering
\includegraphics[width=\textwidth]{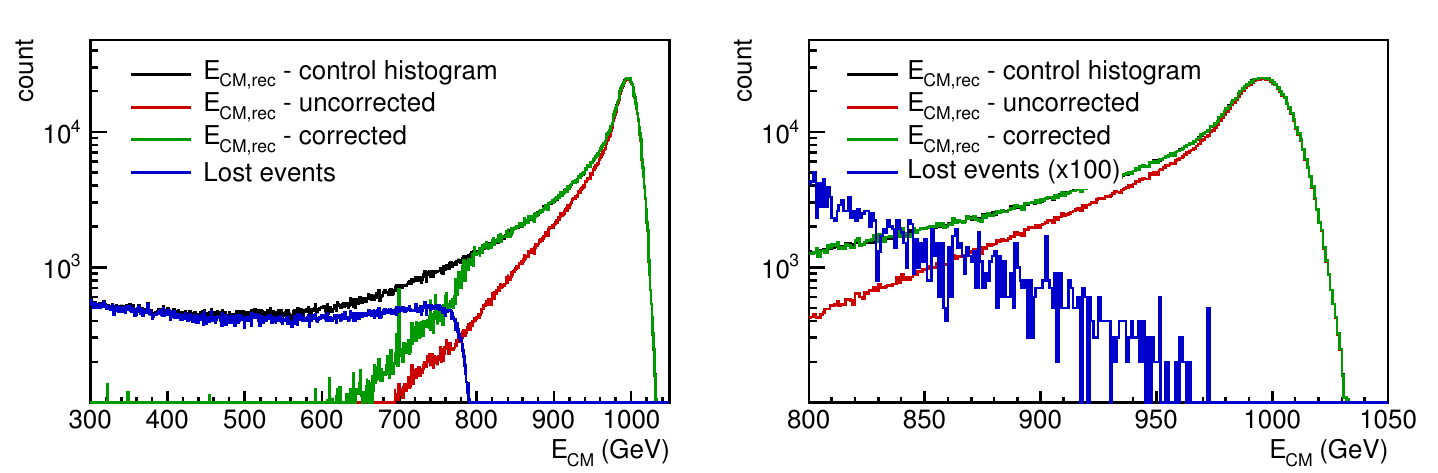}
\caption{\label{fig-BS-corr}Correction of the counting loss due to Beamstrahlung and ISR at 1 TeV. Left: whole spectrum; right: zoom on energies above 800 GeV. Black: Control spectrum without counting loss; red: Reconstructed spectrum with the counting loss; green: Reconstructed spectrum with correction of the counting loss; blue: events inaccessible to the correction}
\end{figure}

This method was tested by Monte Carlo simulation, using Guinea-Pig for beam-beam simulation \cite{schulte1996} and the BHLUMI generator \cite{Jad97} for the final four-momenta of Bhabha events. Simulations were performed for a range of beam-parameter variations, so that 25 different sets were simulated for each of the two energy options. The results of the correction are shown in Fig. \ref{fig-BS-corr} for the 1 TeV ILC case. As can be seen, precise correction is achieved above about 80\% of the nominal CM energy. Below that limit, a large fraction of events is characterized by values of $\beta_{coll}$ which are so large that the effective angular acceptance is zero, so that these events are irrecoverably lost from the detector fiducial volume. A small number of high-$\beta_{coll}$ is also present in the top 20\% of energy, due to ocassional off-axis ISR. The mean relative bias due to such events to the peak integral above 80\% of the nominal CM energy is $(-1.490 \pm 0.007) \times 10^{-3}$ in the 500 GeV case, and $(-1.424 \pm 0.007) \times 10^{-3}$ in the 1 TeV case. 

After correcting for the lost events, the average counting bias in the upper 20\% of energy at 500 GeV is $(+0.39 \pm 0.09) \times 10^{-3}$, while at 1 TeV the bias is $(+0.67 \pm 0.10) \times 10^{-3}$.

To obtain the proper luminosity spectrum, the ISR energy loss is deconvoluted in the next step. The deconvolution is performed numerically using the theoretical form of the distribution of the ISR fractional energy loss. As shown in Ref. \cite{Luk12b}, the uncertainty in the upper 20\% of energy introduced by the deconvolution step is 0.4 permille at 500 GeV, and 0.8 permille at 1 TeV.

The EMD counting loss can be corrected by MC simulation. Under the assumption that the beam parameters can be determined within 20\%, the uncertainty of the correction of the EMD counting loss is 0.5 permille at 500 GeV and 0.2 permille at 1 TeV \cite{Luk12b}.

\subsection{Background from physics processes}

Another major systematic effect in the luminosity measurement originates from the four-fermion neutral current processes of the type $e^+e^- \rightarrow e^+e^-f\bar{f}$. These processes are characterized by the outgoing electron pairs emitted very close to the beam pipe and carrying a large fraction of energy, thus they can be miscounted as Bhabha events. To estimate their influence on the luminosity measurement, the WHIZARD V1.4 generator was used to generate background events at the tree level. The parameters of the generation were tuned to the experimental results from PETRA and LEP. The resulting distributions were compared to the distributions of Bhabha events at the generator level.

The fraction of the 4-fermion background relative to the Bhabha signal is 6.0 permille at 500 GeV and 2.2 permille at 1 TeV \cite{Mila13}. A somewhat smaller fraction of hadronic events is also present. This background can be reduced by applying specific selection cuts, taking care to preserve the invariance of the Bhabha counting method with respect to the longitudinal boost. Cuts on the CM energy, as well as on the acoplanarity ($\Delta \phi = \phi_1 - \phi_2 < \Delta\phi_{max}$) possess such invariance. If only events with $E_{CM}$ above 80\% of the nominal energy are counted, and with the acoplanarity cut $\Delta\phi < 5\deg$, the fraction of background events, leptonic and hadronic included, drops to 2.2 permille at 500 GeV, and 0.8 permille at 1 TeV \cite{Mila13}. 

At LEP, such background was corrected with a relative precision of 20\%. Arguments have been put forward, that a correction with a precision of 40\% will be possible at future linear colliders \cite{ieee1}. This topic will be further addressed in the future.

\newpage

\subsection{Table of systematics in luminosity measurement}

The systematic uncertainties in the luminosity measurement at ILC from different sources are listed in Tab. \ref{tab-all-unc}.

\begin{table}[h]
\caption{\label{tab-all-unc}Systematic uncertainties in the ILC luminosity measurement.}
\begin{center}
\begin{tabular}[ c ]{@{}       |p{8cm}               | c   |          c |  } \hline
{\bf Source of uncertainty}       & \multicolumn{2}{|c|}{$\Delta L/L$ ($10^{-3}$)}       \\
                                        &   500 GeV  &  1 TeV \\
\hline\hline
Bhabha cross section \cite{Arb96}       &     0.54   &  0.54    \\ \hline
Polar-angle resolution \cite{ieee1}     &     0.16   &  0.16    \\ \hline
Polar-angle bias \cite{ieee1}           &     0.16   &  0.16    \\ \hline
Energy resolution \cite{ieee1}          &      0.1   &  0.1     \\ \hline
Energy scale \cite{ieee1}               &      1     &    1     \\ \hline
Beam polarization  \cite{ieee1}         &      0.19  &  0.19    \\ \hline
Physics background \cite{Mila13}        &      2.2   &  0.8     \\ \hline
Beam-beam effects \cite{Luk12b}         &      0.9   &  1.5     \\ \hline
\hline
Total                                   &      2.6   &  2.1     \\ \hline
\end{tabular}
\end{center}
\end{table}

\section{ Sensor Damage from Electromagnetically-Induced Radiation}

Simulation of the production of electron-positron pairs, beamstrahlung, and
environmental radiation suggest electromagnetic radiation fluxes of order
1 MGy per year for the forward most calorimeter (BeamCal) of either of the
Linear Collider Detector designs. This dose will not arise from direct irradiation
of the calorimeter sensors by the electron, positron and gamma flux, but rather
at the maximum of the electromagnetic showers induced by the incident electromagnetic
radiation. Damage to crystalline sensors can usually be parameterized in terms of
the total non-ionizing energy loss (NIEL) of the incident radiation in the
sensor~\cite{NIEL}. Per GeV of incident particle energy, NIEL is significantly
less for electromagnetic particles (electrons, positrons, and photons) than
for hadronic particles, and particularly for neutrons, for which energy-loss
mechanisms almost always result from scattering from nuclei in the sensor crystal.
In addition, damage rates less than those expected from NIEL scaling have been
observed for incident electrons~\cite{summers}, particular for the case of p-type
crystalline silicon. Thus it may be quite possible that damage to silicon sensors
in a fully-developed electromagnetic shower may be dominated by the small
hadronic component of the beam, and an experiment designed to qualify the
radiation hardness of candidate sensor technologies must bear this in mind.

Three dominant processes are thought to contribute to hadron production in
electromagnetic showers. A lateral oscillatory resonance, known as the `giant dipole
resonance', is induced by photons in the 10-20 MeV range. The resonance
de-excites via the evaporation of neutrons, yielding a flux of neutrons that is
isotropic relative to the excited nucleus. Nuclear compton scattering is
dominated by the $\Delta$ resonance at an incident energy of approximately
340 MeV, leading to a relatively isotropic flux of protons, neutrons, and pions.
Finally, photoproduction turns on at approximately 200 MeV, leading to a somewhat
forward flux of pions. However, since much of the hadronic flux is isotropic
(giant dipole resonance) or nearly so (nuclear compton scattering), to
rigorously explore radiation damage from electromagnetic showers, it is likely
necessary to embed the sample within a radiator. In order to assure a realistic
flux of shower-induced hadronic particles, the sample sensor should
be placed within the radiator at shower maximum, and be surrounded on both the
upstream and downstream side by the radiator material. Ideally, radiation damage
studies would also be carried out with the radiator material far from the
sensor sample, so that it may be understood whether, for a given sample technology,
radiation damage in electromagnetic showers is dominated by the ballistic electromagnetic
component or the more isotropic hadronic component.

The Santa Cruz Institute for Particle Physics (SCIPP) at UC Santa Cruz, in collaboration with the
Stanford Linear Accelerator Center (SLAC) is preparing apparatus for the
SLAC End Station Test Beam (ESTB) that can carry out such studies for
candidate sensor technologies. At up to 5 Hz, the ESTB can deliver charges of up to 0.25 nC,
at energies up to 4 GeV (15 GeV in 2014 and beyond). For a sensor at shower max for 3.5 GeV
electrons incident on a tungsten radiator, moved continuously in a ~1 cm square pattern to
achieve even illumination, the rate of accumulation of electromagnetic fluence is given
approximately by
$$ 10 {\rm MGy} \simeq \frac{600}{I_{beam}(nA) \cdot E_{beam} (GeV)} \;\; {\rm hours}.$$
Thus, for a 1 nA 3 GeV beam, a 100 MRad exposure can be accumulated in approximately one day.
Both during and after irradiation, sensors will be maintained at $-10^o$ Celsius to avoid annealing effects.

After irradiation, exposed sensors will be tested for charge-collection efficiency with
an apparatus that searches for coincidences of energy depositions, between the exposed sensor
and a small block of scintillator, from 2.3 MeV beta particles arising from a Strontium 90 source.
This `charge-collection' apparatus, developed by SCIPP for ATLAS radiation damage studies,
has recently been modularized to allow a quick assessment of radiation damage for a large
number of potential samples.

SCIPP has both oxygenated float-zone and magnetic Czochralski sensors currently in hand,
with both p-bulk and n-bulk polarities available for both technologies. A summer-2013
run has been designated T-506 by the SLAC accelerator management, entailing a four-week run from
mid-June to mid-July. The T-506 collaboration hopes to explore the radiation sensitivity
of these available technologies for exposures as high as 1 MGy during this running period.
Later running periods will offer the opportunity to do follow-up studies, including
higher doses allowed by the increased beam energy, as well as damage studies for other sensor technologies of
interest to the FCAL collaboration.

\section{Summary and Future Plans}

A design for the instrumentation of the very forward region of a detector at the International
Linear collider and at CLIC is presented. 
Two calorimeters are foreseen, LumiCal to measure precisely the
luminosity and BeamCal, supplemented by a pair monitor, for a fast luminosity estimate  and 
a measurement of beam parameters.
The performance of the two calorimeters  
has been estimated by Monte Carlo simulations for the chosen geometry
of compact cylindrical sandwich calorimeters with tungsten absorbers and finely segmented sensors.
A measurement of the luminosity with a precision of 10$^{-3}$ and 10$^{-2}$
at the ILC and CLIC, respectively, is feasible.
BeamCal will be able to measure beam parameters bunch-by-bunch and assist beam-tuning via a fast feedback
system.
Both calorimeters will improve the hermeticity of the detectors at ILC and CLIC and improve the physics potential in new particle searches.

Due to beamstrahlung, the nominal beam energy of individual events will have a tail to lower energies
leading to a spectrum of centre-of-mass energies, the luminosity spectrum.
A method has been developed to reconstruct the luminosity spectrum from data.
The impact of beam-beam effects and physics background on the luminosity measurement is investigated 
and found to be containable with systematic uncertainties not deteriorating the required precision.   

Requirements on the sensor granularity and the parameters of the FE
and ADC ASICs, are derived from Monte Carlo simulations.  
Prototypes of sensors were manufactured 
for BeamCal and LumiCal.
To approach the required radiation hardness in BeamCal, sensors made of GaAs and CVD diamond
were investigated in low-energy electron beams. Large area GaAs sensors are currently the baseline
for BeamCal. Only at very small radii, adjacent to the beam-pipe, CVD diamond sensors are an option.
For LumiCal, silicon sensors will be used.

Front-end ASICs and ADC ASICs have been developed, produced and tested for the ILC applications. 
Their measured performance fulfills the specifications
derived from simulations.
The development of ASICs for the very forward calorimeters at CLIC is still in a preparatory phase.

Using FE and ADC ASICs produced in 350 nm AMS technology, sensor 
plane segments for LumiCal and BeamCal have been studied in a 4.5 GeV electron beam. 
The full functionality was demonstrated with excellent performance. Signal-to-noise ratios
equal to or better than 20 were obtained for single particles. Effects at the edges of the sensors 
were studied in detail and found to be only of little impact on the measured signals.  

The next step will be to build a full calorimeter prototype 
to verify the predictions from simulations with respect to energy resolution and 
polar angle measurements. Effort will be invested in connectivity technologies 
to reduce the thickness of sensor planes to approach a highly compact calorimeter with small Moliere 
radius. 
A new generation of  dedicated ASICs for LumiCal, using 130 nm CMOS technology, is 
under development to reduce power dissipation and space for the on-board electronics. 
The Bean-ASIC for BeamCal will be
extended to a multi-channel version with a digital memory array on chip and a fast analog adder for 
a fast feedback to the accelerator. 

In addition to the sensor and electronics development and the integration and alignment studies, 
the design will be further refined by more Monte Carlo studies, also on the 
potential contributions of the very forward calorimeters to the physics measurements.

\section{Acknowledgments}

This work is supported by the Commission of the European Communities
under the 6$^{th}$ Framework Program "Structuring the European
Research Area", contract number RII3-026126 and by FP7 AIDA.
Tsukuba University is supported in part by the Creative 
Scientific Research Grant No. 18GS0202 of the Japan 
Society for Promotion of Science.
The AGH-UST is supported by the Polish Ministry of Science and Higher
Education under contracts: Nr. 372/6.PRUE/2007/7, Nr. 1246/7.PR UE2010/7, Nr. 2156/7. PR UE/2011/2.  
The INP PAN is supported by
    the Polish Ministry of Science
    and Higher Education
    under contract Nr. 2369/7.PR/2012/2.
IFIN-HH is supported by the Romanian Ministry of Education, Research and Innovation 
and the Authority CNCSIS under contract IDEI-253/2007.
The VINCA group is benefiting from the project "Physics and Detector R$\&$D in 
HEP Experiments" supported by the Ministry of Science of the Republic of Serbia.
J. Aguilar, P. Ambalathankandy and O. Novgorodova 
are supported by the 7th Framework Programme
"Marie Curie ITN", grant agreement number 214560.
The Tel Aviv University is supported by the Israel Science Foundation and the German-Israeli Foundation.

\appendix

\section*{APPENDIX}

\begin{center}
 {\Large{ {List of Publications,  Notes, Proceedings, Talks and Theses}}} 
\end{center}
%\author{\bf {The FCAL Collaboration}}
%\date{June 2013} % Activate to display a given date or no date (if empty),
         % otherwise the current date is printed 

%\begin{document}
%\maketitle

\section{Publications}

\begin{enumerate}

\item
S. Lukic, I. Bozovic-Jelisavcic, M. Pandurovic, I. Smiljanic,  \\
{\it Correction of beam-beam effects in luminosity measurement in the forward region at CLIC,}\\
JINST {\bf 8} (2013) P050008.

\item
E. Alvarez, D. Avila, H. Campillo, A. Dragone, and A. Abusleme, \\
{\it Noise in Charge Amplifiers – A gm/ID Approach,} \\
IEEE Transac. Nucl. Sci. {\bf 59} (2012) 2457.

\item
K. Afanaciev et al., \\
{\it Investigation of the radiation hardness of GaAs sensors in an electron beam, }\\
JINST {\bf 7} (2012) P11022.

\item
A. Abusleme, A. Dragone, G. Haller, and B. Wooley, \\
{\it BeamCal Instrumentation IC: Design, Implementation, and Test Results,} \\
IEEE Transac. Nucl. Sci. {\bf 59} (2012) 589.

\item
M. Idzik, K. Swientek, T. Fiutowski, Sz. Kulis, D. Przyborowski, \\
{\it A 10-Bit Multichannel Digitizer ASIC for Detectors in Particle Physics Experiments, }\\
IEEE Transac. Nucl. Sci. {\bf 59} (2012) 294.

\item
D. Przyborowski; M. Idzik, \\
{\it Development of Low-Power Small-Area L-2L CMOS DACs for multichannel readout systems,} \\
JINST {\bf 7} (2012) C01026.

\item
Sz. Kulis, A. Matoga, M. Idzik, K. Swientek, T. Fiutowski, D. Przyborowski, \\
{\it A general purpose multichannel readout system for radiation detectors,} \\
JINST {\bf 7} (2012) T01004.

\item
M. Idzik, K. Swientek, T. Fiutowski, S. Kulis, P. Ambalathankandy, \\
{\it A power scalable 10-bit pipeline ADC for Luminosity Detector at ILC, }\\
JINST {\bf 6} (2011)  P01004.

\item
M. Orlandea, C. Coca, L. Dumitru, E.Teodorescu, \\
{\it High-performance computing system for High Energy Physics,} \\
Rom. J. Phys, {\bf 56} (2011) 359.

\item
H. Abramowicz et al., \\
{\it Forward instrumentation for ILC detectors, }\\
JINST {\bf 5} (2010) P12002.

\item
M Idzik, K Swientek, Sz. Kulis, \\
{\it Development of pipeline ADC for the Luminosity Detector at ILC, }\\
JINST {\bf 5} (2010) P04006.

\item
C. Coca, W. Lohmann, M. Orlandea, A. Sapronov, E. Teodorescu, \\
{\it Expected electromagnetic and neutron doses for the BeamCal at ILD, }\\
Rom. J. Phys, {\bf 55} (2010) 687.

\item
D. Przyborowski, M. Idzik, \\ 
{\it A 10-bit Low-Power Small-Area High-Swing CMOS DAC, }\\
IEEE Trans.Nucl.Sci. {\bf 57} (2009) 292.

\item
M. Idzik, Sz. Kulis, D. Przyborowski, \\
{\it Development of front-end electronics for the luminosity detector at ILC,} \\
Nucl.Instrum.Meth. {\bf A 608} (2009) 169.

\item
Ch. Grah, K. Afanaciev, I. Emeliantchik, H. Henschel, A. Ignatenko, E. Kouznetsova, W. Lange, W. Lohmann, M. Ohlerich and R. Schmidt, \\
{\it Polycrystalline CVD Diamonds for the Beam Calorimeter of the ILC, }\\
IEEE Trans.Nucl.Sci. {\bf 56} (2009) 462.

\item
Ch Grah and A Sapronov, \\
{\it Beam parameter determination using beamstrahlung photons and incoherent pairs,} \\
JINST {\bf 3} (2008) P10004.

\item
C. Rimbault, P. Bambade, K. Moenig and D. Schulte, \\
{\it Impact of beam-beam effects on precision luminosity measurements at the ILC, }\\
JINST {\bf 2} (2007) P09001.

\item
I.Bozovic-Jelisavcic, M. Pandurovc, I. Smiljanic, \\
{\it Physics background as a systematic effect in luminosity measurement at
International Linear Collider,} \\
FACTA UNIVERSITATIS Series: Physics, Chemistry and Technology {\bf 5} (2007) 19.

\item
H. Abramowicz et al., \\
{\it Instrumentation on the very forward region of a linear collider detector, }\\
IEEE Trans.Nucl.Sci. {\bf 51} (2004) 2983.

\end{enumerate}

\section{Notes}

\begin{enumerate}

\item
J. Aguilar, P. Ambalathankandy, T. Fiutowski, M. Idzik, Sz. Kulis, D. Przyborowski, K. Swientek et al., \\
{\it Infrastructure for Detector Research and Development towards the International Collider.} \\
 arXiv:1201.4657v1 [physics.ins-det], 2012, \\

\item
Sz. Kulis, M. Idzik, \\
{\it Study of readout architectures for triggerless high event rate detectors at CLIC, }\\
CERN LCD-Note-2011-015.

\item
S. Poss and A. Sailer, \\
{\it Differential Luminosity Measurement using Bhabha Events, }\\
LCD-Note-2011-040.

\item
D. Dannheim and A. Sailer, \\
{\it Beam-Induced Backgrounds in the CLIC Detectors,}\\
LCD-Note-2011-021.

\item
S. Lukic, I. Smiljanic, \\
{\it Correction of beam-beam effects in luminosity measurement at ILC, }\\
LC note PRELLC-PHSM-2012-001, arXiv:1211.6869.

\item
E. Teodorescu and A. Sailer, \\
{\it Radiation Dose to the QD0 Quadrupole in the CLIC Interaction Region,} \\
LCD-Note-2010-013.

\item
A. Sailer, \\
{\it Simulation of Beam-Beam Background at CLIC, }\\
LCD-Note-2010-007.

\item
J.A. Aguilar, B. Pawlik, \\
{\it Monte Carlo evaluation of tile gap effect on energy solution in LumiCal, }\\
EUDET-Memo-2010-35.

\item
S. Kulis et al., \\
{\it Test beam studies of the LumiCal prototype,}\\
EUDET-Memo-2010-09.

\item
J. Aguilar, W. Daniluk, E. Kielar, J. Kotuła et al. \\
{\it Laser Alignment System for LumiCal, }\\
EUDET-Memo-2010-08.

\item
W. Daniluk, E. Kielar, J. Kotuła et al., \\
{\it Redesign of LumiCal mechanical structure, }\\
EUDET-Memo-2010-06.

\item
W. Lohmann et al., \\
{\it VFCAL Annual Report 2010, }\\
EUDET-Memo-2010-03.

\item
H. Abramowicz, R. Ingbir, S. Kananov, A. Levy, I. Sadeh, \\
{\it A Luminosity Calorimeter for CLIC,} \\
LCD-Note-2009-002.

\item
J. Błocki, W. Daniluk, E. Kielar et al., \\
{\it LumiCal new mechanical structure, }\\
EUDET-Memo-2009-10.

\item
M. Idzik, K. Swientek, T. Fiutowski, Sz. Kulis, D. Przyborowski, P. Ambalathankandy, \\
{\it Readout electronics for LumiCal detector, }\\
EUDET-Report-2009-09.

\item
J. Błocki, W. Daniluk, E. Kielar et al., \\
{\it Silicon Sensors Prototype for LumiCal Calorimeter, }\\
EUDET-Memo-2009-07.

\item
H. Abramowicz, R.Ingbir, S.Kananov et al., \\
{\it A Luminosity Calorimeter for CLIC, }\\
EUDET-Memo-2009-06.

\item
W. Wierba et al., \\
{\it Lumical mechanical design proposal and integration with ILD, }\\
EUDET-Memo-2008-13.

\item
M. Idzik et al., \\
{\it Status o lumical readout electronics, }\\
EUDET-Report-2008-08.

\item
W. Daniluk et al., \\
{\it Laser alignement system for lumical, }\\
EUDET-Report-2008-05.

\item
H. Abramowicz et al., \\
{\it Geant4 simulation of the electronic readout constraints for the luminosity detector of the ILC, }\\
EUDET-Memo-2007-17.

\item
M. Idzik, K. Swientek, Sz. Kulis, W. Dabrowski, L. Suszycki, B. Pawlik, W. Wierba, L. Zawiejski, \\
{\it The Concept of LumiCal Readout Electronics.} \\
EUDET-Memo-2007-13.

\item
J. Blocki et al., \\
{\it Proposed design of the silicon sensors for the LumiCal, }\\
EUDET-Memo-2007-09.

\item
H. Abramowicz et al., \\
{\it A luminosity detector for the international linear collider, }\\
Linear Collider notes, LC-DET-2007-006.

\item
H.Abramowicz, R.Ingbir, S.Kananov, A.Levy, \\
{\it Monte Carlo study of a luminosity detector for the International Linear Collider, }\\
physics/0508074.

\item
R. Dollan et al., \\
{\it Beam calorimeter technologies, }\\
physics/0507002.

\item
W. Lohmann, \\
{\it R\&D for a Detector at the International Linear Collider, }\\
DESY-05-141,August 2005.

\item
W. Lohmann, \\
{\it Instrumentation of the Very Forward Region of a Linear Collider Detector, }\\
DESY-05-142,August 2005.

\item
A. Stahl, \\
{\it Luminosity Measurement via Bhabha Scattering: Precision Requirements for the Luminosity Calorimeter,} \\
LC-DET-2005-004, 2005.

\item
A. Stahl, \\
{\it Diagnostics of Colliding Bunches from Pair Production and Beam Strahlung at the IPregion of a linear collider detector, }\\
LC-DET-2005-003, 2005.

\end{enumerate}

\section{Proceedings}

\begin{enumerate}

\item
O. Novgorodova,  [on behalf of the FCAL Collaboration], \\
{\it Forward Calorimeters Test Beam Results for Future Linear Colliders, }\\
Proceedings of the XXXVI International Conference on High Energy Physics, 4-11 July 2012, Melbourne, Australia, POS (ICHEP 2012) 552, 2012.

\item
I. Bozovic Jelisavcic [on behalf of the FCAL Collaboration], \\
{\it Physics and detector studies with the very forward calorimeters at a future linear collider,} \\
Proceedings of the XXXVI International Conference on High Energy Physics, 4-11 July 2012, Melbourne, Australia, PoS (ICHEP 2012),  p.540.

\item
I. Bozovic Jelisavcic [on behalf of the FCAL Collaboration], \\
{\it Design and R\&D of very forward calorimeters for detectors at future $e^+e^-$ collider,}\\
The 2011 Europhysics Conference on High Energy Physics - HEP2011, Grenoble, Rhone-Alpes, July 2011, Proceedings of Science, PoS(EPS-HEP2011), p.210, arXiv:1201.6495.

\item
I. Bozovic-Jelisavcic, M. Pandurovic, I. Smiljanic, T. Jovin, I .Sadeh, [on behalf of the FCAL Collaboration], \\
{\it Forward region studies for ILC, }\\
Proceedings of the 7th International Conference of the Balkan Physical Union, Alexandroupolis, Greece, 9-13 September 2009, AIP Conf.Proc.1203 (2010),  p.49, ISBN 978-0-7354-0740-4.

\item
O. Novgorodova, K.Afanaciev, J.Aguilar, H.Henschel, M. Idzik, A.Ignatenko, Sz.Kulis, S.Kollowa, W.Lange, I.Levy, W.Lohmann, S.Schuwalow,  [on behalf of the FCAL Collaboration], \\
{\it Forward Calorimeters for the Future Electron-Positron Linear Collider Detectors, }\\
Proceedings of the XIXth International Workshop on High Energy Physics and Quantum Field Theory - QFTHEP2010 Golitsyno, Moscow, Russia
September 08–15 2010, PoS (QFTHEP2010) 030.

\item
I. Bozovic-Jelisavcic, H. Abramowicz, P. Bambade, T. Jovin, M. Pandurovic, B. Pawlik, C. Rimbault, I. Sadeh, I. Smiljanic, [on behalf of the FCAL Collaboration], \\
{\it Luminosity Measurement at ILC, }\\
Proceedings of the International Linear Collider Workshop 2010 LCWS10 \& ILC10, arXiv:1006.2539v1 [physics.ins-det].

\item
O. Novgorodova [on behalf of the FCAL Collaboration], \\
{\it Studies on the Electron Reconstruction Efficiency for the Beam Calorimeter of an ILC Detector,} \\
Proceedings of the International Linear Collider Workshop 2010 LCWS10 \& ILC10, arXiv:1006.3402v1 [physics.ins-det].

\item
E. Teodorescu, [on behalf of the FCAL Collaboration], \\
{\it Background Simulations in BeamCal at the ILC, }\\
Proceedings of the Trans-European School of High Energy Physics. Zakopane, Poland. July
7-15, 2009, p. 205.

\item
M. Idzik, Sz. Kulis, J. Gajewski, D. Przyborowski, [on behalf of the FCAL Collaboration], \\
{\it Development of front-end electronics for the luminosity detector at ILC, }\\
Proceedings of the 15th International Conference, Mixed Design of Integrated Circuits and Systems MIXDES 2008, Poznan Poland 19-21 June 2008.

\item
M. Pandurovic, I. Bozovic-Jelisavcic, [on behalf of the FCAL Collaboration], \\
{\it Systematic Effects in Luminosity Measurement at ILC, }\\
Analele Universitatii De Vest Din Timisoara Seria Fizica Vol. 50/2007, p. 63, ISSN 1224-9718, 2008.

\item
 I. Bozovic-Jelisavcic, [on behalf of the FCAL Collaboration], \\
{\it Forward Calorimetry at ILC, }\\
Analele Universitatii De Vest Din Timisoara Seria Fizica Vol. 50/2007, p. 23, ISSN 1224-9718, 2008.

\item
M. Pandurovic, I. Bozovic Jelisavcic [on behalf of the FCAL Collaboration], \\
{\it Physics Background as a Systematic Effect in Luminosity Measurement at ILC, }\\
Proceedings of the International Linear Collider Workshop LCWS 2007 and ILC 2007 Vol 2, p.710, ISBN 978-3-935702-27-0 (2008).

\item
W.Wierba,  [on behalf of the FCAL Collaboration], \\
{\it Very forward calorimeters readout and machine interface, }\\
PRAMANA – Journal of Physics, Indian Academy of Sciences, VOL. 69  (2007). Proccedings of the LCWS06 Bangalore Workshop.

\end{enumerate}

\section{Talks}
\begin{enumerate}

\item
 S. Lukic, [on behalf of the FCAL Collaboration], \\
{\it Luminosity measurement at 500 GeV and 1 TeV ILC, }\\
ECFA LC2013, DESY, Hamburg, May 27 - 31, 2013.

\item
W. Wierba, [on behalf of the FCAL Collaboration], \\
{\it The forward calorimetry for future linear collider.} \\
CHEF 2013 Paris, France, 22- 25 April, 2013.

\item
W. Lohmann, [on behalf of the FCAL Collaboration], \\
{\it ILC Detector R\&D,} \\
ILC PAC KEK 2012 Tokyo, Japan, 13-14 December, 2012.

\item
S. Kulis, [on behalf of the FCAL Collaboration], \\
{\it State-of-the-art in Forward Calorimetry and other Miscellaneous Detector Applications,} \\
Nuclear Science Symposium 2012, 29/10 - 3/11 2012, Anaheim, California, USA.

\item
I. Smiljanic, [on behalf of the FCAL Collaboration], \\
{\it Correction methods for counting losses induced by the beam-beam effects in luminosity measurement at ILC, }\\
LCWS12, Arlington, Texas, USA, October 2012.

\item
A. Sailer, Stéphane Poss, [on behalf of the FCAL Collaboration], \\
{\it Measurement of the Differential Luminosity at 3 TeV CLIC, }\\
Linear Collider Workshop, Arlington, Texas, October 2012. 

\item
A. Sailer, [on behalf of the FCAL Collaboration], \\
{\it Electron Tagging with the BeamCal at 3 TeV CLIC, }\\
Linear Collider Workshop, Arlington, Texas, October 2012. 

\item
E. Teodorescu (on behalf of the FCAL Collaboration), \\
{\it Detector studies with the very forward calorimeters at a future linear collider,}\\
CNF-2012, National Physics Conference, 8 - 10 July 2012, Constanta, Romania.

\item
S. Lukic, [on behalf of the FCAL Collaboration], \\
{\it Beam-beam effects in luminosity measurement for ILC and CLIC, }\\
Joint ACFA Physics / Detector Workshop and GDE meeting on Linear Collider - KILC12, 23-27 April 2012, Daegu, Korea.

\item
M. Idzik, [on behalf of the FCAL Collaboration], \\
{\it Development of ASICs for forward calorimetry in future linear collider, }\\
A Joint Instrumentation Seminar of the Particle Physics and Photon Science communities at DESY, Hamburg University and XFEL, DESY, Hamburg Germany February 17, 2012,  http://instrumentationseminar.desy.de/e70397/e95574/.

\item
W. Lohmann, [on behalf of the FCAL Collaboration], \\
{\it Performance of a fully instrumented sensor plane for FCAL'}, \\
International Workshop on future Linear Colliders, University of Texas at Arlington, 2012.

\item
O. Novgorodova, [on behalf of the FCAL Collaboration], \\
{\it Test Beam Resullts from the Forward Calorimeters'}, \\
KILC12, Daegu, Korea.

\item
W. Lohmann, [on behalf of the FCAL Collaboration], \\
{\it Forward Instrumentation of ILC and CLIC Detectors',} \\
KILC12, Daegu, Korea.

\item
S. Kulis, [on behalf of the FCAL Collaboration], \\
{\it ILD Forward Region ,}\\
International Workshop on Future Linear Colliders 2011, 26-30 September 2011, Granada, Spain.

\item
S. Kulis, [on behalf of the FCAL Collaboration], \\
{\it Development of readout electronics and test-beam results of FCAL detector prototype, }\\
International Workshop on Future Linear Colliders 2011, 26-30 September 2011, Granada, Spain.

\item
D. Przyborowski, M. Idzik, [on behalf of the FCAL Collaboration], \\
{\it Development of Low-Power Small-Area L-2L CMOS DACs for Multichannel Readout Systems,}\\
TWEP 2011 Topical Workshop on Electronics for Particle Physics, 26-30 September 2011 Vienna, Austria.

\item
J. Moron, M. Firlej, M. Idzik, [on behalf of the FCAL Collaboration], \\
{\it Development of low power Phase-Locked Loop (PLL) and PLL-based serial transceiver,}\\
TWEP 2011 Topical Workshop on Electronics for Particle Physics, 26-30 September 2011 Vienna, Austria.

\item
M. Idzik, [on behalf of the FCAL Collaboration], \\
{\it Development of ASICs for LumiCal readout within FCAL and Power pulsing issues, }\\
Linear Collider Power Distribution and Pulsing workshop, 9-10 May 2011, LAL Orsay France.

\item
W. Wierba, [on behalf of the FCAL Collaboration], \\
{\it FCAL Developments towards DBD and beyond,}\\
ILD Worshop 2011, 23-25 May 2011, LAL, Paris, France.

\item
W. Lohmann,  [on behalf of the FCAL Collaboration], \\
{\it FCAL update',} \\ 
2011 Linear Collider Workshop of the Americas (ALCPG11), University of Oregon.

\item
O. Novgorodova,  [on behalf of the FCAL Collaboration], \\
{\it Testbeam Results of Beam Calorimeter GaAs Prototype, } \\
2011 Linear Collider Workshop of the Americas (ALCPG11), University of Oregon.

\item
W. Lohmann,  [on behalf of the FCAL Collaboration], \\
{\it ILC Detector R\&D: Its Impact,} \\
International Workshop on Future Linear Collider'', Granada, Spain 2011.

\item
A. Rosca,  [on behalf of the FCAL Collaboration], \\
{\it Electron Reconstruction in BeamCal, } \\ 
International Workshop on Future Linear Collider, Granada, Spain, 2011.  

\item
J. Moron, M. Firlej, M. Idzik, [on behalf of the FCAL Collaboration], \\
{\it Design and measurements of CMOS transceiver for fast serial readout and design of fast PLL general purpose block, }\\
Workshop on timing detectors: electronics, medical and particle physics applications, 29 November - 01 December 2010 Cracow Poland.

\item
S. Kulis, M. Idzik, [on behalf of the FCAL Collaboration], \\
{\it Triggerless readout with time and amplitude reconstruction of event based on deconvolution algorithm, }\\
Workshop on Timing Detectors, 29 November - 01 December 2010, Cracow, Poland.

\item
S. Kulis, M. Idzik, K. Swientek, [on behalf of the FCAL Collaboration], \\
{\it Progress in development of LumiCal readout electronics and testbeam results of a prototype detector sector, }\\
IWLC2010 International Workshop on Linear Colliders 2010, 18-22 October 2010, CERN, Geneva, Switzerland.

\item
A. Sailer, [on behalf of the FCAL Collaboration], \\
{\it Simulation Studies regarding Beam-Beam Background in a CLIC Detector, }\\
International Workshop on Linear Colliders (IWLC2010), 18-22 October 2010, CERN, Geneva, Switzerland.

\item
A. Sailer, [on behalf of the FCAL Collaboration], \\
{\it Simulation of Beam-Beam Background at CLIC, }\\
LCWS2010, Beijing, China, 29 March 2010.

\item
B. Pawlik (on behalf of FCAL Collaboration), \\
{\it LumiCal simulation status, }\\
ILD Workshop, 27-30 January, 2010, Paris. 

\item
L. Zawiejski (on behalf of FCAL Collaboration), \\
{\it Forward detectors, }\\
ILD Workshop, 27-30 January, 2010, Paris.

\item
M. Idzik, K. Swientek, Sz. Kulis, [on behalf of the FCAL Collaboration], \\
{\it Design and measurements of 10 bit pipeline ADC for the Luminosity Detector at ILC, }\\
Topical Workshop on Electronics for Particle Physics, TWEPP-09, Paris, France, 21-25 Sep, 2009.

\item
D. Przyborowski, M. Idzik, [on behalf of the FCAL Collaboration], \\
{\it Development of a General Purpose Low-power Small-area 10 bit Current Steering CMOS DAC, }\\
16th International Conference "Mixed Design of Integrated Circuits and Systems" MIXDES 2009, ód, Poland, 25-27 June 2009.

\item
M. Idzik, K. Swientek, Sz. Kulis, [on behalf of the FCAL Collaboration], \\
{\it Development of Pipeline ADC for the Luminosity Detector at ILC , }\\
Mixed Design of Integrated Circuits and Systems MIXDES 2008, Poznan Poland 19-21 June 2008.

\item
B. Pawlik (on hehalf of FCAL Collaboration), \\
{\it Forward CALorimetry Status, }\\
ECFA ILC Workshop, June 9-12, 2008, Warsaw, Poland.

\item
W. Wierba (on behalf of FCAL Collaboration), \\
{\it LumiCal Mechanical Design and Laser Alignment System.}\\
ILC Interaction Region Engineering Design Workshop, SLAC, Menlo Park,  California, USA, 17-21.09.2007.

\item
M. Idzik, K. Swientek, S. Kulis, (on behalf of FCAL Collaboration), \\
{\it Readout electronics for LumiCal detector, }\\
The Linear Collider Workshop 2007 and the International Linear Collider meeting 2007 DESY Hamburg, Germany, May 30 - June 3, 2007.

\item
I. Bozovic-Jelisavcic, M. Pandurovic, (on hehalf of FCAL Collaboration), \\
{\it Four-fermion C processes as a background in the luminosity calorimeter at ILC,} \\
International Linear Collider Workshop, Valencia, Spain, 6-10 November, 2006.

\end{enumerate}

\section{Theses}

\begin{enumerate}

\item
{\it Development of prototype luminosity detector modules for future experiments on linear colliders}, {\bf S. Kulis, PhD Thesis (2013)}, AGH-UST, Cracow, Poland

\item
{\it Radiation and Background Levels in a CLIC Detector due to Beam-Beam Effects - Optimisation of Detector Geometries and Technologies,} {\bf A. Sailer, PhD Thesis (2013) },Humboldt University,  Berlin, Germany.

\item
{\it Characterisation and Application of Radiation Hard Sensors at ILC and LHC,} {\bf O. Novgorodova, PhD Thesis (2013),} Brandenburg University of Technology, Cottbus, Germany.

\item
{\it Aufbau und Inbetriebnahme eines Sensormessplatzes zur Bestimmung der Ladungssammeleffizienz,} {\bf O. Anton, BSc Thesis (2013),} Brandenburg University of Technology, Cottbus, Germany.

\item
{\it Vermessung der Nachweiseigenschaften neuartiger Diamantsensoren fuer ionisierende Strahlung, } {\bf O. Reetz, BSc Thesis (2012), } Brandenburg University of Technology, Cottbus, Germany.

\item
{\it Aufbau und Test einer Detektorebene mit GaAs Sensor,} {\bf S. Kollowa, Diploma Thesis (2012), }Brandenburg University of Technology, Cottbus, Germany.

\item 
{\it Detector development for the instruments in the forward region of future linear colliders,} {\bf I. Levy, MSc Thesis (2012), } Tel Aviv University, Israel.

\item
{\it  Luminosity measurement at the Compact Linear Collider}, {\bf R. Schwartz, MSc Thesis (2012), } Tel Aviv University, Israel.

\item
{\it Luminosity detector at ILC: Monte Carlo Simulations and analysis of test beam data,} {\bf J. Aguilar, MSc Thesis (2012), }AGH-UST, Cracow, Poland.

\item
 {\it Rare decays in high energy interactions and performance studies on FCAL region at a linear collider}, {\bf E. Teodorescu, PhD Thesis (2011),} University of Bucharest, Bucharest, Romania.

\item
{\it Physics background in luminosity measurement at ILC and measurement of the proton b-content at H1 using multivariate method}, {\bf M. Pandurovic, PhD Thesis (2011), }University of Belgrade, Serbia.

\item
{\it Investigations of the Physics Potential and Detector Development for the ILC},  {\bf M. Ohlerich, PhD Thesis (2010),} Brandenburgische Technische Universitaet Cottbus, Cottbus, Germany.

\item
{\it Studies on the Measurement of Differential Luminosity using Bhabha Events at the International Linear Collider},  {\bf A. Sailer, MSc Thesis (2009), } Humboldt University, Berlin, Germany.

\item
{\it Simulations of Electromagnetic and Hadronic Background in BeamCal at the ILC}, {\bf E. Teodorescu, MSc Thesis (2008),} University of Bucharest, Bucharest, Romania.

\item
{\it Luminosity Measurement at the International Linear Collider},  {\bf I. Sadeh, MSc Thesis (2008),} Tel Aviv University, Israel.

\item
{\it Optimisation of a Testbeam Setup and Background Estimates for Detectors at the ILC using Monte Carlo Simulations}, {\bf R. Schmidt, MSc Thesis  (2007),} Brandenburgische Technische Universitaet Cottbus, Cottbus, Germany.

\item
{\it Design Studies and Sensor Tests for the Beam Calorimeter of the ILC Detector}, {\bf E. Kuznetsova, PhD Thesis (2007), }Humboldt University, Berlin, Germany.

\item
{\it A Luminosity Detector for the ILC}, {\bf R. Ingbir, MSc Thesis (2006),} Tel Aviv University, Israel.

\item
{\it Beam diagnostic Laser-Wire and Fast Luminosity Spectrum Measurement at the International Linear Collider,} {\bf F. Poirier, PhD Thesis (2005),} University of London, London, UK.

\end{enumerate}


\begin{thebibliography}{99}

%%%%%%%%%%%%%%%%%%%%%%%%%%%%%%% INTRODUCTION %%%%%%%%%%%%%%%%%%%%%%%%

\bibitem{ILC_pub}
{ \it International Linear Collider Global Design Effort and World Wide Study,} FERMILAB-DESIGN-2007-03, arXiv:0712.1950[physics.acc-ph]; \\
{\it ILC Reference Design Report Volume 3 -Accelerator}, arXiv:0712.2361[physics.acc-ph]; \\
{\it The International Linear Collider,} Technical Design Report, to be
published, 2013.



\bibitem{clic_info}
{ \it The Compact Linear Collider Study},
http://clic-study.web.cern.ch/clic-study/. \\
{\it The CLIC Programme: towards a staged e$^+$e$^-$ Linear Collider exploring the
Terascale,} 2012, eds. P. Lebrun, L. Linssen, A. Lucaci-Timoce, D. Schulte,
F. Simon, S. Stapnes, N. Toge, H. Weerts, J.D. Wells,
ANL-HEP-TR-12-51, KEK Report 2012-2, MPP-2012-115,arXiv:1209.2543.


\bibitem{ILD_pub}
T. Abe \etal, {\it The International Large Detector: Letter of Intent},
FERMILAB-LOI-2010-01, FERMILAB-PUB-09-682-E, DESY-2009-87, KEK-REPORT-2009-6, arXiv:1006.3396, 
(2010).

\bibitem{SiD_pub}
E.L. Berger \etal,
{ \it SiD Letter of Intent} (2009)
https://confluence.slac.stanford.edu/display/SiD/home.

\bibitem{ieee1}
H.~Abramowicz \etal,
{ \it	
Forward Instrumentation for ILC Detectors}, 
{JINST}  {\bf 5}  (2010)  P12002.

\bibitem{drugakov}
P.~Bambade, V. Drugakov and W. Lohmann,
{\it The impact of Beamcal performance at different ILC beam parameters
  and crossing angles on stau searches},
{Pramana  J. Phys.} {\bf 69} (2007) 1123.
\bibitem{new_Bhabha}
R. Bonciani and A. Ferroglia,
{\it Bhabha Scattering at NNLO},
{Nucl. Phys. Proc. Suppl. 181-182}  (2008)  259; \\  
T. Becher and K. Melnikov, {\it Two-loop QED corrections to Bhabha scattering},
JHEP {\bf 0706} (2007) 084; \\
S. Actis, M. Czakon, J. Gluza and T. Riemann, 
{\it Fermionic NNLO contributions to Bhabha scattering},
{Acta Phys. Polon.} {\bf B 38} (2007) 3517; \\
A.A. Penin,
{\it Two-loop photonic corrections to massive Bhabha scattering},
{Nucl. Phys.} {\bf B 734} (2006) 185; \\
M. Czakon, J. Gluza and T. Riemann,
{\it The Planar four-point master integrals for massive two-loop Bhabha scattering},
Nucl. Phys. {\bf B 751} (2006) 1.



%\bibitem{klaus}
%K.~M\"onig,
%{\it Physics needs for the forward region},
%{V. Workshop: Instrumentation of the Forward Region of a Linear
%Collider Detector}, August, 26--28 (2004) DESY, Zeuthen, Germany,
%http://www-zeuthen.desy.de/lcdet/Aug\_04\_WS/aug\_04\_ws.html.

\bibitem{grah1}
Ch. Grah and A.~Sapronov,
{\it Beam parameter determination using beamstrahlung photons and
  incoherent pairs},
{JINST} {\bf 3} (2008) P10004.

%%%%%%%%%%%%%%%%%%%%%%%%%%%%%%%%%%%%%%%%%%%%%%%%%%%%%%%%%%%%%%%%%%%%%%%%%%%%%%%%%%%%%%%%%%%%%%

%%%%%%%%%%%%%%%%%%%%%   LUMICAL DESIGN %%%%%%%%%%%%%%%%%%%%%%%%%%%%%%%%%%%%%%%%%%%%%%%%%%%%

\bibitem{BHWIDE}
S.~Jadach, W. Placzek and B.F.L. Ward,
{\it BHWIDE 1.00: O(alpha) YFS exponentiated Monte Carlo for 
       Bhabha  scattering at wide angles for LEP1/SLC and LEP2},
{Phys. Lett. } {\bf B 390} (1997) 298.

\bibitem{g4}
S.~Agostinelli \etal,
{\it Geant4- a simulation toolkit}, 
{Nucl. Inst. and Meth. } {\bf A 506} (2003) 250.

\bibitem{mokka}
{\it MOKKA, A simulation program for linear collider detectors},
http://polzope.in2p3.fr:8081/MOKKA/.

\bibitem{bib17}
T.C. Awes, F.E. Obenshain, F. Plasil, S. Saini, S.P. Sorensen and G.R. Young, 
{\it A simple method of shower localization and
  identification in laterally segmented calorimeters}, 
{Nucl. Inst. Meth. } {\bf A 311} (1992) 130.

\bibitem{bib23}
R~Ingbir, {\it A Luminosity Detector for the ILC}, MSc Thesis (2006) Tel Aviv University, Israel.

\bibitem{bib20}
I.~{S}adeh, 
{\it Luminosity measurement at the International Linear Collider},
http://alzt.tau.ac.il/~sadeh/mscThesis/iftachSadeh\_mscThesis.pdf, 2008.


\bibitem{pawlik}
J. A. Aguilar and B. Pawlik, 
{\it Monte Carlo evaluation of tile gap effect on energy resolutio in LumiCal}.
{EUDET-Memo-2010-35}, 2010.


\bibitem{bib25}
H.~{A}bramowicz \etal,
{\it Revised requirements on the readout of the luminosity calorimeter}.
{EUDET-Memo-2008-08}, 2008.
http://www.eudet.org.

%%%%%%%%%%%%%%%%%%%%%%%%%%%%%%%%%%%%%%%%%%%%%%%%%%%%%%%%%%%%%%%%%%%%%%%%%%%%%%%%%%%%%%%%%%%%%%%

%%%%%%%%%%%%%%%%%%%%%DESIGN BEAMCA: %%%%%%%%%%%%%%%%%%%%%%%%%%%%%%%%%%%%%%%%%%%%%%%%%%%%%
%\bibitem{nominal_set}
%J. Brau \etal, {\it ILC Reference Design Report},
%http://arxiv.org/abs/0712.1950, 2007.

%\bibitem{TDR_set}
%{\it The International Linear Collider, Technical Design Report}, to be published, 2013.

\bibitem{guinea}
D.~Schulte,
{\it Beam-beam simulations with guinea-pig},
{CERN-PS-99-014LPCLIC-Note 387}, 1998.

\bibitem{anti-DID}
A. Seryi, T. Maruyama and B. Parker,
{\it IR optimization and anti-DID},
{SLAC-PUB-11662}, 2006.

\bibitem{a_seiler}
A. Sailer, {\it Radiation and Background Levels in a CLIC Detector due to Beam-Beam Effects - Optimisation of Detector Geometries and Technologies,} PhD Thesis (2013), Humboldt University, Berlin, Germany, CERN-THESI~S-2012-223.

\bibitem{bertini}
A. Heikkinen and N. Stepanov,
{\it Bertini Intra-nuclear Cascade implementation in Geant4},
{Proceedings of CHEP03}, La Jolla, California, 2003, nucl-th0306008.

\bibitem{RomJourPhys}
C. Coca, W.  Lohmann, M. Orlandea, A. Sapronov, E. Teodorescu, 
{\it Expected electromagnetic and neutron doses for the BeamCal at ILD,} \\
Rom. J. Phys. {\bf 55} (2010) 687 (2010); \\
E. Teodorescu, {\it Rare decays in high energy interactions and performance studies on FCAL region at a linear collider}, PhD Thesis (2011), University of Bucharest.

\bibitem{tauchi1}
 T. Tauchi and K. Yokoya,
{\it Nanometer beam-size measurement during
collisions at linear colliders},
Phys. Rev. {\bf E 51} (1995) 6119.

\bibitem{tauchi2}
T. Tauchi, K. Yokoya and P. Chen,
{\it Pair creation from beam-beam
interaction in linear colliders},
{Part. Accel.} {\bf 41} (1993) 29.

\bibitem{ito}
K.Ito,
{\it Study of Beam Profile Measurement at Interaction Point in International Linear Collider},
{arXive: 0901.4151[physics.ins-det]}.

%\bibitem{elagin_snowmass}
%A. Elagin,
%{\it The optimized sensor segmentation for the very forward calorimeter},
%in proceedings of
%the 2005 International Linear Collider Physics and Detector Workshop,
%2005 Snowmass, Colorado, ECONF C0508141:ALCPG0719.

\bibitem{LumiCal_mechanics}
J. Blocki \etal,
{\it LumiCal new mechanical structure},
{EUDET-Memo-2009-10}, 2008, 
http://www.eudet.org/e26/e28.

\bibitem{laser_alignment}
J. Blocki \etal,
{\it Laser alignment system for LumiCal},
{EUDET-Report-2008-05}, 2008, 
ttp://www.eudet.org/e26/e27/.

%%%%%%%%%%%%%%%%%%%%%%%%%%%%%%%%%%%%%%%%%%%%%%%%%%%%%%%%%%%%%%%%%%%%%%%%%%%%%%%%%%%%%%%%%%%%

%%%%%%%%%%%%%%%%%%%%%%% PROTOTYPING %%%%%%%%%%%%%%%%%%%%%%%%%%%%%%%%%%%%%%%%%%%%%%%%%%%%%%%%%%

%%%%%%%%%%%%%%%%%%%%%%%%%%%%%%%%   SENSORS    %%%%%%%%%%%%%%%%%%%%%%%%%%%%

\bibitem{jinst_gaas}
K.~Afanaciev \etal,
{\it Investigation of the radiation hardness of GaAs sensors in an electron beam},
JINST {\bf 7} (2012) P11022.

\bibitem{ieee2}
Ch.~Grah \etal,
{\it Radiation hard sensor for the BeamCal of the ILD detector},
Proceedings of the IEEE conference, October 27 -- November 3, 2007, 
Honolulu, USA.

\bibitem{ieee3}
Ch.~Grah \etal,
{\it Polycrystalline CVD Diamonds for the Beam Calorimeter of the
  ILC},
{IEEE Trans. Nucl. Sci.} {\bf 56} (2009) 462.

%%%%%%%%%%%%%%%%%%%%%%%%%%%%%%%%%%%%%%%%%%%%%%%%%%%%%%%%%%%%%%%%%%%%%%%%%%%%

%%%%%%%%%%%%%%%%%%%%%%%%%%%%%%%%%    ASICs   %%%%%%%%%%%%%%%%%%%%%%%%%%%%%%%%%%%%%


\bibitem{angel1} 
E. Alvarez, A. Abusleme, 
{\it Noise Power Normalisation: Extension of gm/ID Technique For Noise Analysis}, {Electronics Letters}, {\bf 48} (2012) 430-43


\bibitem{angel2} 
E. Alvarez, D. Avila, H. Campillo, A. Dragone, A. Abusleme, 
{\it Noise in Charge Amplifiers - A gm/ID Approach}, {IEEE Transactions on Nuclear Science}, {\bf 59} (2012) 2457-2462


\bibitem{angel3} 
D. Avila, E. Alvarez, A. Abusleme, 
{\it Noise Analysis in Pulse-Processing Discrete-Time Filters}, {IEEE Transactions on Nuclear Science}, {\bf under review} 


\bibitem{VFCAL_memo} 
M. Idzik, H. Hanschel, W. Lohmann et al., {\it Status of VFCAL}, EUDET-memo-2008-01, 2008.


\bibitem{beam} 
T. Behnke, S. Bertolucci, R.D. Heuer, R. Settles, {\it TESLA Technical Design Report, PART IV, A Detector for TESLA}, March 2001.


\bibitem{pzc}  
R.A. Boie, A.T. Hrisoho, P. Rehak, {\it Signal shaping and tail 
cancellation for gas proportional detectors at high counting rates}, 
{Nucl. Instr. and Meth.}, {\bf 192}  (1982) 365.



\bibitem{lumi_frontend} 
M. Idzik, Sz. Kulis, D. Przyborowski,
{\it Development of front-end electronics for the luminoisty detector at ILC}, 
{Nucl. Instr. and Meth.}, {\bf A~608} (2009) 169.


\bibitem{lumi_adc} 
 M. Idzik, K. Swientek, T. Fiutowski, Sz. Kulis, D. Przyborowski,
{\it A 10-bit multichannel digitizer ASIC for detectors in particle physics experiments}, 
{IEEE Trans. Nucl. Sci.}, {\bf 59} (2012) 294.




\bibitem{fcal_readout_board}
S. Kulis, A. Matoga, M. Idzik, K. Swientek, T. Fiutowski, D. Przyborowski, 
{\it A general purpose multichannel readout system for radiation detectors}
JINST {\bf 7} (2012) T01004.


%%%%%%%%%%%%%%%%%%%%%%%%%%%%%%%%%%%%%%%%%%%%%%%%%%%%%%%%%%%%%%%%%%%%%%%%%%%%%%%%

%%%%%%%%%%%%%%%%%%%%%%%%%%    ALIGNMENT   %%%%%%%%%%%%%%%%%%%%%%%%%%%%%%%%%%%%%%%%%

\bibitem{lc_p7} W. Daniluk \etal, {\it Laser Alignment System for LumiCal}, EUDET-Report-2008-05, 2008.
\bibitem{lc_p8} J. Aguilar \etal, {\it Laser Alignment System for LumiCal}, EUDET-Memo-2010-08, 2010.
\bibitem{lc_p9} K. Korcsak-Gorzo \etal, {\it The optical alignment system of the ZEUS microvertex detector}, 
Nucl. Instrum. Meth. {\bf A~580} (2007) 1227.
\bibitem{lc_p10} A.F. Fox-Murphy \etal, {\it Frequency scanned interferometry (FSI): the bassis of a survey system
for ATLAS using fast automated remote interferometry}, Nucl. Instrum. Meth. {\bf A~383} (1996) 229.
\bibitem{lc_p11} J. R. Green, PhD Thesis, {\it Development of a Prototype Frequency Scanning Interferometric Absolute Distance 
Measurement System for Survey \& Alignment of the International Linear Collider}, University of Oxford, 2007. 
\bibitem{lc_p12} Hai-Jun Yang, Sven Nyberg, Keith Riles, {\it High-precision absolute distance measurement
using dual-laser frequency scanned interferometry under realistic conditions}, 
 Nucl. Instrum. Meth. {\bf A~575} (2007) 395.
\bibitem{lc_p13} M. Warden \etal, {\it MONALISA: A precise system for accelerator component position monitoring},
EUROTeV-Report-2008-08, 2008.

%%%%%%%%%%%%%%%%%%%%%%%%%%%%%%%%%%%%%%%%%%%%%%%%%%%%%%%%%%%%%%%%%%%%%%%%%%%%%%%%%%%%%%%%%%%%%%%%%%%%

%%%%%%%%%%%%%%%%%%%%% TESTBEAM RESULTS %%%%%%%%%%%%%%%%%%%%%%%

\bibitem{Itamar} I. Levy, {\it Detector development for the instruments in the forward region of future linear colliders}, MSc Thesis, Tel Aviv University, Tel Aviv, 2012.
\bibitem{Kulis_phd}
S.~Kulis, {\it Development of prototype luminosity
detector modules for future experiments on linear colliders}, Ph.D. Thesis, AGH University of Science and Technology, Cracow, 2012.

%%%%%%%%%%%%%%%%%%%%%%%%%%%%%%%%%%%%%%%%%%%%%%%%%%%%%%%%%%%%%%%%%%%%



%%%%%%%%%%%%%%%%%%%%%%%%%%%%%%%%    LUMINOSITY SPECTRUM    %%%%%%%%%%%%%%%%%%%%%%%%%%%%%%%%%%%%%%%%%%%%%%%%

\bibitem{chen_beam}
P.~Chen, {\it An introduction to beamstrahlung and disruption}.
in Frontiers of Particle
  Beams, vol. 296 of Lecture Notes in Physics (M.~Month and S.~Turner, eds., Berlin Springer Verlag),
  pp. 495--532. 1988, SLAC-PUB-4379.

\bibitem{Chen:242895}
P.~Chen, {\it  Beamstrahlung and the QED, QCD backgrounds in linear colliders},
in 9th International Workshop on Photon-photon Collisions. La
  Jolla, CA, USA, 1992, SLAC-PUB-5914.

\bibitem{Schroeder:216344}
D.~V. Schroeder.
{\it Beamstrahlung and {QED} backgrounds at future linear
  colliders},
 Ph.D. Thesis,  Stanford University, Stanford, CA, 1990.


\bibitem{schulte1996}
D.~Schulte.
{\it Study of Electromagnetic and Hadronic Background in the
  Interaction Region of the TESLA Collider},
 Ph.D. hesis, University of Hamburg, 1996.

\bibitem{sailer}
A.~Sailer.
{\it Studies on the Measurement of Differential Luminosity using
  Bhabha Events at the International Linear Collider},
Diploma thesis, Humboldt-Universit\"at zu Berlin, 2009.

\bibitem{LCD-2011-040}
S.~Poss and A.~Sailer.
{\it Differential luminosity measurement using Bhabha events},
LCD-Note http://edms.cern.ch/document/1175011 {2011-040}, CERN, 2011.

\bibitem{moenig:DiffLumi}
K.~M\"onig,
{\it Measurement of the differential luminosity using Bhabha events in
  the forward tracking region at TESLA}, LC-PHSM-2000-060, 2000.

%\bibitem{Jadach:309289}
%S.~Jadach, W.~Placzek, and B.~F.~L. Ward.
%{\it  {BHWIDE} 1.00: $\mathcal{O}(\alpha)$ {YFS} exponentiated {M}onte
%  {C}arlo for {B}habha scattering at wide angles for {LEP1}/{SLC} and {LEP2}},
%Tech. rep., CERN, Geneva, UTHEP-95-1001, oai:cds.cern.ch:309289, hep-ph/9608412, 1996.

%%%%%%%%%%%%%%%%%%%%%%%%%%%%%%%%%%%%%%%%%%%%%%%%%%%%%%%%%%%%%%%%%%%%%%%%%%%%%%%%


%%%%%%%%%%%%%%%%%%%%%%%%%%%%%%%%  BACKGROUND  %%%%%%%%%%%%%%%%%%%%%%%%%%%%%%%%%%%%%%%%%%%%%%%%%%%%%%

%\bibitem{Abr10}
%H.~Abramowicz et~al., {\it Forward instrumentation for {ILC} detectors},  {\em
 % Journal of Instrumentation} {\bf 5} (2010) P12002.

\bibitem{Opal00}
{OPAL Collaboration}, {\it {{Precision luminosity for Z0 line shape
  measurements with a silicon tungsten calorimeter}}},  {\em Eur.Phys.J.} {\bf
  C~14} (2000) 373,
  [\href{http://xxx.lanl.gov/abs/{arXiv}:hep-ex/9910066}{{\tt
  {arXiv}:hep-ex/9910066}}]. CERN-EP-99-136.

\bibitem{CLIC}
M.~Aicheler, P.~Burrows, M.~Draper, T.~Garvey, P.~Lebrun, K.~Peach, N.~Phinney,
  H.~Schmickler, D.~Schulte, and N.~Toge, eds., {\em {A Multi-TeV Linear
  Collider based on CLIC Technology: CLIC Conceptual Design Report}}.
CERN-2012-007,  CERN, 2012.

\bibitem{ILC}
J.~Brau, Y.~Okada, and N.~Walker, eds., {\em {International Linear Collider -
  Reference Design Report}}, vol.~1: {Executive Summary}.
\newblock {ILC} Global Design Effort and {World} Wide Study, 2007.

\bibitem{Yok91}
K.~Yokoya and P.~Chen, {\it Beam-beam phenomena in linear colliders},  in {\em
  Frontiers of Particle Beams: Intensity Limitations}, vol.~400 of {\em Lecture
  Notes in Physics}, pp.~415--445, Springer Verlag, 1991.
\newblock US-CERN School on Particle Accelerators.

\bibitem{Luk12b}
S.~Luki\'c and I.~Smiljani\'c, {\it Correction of beam-beam effects in
  luminosity measurement at {ILC}},
  \href{http://xxx.lanl.gov/abs/{arXiv}:1211.6869}{{\tt {arXiv}:1211.6869}}.

\bibitem{Luk12}
S.~Luki\'c, {\it {Correction of systematic uncertainties due to beam-beam
  effects in luminosity measurement in LumiCal at CLIC}},  JINST {\bf 8} (2013) P050008.

%\bibitem{Sch96}
%D.~Schulte, {\em Study of Electromagnetic and Hadronic Background in the
 % Interaction Region of the TESLA Collider}.
%\newblock PhD thesis, Hamburg University, 1996.

\bibitem{Jad97}
S.~Jadach, W.~Placzek, E.~Richter-Was, B.~Ward, and Z.~Was, {\it Upgrade of the
  {Monte Carlo} program {BHLUMI} for {Bhabha} scattering at low angles to
  version 4.04},  {\em Computer Physics Communications} {\bf 102} (1997) 229.

\bibitem{Mila13}
M.~Pandurovi\'c and I.~Bo\v{z}ovi{\'c}-Jelisav\v{c}i{\'c}, {\it Physics
  background at {ILC} at {500~GeV} and {1~TeV}},
  \href{http://xxx.lanl.gov/abs/{arXiv}:1301.2494}{{\tt {arXiv}:1301.2494}}. LC
  note under review PREL-LC-DET-2012-011.

\bibitem{Arb96}
A.~Arbuzov et~al., {\it The present theoretical error on the {Bhabha}
  scattering cross section in the luminometry region at {LEP}},  {\em Physics
  Letters B} {\bf 383} (1996) 238 -- 242.

%%%%%%%%%%%%%%%%%%%%%%%%%%%%%%%%%%%%%%%%%%%%%%%%%%%%%%%%%%%%%%%%%%%%%%%%%%%%%%%%%%%%%%

%%%%%%%%%%%%%%%%%%%%%%%%%%%%%%%  FUTURE  %%%%%%%%%%%%%%%%%%%%%%%%%%%%%%%%%%%%%%%%%%%%%%%%

\bibitem{NIEL}
A. Vasilescu and G. Lindstroem, {\it Displacement Damage in Silicon, On-Line
Compilation}, http://hepweb03.phys.sinica.edu.tw/opto/Irradiation/Documents/NIEL$_{-}$scaling/gunnar.htm,
August, 2000.

\bibitem{summers}
G. P. Summers et al., {\it Damage Correlations in Semiconductors Exposed to
Gamma, Electron, and Proton Radiations}, IEEE Trans. on Nucl. Sci. 40, 1372 (1993).

%%%%%%%%%%%%%%%%%%%%%%%%%%%%%%%%%%%%%%%%%%%%%%%%%%%%%%%%%%%%%%%%%%%%%%%%%%%%%%%%%%%%%%%%
\end{thebibliography}
\end{document}